\documentclass[aps,
    prd,
    amssymb,
    amsmath,
    amsfonts,
    eqsecnum,
    showpacs,
    nofootinbib,
    floatfix,
    twocolumn,
    twoside,
    a4paper,
    superscriptaddress,
    longbibliography,
]{revtex4-1}

\usepackage{graphicx}\usepackage{dcolumn}\usepackage{bm}\usepackage[colorlinks=true, linkcolor=blue, urlcolor=blue, citecolor=blue]{hyperref}\usepackage{orcidlink}
\usepackage{breqn}
\usepackage{subcaption}
\usepackage{xspace}
\usepackage{booktabs}
\usepackage{multirow}
\usepackage{rotating}

\usepackage[normalem]{ulem}

\graphicspath{{figures/}}

\newcommand{\phTE}{\textsc{IMRPhenomTEHM}\xspace}
\newcommand{\phTHM}{\textsc{IMRPhenomTHM}\xspace}

\newcommand{\phXHM}{\textsc{IMRPhenomXHM}\xspace}

\newcommand{\phenomxpy}{\textsc{Phenomxpy}\xspace}
\newcommand{\scipy}{\textsc{SciPy}\xspace}
\newcommand{\cupy}{\textsc{CuPy}\xspace}

\newcommand{\bilby}{\textsc{Bilby}\xspace}
\newcommand{\ptemcee}{\textsc{Ptemcee}\xspace}

\newcommand{\UIB}{
Departament de F\'isica, Universitat de les Illes Balears,
IAC3, \\ Carretera Valldemossa km 7.5, E-07122 Palma, Spain
}
\newcommand{\ICE}{Institut de Ci\`encies de l'Espai (ICE, CSIC), 
Campus UAB, \\ Carrer de Can Magrans s/n, 08193 Cerdanyola del Vall\`es, Spain
}

\begin{document}

\title{Accelerating the time-domain LISA response model \\ with central finite differences and hybridization techniques}

\author{Jorge Valencia\,\orcidlink{0000-0003-2648-9759}}
\email{jorge.valencia@uib.es}
\affiliation{\UIB}

\author{Sascha Husa\,\orcidlink{0000-0002-0445-1971}}
\affiliation{\ICE}
\affiliation{\UIB}

\date{\today}

\begin{abstract}
    Accurate and efficient modeling of the Laser Interferometer Space Antenna (LISA) response is crucial for gravitational-wave (GW) data analysis. 
    A key computational challenge lies in evaluating time-delay interferometry (TDI) variables, which require projecting GW polarizations onto the LISA arms at different retarded times. 
    Without approximations, the full LISA response is computationally expensive, and traditional approaches, such as the long-wavelength approximation, accelerate the response calculation at the cost of reducing accuracy at high frequencies. In this work, we introduce a novel hybrid time-domain response for LISA that balances computational efficiency and accuracy across the binary's evolution.
    Our method is applicable to massive black hole binaries and implements a fast low-frequency approximation during the early inspiral—where most of these binaries spend most of the time in the sensitive frequency band of LISA—while reserving the computationally intensive full-response calculations for the late inspiral, merger, and ringdown phases. The low-frequency approximation (LFA) is based on Taylor expanding the response quantities around a chosen evaluation time such that time delays correspond to central finite differences.
    Our hybrid approach supports CPU and GPU implementations, TDI generations 1.5 and 2.0, and flexible time-delay complexity, and has the potential to accelerate parts of the global fit and reduce energy consumption. We also test our LFA and hybrid responses on eccentric binaries, and we perform   
    parameter estimation for a ``golden'' binary.
    Additionally, we assess the efficacy of our low-frequency response for ``deep alerts'' by performing inspiral-only Bayesian inference.
\end{abstract}

\maketitle

\tableofcontents

\section{Introduction}\label{sec:introduction}

The Laser Interferometer Space Antenna (LISA)~\cite{LISA,redbook} is a planned space-borne gravitational-wave (GW) observatory that will operate in the millihertz frequency band ($0.1\,\rm{mHz}\,–\,1\,\rm{Hz}$), a regime inaccessible to current ground-based detectors such as Advanced LIGO~\cite{LIGOScientific:2014pky}, Advanced Virgo~\cite{VIRGO:2014yos}, and KAGRA~\cite{KAGRA:2018plz},
which are limited by terrestrial noise sources such as seismic and Newtonian noise at low frequencies~\cite{seismic_noise, PhysRevD.30.732, PhysRevD.58.122002}. 
While ground-based observatories have successfully detected GWs from stellar-mass compact binary coalescences (CBCs) in the $10\,\rm{Hz}\, – \,1\,\rm{kHz}$ range~\cite{gwtc1,gwtc2,gwtc2.1,gwtc3}, LISA will observe an entirely different population of sources, including massive black hole binaries (MBHBs), extreme-mass-ratio inspirals, galactic binaries, and stochastic backgrounds (see~\cite{LISA_astro_paper} for details).
Overall LISA will present challenges in instrumentation~\cite{LISA_instrumentation, PhysRevD.107.083019, Armano:2024fvz, Armano:2024bvh, PhysRevLett.126.131103, PhysRevD.106.082001, PhysRevD.99.082001, PhysRevD.99.122003}, waveform modeling~\cite{LISA_wav_white_paper}, and data analysis (see e.g.~\cite{Deng:2025wgk, Katz:2024oqg, Littenberg:2023xpl, Strub:2024kbe, Vallisneri:2008ye,PhysRevD.100.022003,Burke:2025bun}).

The fundamental differences between LISA and ground-based detectors 
go beyond their target frequency bands. 
A key difference arises from the detector response. 
For current ground-based interferometers, the response is given by a linear combination of the two GW polarizations, 
$h_{+}$ and $h_{\times}$, weighted by antenna pattern functions. 
For CBCs and the current generation of detectors, these patterns are typically treated as static and depend only on the sky location and the polarization angle. 
For LISA, the detector response to GWs consists of two steps: an initial projection of the GW polarizations onto the time-evolving constellation along the six links connecting the spacecraft (accounting for both forward and backward laser propagation), followed by the construction of time-delay interferometry (TDI)~\cite{PhysRevD.59.102003,PhysRevD.65.082003,Tinto2014} observables. 
TDI is a noise reduction algorithm that combines the six individual link projections with appropriate time shifts to reduce the otherwise overwhelming laser-frequency noise.
Depending on the noise-reduction capabilities and the time dependence of the LISA armlengths, there are different TDI generations, in particular 1.5 (or first generation) and 2.0 (or second generation), but also more general techniques such as TDI-$\infty$~\cite{PhysRevD.103.082001}. 
Both the LISA response to GWs and noise sources vary according to these TDI generations. 
However, the choice of TDI algorithm should not modify the probability of finding an astrophysical signal 
in the data, which leads to an equivalence (up to their noise-canceling property) between 1.5 and 2.0 TDI, as we will further discuss later on.
Ultimately, the LISA GW response can be expressed as a linear combination of time-shifted $h_{+}$ and $h_{\times}$ weighted by antenna patterns that depend on time, sky position, and polarization angle. 

As opposed to ground-based observatories, with sparse-in-time GW detections, we expect the data measured by LISA to be signal-dominated.
This means that there might not be time segments free of detectable GW activity, and GW signals will overlap both in time and frequency.
This complex analysis will require a global fit across all parameters describing the noise and the waveform models used to reproduce the GW signals buried within the data stream.
Several studies~\cite{Deng:2025wgk, Katz:2024oqg, Littenberg:2023xpl, Strub:2024kbe, Vallisneri:2008ye} 
have developed global-fit pipelines which have been tested on the LISA Data Challenge (LDC) 
datasets~\cite{first_mock_data_challenges,spritz_dataset,second_mock_data_challenges,sangria_dataset,sangria}. These datasets are periodically updated to incorporate a broader range of source types and to progressively increase the complexity of the data stream.
Such upgrades are intended to better approximate realistic LISA observations by including features such as data gaps, non-stationary and non-Gaussian noise, and instrumental glitches.

To accelerate LISA parts of data-analysis pipelines, various simplifying assumptions are commonly adopted in the literature. 
In terms of time-delay modeling, one sometimes assumes equal and constant delays between spacecraft, especially in frequency-domain codes~\cite{PhysRevD.103.083011, Deng:2025wgk, deng2025fastdetectionreconstructionmerging, Toubiana:2020cqv}, instead of accounting for the more accurate unequal and time-dependent delays, already implemented in time-domain codes~\cite{lisa_gw_response, lisa_data_challenge_working_group_2022_7332221, PhysRevD.106.103001}.
From the detector response perspective, some studies~\cite{Deng:2025wgk, deng2025fastdetectionreconstructionmerging} adopt the long-wavelength approximation (LWA),
which consists of ignoring the light travel time between spacecraft when the GW wavelength is 
significantly longer than the detector's armlength \mbox{$L=2.5\times 10^9\,\rm{m}$}.
In this regime, inter-spacecraft time delays become perturbatively small, 
allowing time-delayed quantities to be systematically expanded in terms of time derivatives 
(see e.g.~\cite{Armstrong_1999, Babak:2021mhe}).
The choice of the expansion point can affect the accuracy of the method, 
as we will discuss later on the paper.
Overall, independently of the expansion point, this approximation leads to a significant 
simplification of the TDI expressions and accelerates the 
projection process compared to the full response.
However, since LISA will not operate exclusively in this limit, this approximation loses accuracy 
toward the late inspiral, merger, and ringdown phases.
Therefore, although valid for inspiral-only Bayesian inference analysis, 
employing this response for a full inspiral-merger-ringdown study 
will lead to biases in the recovery of the astrophysical 
parameters~\cite{PhysRevD.70.082002, AlbertoVecchio_low_freq}.

In this work, we present a hybrid time-domain LISA response model that employs a modified 
low-frequency approximation based on the LWA during the early inspiral phase, 
and transitions smoothly to the full response
for the late inspiral, merger, and ringdown. 
The proposed algorithm benefits from the speedup of low-frequency/long-wavelength approximations 
during the early inspiral, 
where most binaries spend the majority of their evolution in the sensitive frequency band of LISA, 
while reserving the more computationally costly full-response calculations 
for the shorter-duration, high-frequency portion of the signal.
Our modified low-frequency approximation (LFA) exploits the inherent structure of time-delayed 
measurements in the LISA response to emulate a finite difference scheme.
In particular, we employ central finite differences,
which leads to explicit analytical TDI expressions in terms of derivatives of $h_+$ and $h_\times$, that are able to justify up-to-now empirical \emph{ad hoc} modifications of the low-frequency response introduced in previous studies~\cite{Deng:2025wgk, deng2025fastdetectionreconstructionmerging}.
As by-products of this work we have explored the equivalence between first- and second-generation TDI observables in the limit of low-frequencies, which explains observations previously reported in~\cite{Garcia-Quiros:2025usi, PhysRevD.111.044039}.

We will also briefly discuss the equivalence between different response descriptions from a general point of view. Already for ground-based detectors, one can ask whether e.g. parameter estimation could be performed with an alternative description of the GW signal, and~\cite{PhysRevX.13.041048} discusses the sense in which replacing the GW strain by the Newman-Penrose scalar $\psi_4$~\cite{Newman1962, PhysRevD.59.124022}, which corresponds to two-time derivatives of the strain, leads to equivalent results. 
In App.~\ref{sec:app:lisa_vs_groundbased} we derive the key results of~\cite{PhysRevX.13.041048} in a simpler way and relate them to the difference between the standard ground-based analysis based on the strain and an analysis based on TDI variables. 
We also compare the low-frequency response
for LISA and current ground-based detectors, to gain intuitive understanding regarding the relative weighting of waveform content in different frequency regimes.

This paper is organized as follows.
In Sec.~\ref{sec:full_response} we present the full LISA response in the time domain. 
The low-frequency approximation is discussed in 
Sec.~\ref{sec:lwa_response}, and our hybrid response construction is presented in
Sec.~\ref{sec:hybrid_response}.
The accuracy and computational efficiency of our response implementations are tested in
Sec.~\ref{sec:response_model_validation},
and comparisons for parameter estimation (PE) results are present in
Sec.~\ref{sec:parameter_estimation_studies}.
We conclude with a discussion of our results and the implications for future work in
Sec.~\ref{sec:discussion}.

In App.~\ref{sec:app:fourier_conventions} we briefly present the Fourier conventions followed in this paper.
In App.~\ref{sec:app:expansions} we present the general Taylor expansions for the time-domain response quantities along with their low-frequency limit.
The agreement of the results obtained in this part with their Fourier-domain version~\cite{Marsat:2018oam,PhysRevD.103.083011} is shown
in App.~\ref{sec:app:comparison_with_fd_response}.
We compare the response of LISA and ground-based detectors in the low-frequency regime in App.~\ref{sec:app:lisa_vs_groundbased}.
In App.~\ref{sec:app:first_vs_second_tdi} we show the approximate equivalence between 1.5 and 2.0 TDI whitened waveforms.
Finally, in App.~\ref{sec:app:non_monochromatic inspiral} we show the performance of our low-frequency and hybrid responses on eccentric binaries.

\section{Full response}\label{sec:full_response}

\begin{figure}[htb]
    \centering
    \includegraphics[width=\columnwidth]{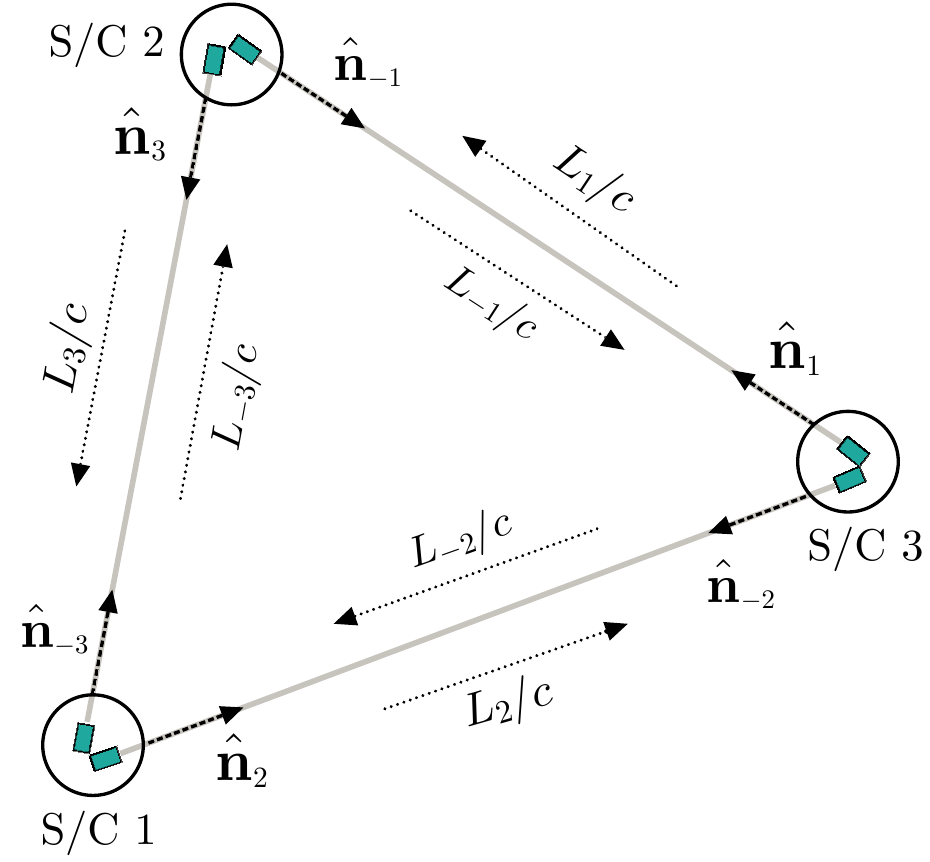}
    \caption{Notation and indexing convention.}
    \label{fig:lisa_sc_conventions}
\end{figure}

This section introduces the instrumental LISA response to GWs \cite{Rubbo:2003ap, Cornish:2001bb, Marsat:2018oam, PhysRevD.103.083011, PhysRevD.106.103001, PhysRevD.67.022001}. 
We follow the conventions proposed in the LISA Data Challenge (LDC) \cite{sangria, Babak:2021mhe} and illustrated in Fig.~\ref{fig:lisa_sc_conventions}, 
and the code infrastructure presented in \cite{Garcia-Quiros:2025usi}.

\subsection{GW projection}

The Solar System Barycenter (SSB) serves as the reference frame for representing the LISA spacecraft (S/C) positions and the sky localization of astrophysical sources.
In this frame, standard spherical coordinates $(\theta,\phi)$ are related to the ecliptic latitude $\beta = \pi/2 - \theta$ and the ecliptic longitude $\lambda=\phi$, which parametrize the source's position. 

We introduce the reference polarization unit vectors, $\hat{\mathbf{u}}$ and $\hat{\mathbf{v}}$, and the GW propagation unit vector $\hat{\mathbf{k}}$ (pointing away from the source), which are expressed in terms of the spherical orthonormal basis $(\mathbf{\hat{e}}_{r}, \mathbf{\hat{e}}_{\phi}, \mathbf{\hat{e}}_{\theta})$ as
\begin{subequations}
    \begin{align}
        \mathbf{\hat{k}} & = -\mathbf{\hat{e}}_{r} =  (-\cos\beta \cos\lambda, -\cos\beta \sin\lambda, \sin\beta)\, , \\
        \mathbf{\hat{u}} & = -\mathbf{\hat{e}}_{\phi} =  (\sin\lambda, \cos\lambda, 0)\, , \\
        \mathbf{\hat{v}} & = -\mathbf{\hat{e}}_{\theta} =  (-\sin\beta \cos\lambda, -\sin\beta \sin\lambda, \cos\beta)\, ,
    \end{align}
\end{subequations}
so that $(\hat{\mathbf{u}}, \hat{\mathbf{v}}, \hat{\mathbf{k}})$ is also an orthonormal triad.

Each pair of spacecraft is connected by two laser beams propagating in opposite directions, forming six links. The response for each link is computed separately and denoted as $y_{slr}$. Here \mbox{$s,r\in\{1,2,3\}$} represent the sender ($s$) and the receiver ($r$) spacecraft, and \mbox{$l\in\{\pm1,\pm2,\pm3\}$} labels the links. We adopt the sign convention for $l$ from \cite{PhysRevD.71.022001}, where $\text{sgn}(l)=\epsilon_{s|l|r}$.

The projection of the GW polarizations into the link $l$ of the constellation is
\begin{equation}\label{eq:H}
    H_l(t) = \xi_l^{+}(t)h_{+}^{\rm{SSB}}(t) + \xi_l^{\times}(t)h_{\times}^{\rm{SSB}}(t)\, ,
\end{equation}
where
\begin{subequations}
    \begin{align}
        h_{+}^{\rm{SSB}}(t) & = h_{+}(t)\cos(2\psi) - h_{\times}(t)\sin(2\psi)\, , \label{eq:hplus_ssb}\\
        h_{\times}^{\rm{SSB}}(t) & = h_{+}(t)\sin(2\psi) + h_{\times}(t)\cos(2\psi)\, ,\label{eq:hcross_ssb}
    \end{align}
\end{subequations}
with $\psi$ the polarization angle and $h_{+,\times}$ modeled in the source frame. The antenna patterns in Eq.~\eqref{eq:H} are given by (see e.g. \cite{PhysRevD.106.103001})
\begin{subequations}
    \begin{align}
        \xi_l^{+}(t) & = (\mathbf{\hat{u}}\cdot\mathbf{\hat{n}}_l(t))^2 - (\mathbf{\hat{v}}\cdot\mathbf{\hat{n}}_l(t))^2\, , \label{eq:xi_plus}\\
        \xi_l^{\times}(t) & = 2(\mathbf{\hat{u}}\cdot\mathbf{\hat{n}}_l(t))(\mathbf{\hat{v}}\cdot\mathbf{\hat{n}}_l(t))\, , \label{eq:xi_cross}
    \end{align}
\end{subequations}
where $\mathbf{\hat{n}}_l(t)$ is the link $l$ unit vector (dashed
arrows in Fig.~\ref{fig:lisa_sc_conventions}).

\subsection{Time-delay interferometry}

The LISA response for the link $l$ represents the laser's relative frequency shift after traveling from S/C $s$ to $r$ and can be expressed as (see, for instance,~\cite{Babak:2021mhe,PhysRevD.106.103001})
\begin{equation}\label{eq:yslr}
     y_{slr}(t) = \frac{H_l(t - \delta_s(t)) - H_l(t-\delta_r(t))}{2(1-\mathbf{\hat{k}}\cdot\mathbf{\hat{n}}_l(t))}\, ,
\end{equation}
where $t$ is the laser's reception time as measured in the receiver's S/C frame. 
The corresponding emission and reception times at the SSB are given by $t-\delta_s(t)$ and $t-\delta_r(t)$, respectively.
Here, $\delta_s(t)$ and $\delta_r(t)$ denote the propagation delays associated with the sender and receiver S/C, and can be approximated as (see e.g. Sec.~IV~B of~\cite{PhysRevD.106.103001})
\begin{subequations}
    \begin{align}
        \delta_s(t) & = \frac{L_l(t)}{c}+\frac{\mathbf{\hat{k}}\cdot\mathbf{R}_s(t)}{c}\, ,\label{eq:delta_s}\\
        \delta_r(t) & = \frac{\mathbf{\hat{k}}\cdot\mathbf{R}_r(t)}{c}\, .\label{eq:delta_r}
    \end{align}
\end{subequations}
The positions of the sender and receiver spacecraft relative to the SSB are given by $\mathbf{R}_s(t)$ and $\mathbf{R}_r(t)$, $L_l(t)$ denotes the laser's optical path along the link $l$, $L_l(t)/c$ is the inter-spacecraft light travel time and $c$ is the speed of light.

Several previous works have explored the impact of realistic ESA orbits~\cite{PhysRevD.106.103001, Garcia-Quiros:2025usi,inference_tdi_inf} in LISA data analysis. Here we adopt an equal-arm-length configuration and compute the S/C positions with the class \texttt{AnalyticOrbit} from the LDC software tools \cite{lisa_data_challenge_working_group_2022_7332221}. 
Note that even with this orbital configuration, the rotation of LISA makes the light travel time time-dependent and, in general, \mbox{$L_l(t)\neq L_{-l}(t)$~\cite{Neil_Cornish_2003, PhysRevD.69.022001, PhysRevD.69.082001,PhysRevD.71.022001}}.

The single-link measurements of Eq.~\eqref{eq:yslr} are orders of magnitude 
below laser noise. 
TDI techniques mitigate this noise by linearly combining the readouts at 
different times~\cite{PhysRevD.59.102003, Armstrong_1999, PhysRevD.62.042002, PhysRevD.65.102002, PhysRevD.69.022001, TintoLRR, PhysRevD.68.061303, Neil_Cornish_2003, PhysRevD.69.082001}. 
Several choices are possible for such a new set of TDI observables, 
and the response to GWs and the remaining noise sources depends on this 
choice~\cite{PhysRevD.71.022001}.
Commonly used combinations are designed to synthesize interferometric 
configurations analogous to Sagnac or Michelson interferometers. 
Sagnac-type observables emulate the interference pattern generated by photons 
propagating in clockwise and counterclockwise directions around the 
constellation. 
In contrast, Michelson-type observables construct a virtual Michelson 
interferometer configuration, effectively suppressing laser frequency noise in 
a 3-laser configuration.
In this work, we restrict our analysis to Michelson-type TDI variables.

Setting $c=1$ and dropping the time dependence on the delays for simplicity, 
the 1.5-generation Michelson combination $X_{1.5}$ is given by (see e.g.~\cite{Babak:2021mhe})
\begin{equation}\label{eq:X1.5}
    \begin{aligned}
        X_{1.5}(t) & = y_{1-32}(t-L_{-2}-L_2-L_3) \\
        & - y_{1-32}(t-L_3) \\
        & + y_{231}(t-L_2-L_{-2}) \\
        & - y_{231}(t) \\
        & + y_{123}(t-L_{-2}) \\
        & - y_{123}(t-L_{-2}-L_{-3}-L_{3}) \\
        & + y_{3-21}(t) \\
        & - y_{3-21}(t-L_{-3}-L_{3})\, , \\
    \end{aligned}
\end{equation}
and the 2.0-generation combination $X_{2.0}$ by (see e.g.~\cite{Babak:2021mhe})
\begin{equation}\label{eq:X2.0}
    \begin{aligned}
        X_{2.0}(t) & = X_{1.5}(t) \\& + y_{1-32}(t-L_{-2}-L_2-2L_3-L_{-3}) \\
        & - y_{1-32}(t-2L_3-L_{-3}-2L_{-2}-2L_2) \\
& + y_{231}(t-L_{-2}-L_2-L_3-L_{-3}) \\
        & - y_{231}(t-L_3-L_{-3}-2L_{-2}-2L_2) \\
& + y_{123}(t-2L_{-2}-L_2-2L_3-2L_{-3}) \\ & - y_{123}(t-L_3-L_{-3}-2L_{-2}-L_2) \\
& + y_{3-21}(t-L_{-2}-L_2-2L_3-2L_{-3}) \\
        & - y_{3-21}(t-L_3-L_{-3}-L_{-2}-L_2)\, .
    \end{aligned}
\end{equation}
We evaluate each of the delays $L_l$ (dotted arrows in Fig.~\ref{fig:lisa_sc_conventions}) in Eqs.~\eqref{eq:X1.5} and~\eqref{eq:X2.0} at time $t$, since we do not expect nested delays to affect the GW projection~\cite{PhysRevD.106.103001}. The variables $Y$ and $Z$ are computed by cyclic permutation of the indices.

The $X, Y, Z$ observables have correlated noise and have to be combined linearly to obtain the ``noise-uncorrelated'' $A, E, T$ variables\footnote{The variables $A, E, T$ can be treated as orthogonal for equal-arm-length configurations. However, in realistic orbit scenarios, these channels are not perfectly orthogonal and have correlated noise.}~\cite{PhysRevD.66.122002}:
\begin{subequations}
    \begin{align}
        A(t) &= \frac{1}{\sqrt{2}}[Z(t) - X(t)]\, , \\
        E(t) &= \frac{1}{\sqrt{6}}[X(t) - 2Y(t) + Z(t)]\, , \\
        T(t) &= \frac{1}{\sqrt{3}}[X(t) + Y(t) + Z(t)]\, .
    \end{align}
\end{subequations}

\subsection{Noise model}\label{subsec:noise_model}

We employ the analytical noise model developed for the Sangria dataset, 
the second round of the LISA Data Challenge (LDC2a)~\cite{sangria,sangria_dataset}, 
as detailed in \cite{lisa_data_challenge_working_group_2022_7332221}.
The PSDs are computed with the Python class \texttt{AnalyticNoise} from the LDC software~\cite{lisa_data_challenge_working_group_2022_7332221}.

The proof-mass, $S_{\rm{pm}}$, and the optical-path, $S_{\rm{op}}$, noises are modeled by
\begin{subequations}
    \begin{align}
        \begin{split}
            S_{\rm{pm}}(f) &= S_{\rm{acc}}\left(2\pi f c\right)^{-2}\left[1+\left(\frac{0.0004\,\rm{Hz}}{f}\right)^2\right]  \\
            & \times \left[1+\left(\frac{f}{0.008\,\rm{Hz}}\right)^4\right]\, ,
        \end{split}\\
        \begin{split}
            S_{\rm{op}}(f) &= S_{\rm{oms}}\left(\frac{2 \pi f}{c}\right)^2 
            \left[1+\left(\frac{0.002\,\rm{Hz}}{f}\right)^4\right]\, ,
        \end{split}
    \end{align}
\end{subequations}
where $\sqrt{S_{\rm{acc}}}=2.4\times 10^{-15} \, \rm{m} \, \rm{s}^{-2} \, \rm{Hz}^{-0.5}$ and $\sqrt{S_{\rm{oms}}}=7.9\times 10^{-12} \, \rm{m} \, \rm{Hz}^{-0.5}$. 
For the galactic foreground noise we follow \cite{PhysRevD.104.043019, lisa_data_challenge_working_group_2022_7332221}:
\begin{equation}\label{eq:sgal}
    \begin{aligned}
        S_{\rm{gal}}(f) &= 2x^2\sin^2(x)
        A f^{-7/3}e^{-(f/f_1)^\alpha} \\
        & \times \left\{1+\tanh \left[(f_{\rm{knee}}-f)/f_2\right]\right\}\, ,
    \end{aligned}
\end{equation}
with $x=2\pi f L/c$. 
We use $A=1.28265531\times 10^{-44}$, $\alpha=1.62966700$, and $f_2=4.81078093\times10^{-4}\,\rm{Hz}$~\cite{lisa_data_challenge_working_group_2022_7332221}.
The parameters $f_1$ and $f_{\rm{knee}}$ are given by~\cite{PhysRevD.104.043019}
\begin{subequations}
    \begin{align}
        \log_{10}(f_1)& =a_1\log_{10}(T_{\rm{obs}})+b_1\, ,\\
        \log_{10}(f_{\rm{knee}})& = a_k\log_{10}(T_{\rm{obs}})+b_k\, .
    \end{align}
\end{subequations}
We use $(a_1,b_1)=(-0.223499956, -2.70408439)$, $(a_k,b_k)=(-0.360976122, -2.37822436)$ and $T_{\rm{obs}}=1\,\rm{year}$, as in \cite{lisa_data_challenge_working_group_2022_7332221}. 

\begin{figure}[htb]
    \centering
    \includegraphics[width=\columnwidth]{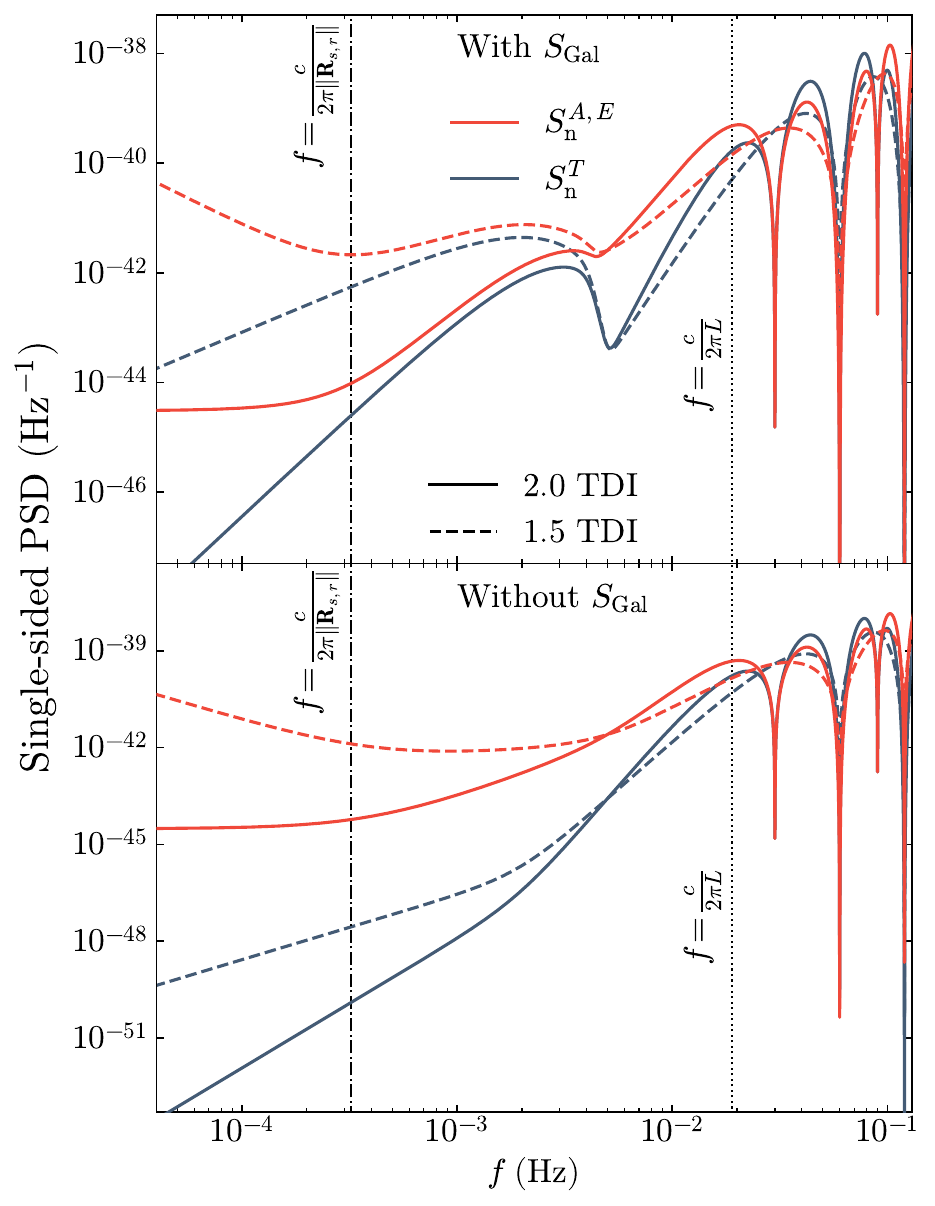}
    \caption{Single-sided analytical PSDs for the $A$, $E$ (red), and $T$ (gray) channels for 1.5 TDI (dashed) and 2.0 TDI (solid). The upper panel includes the contribution from the galactic foreground noise $S_{\rm{Gal}}$, as defined in Eq.~\eqref{eq:sgal}, whereas the lower panel excludes this contribution. The vertical lines represent the reference frequencies \mbox{$f=c/(2\pi L)=0.019\,\rm{Hz}$} (dotted) and \mbox{$f= c/(2\pi \|\mathbf{R}_{s,r}\|)= 3.2\times 10^{-4}\,\rm{Hz}$} (dash-dotted).}
    \label{fig:aet_psds}
\end{figure}

For equal-arm-length orbits, the PSDs for the $A,E,T$ channels are determined from the proof-mass, optical-path, and galactic foreground noises as \cite{lisa_data_challenge_working_group_2022_7332221}
\begin{subequations}
    \begin{align}
        \begin{split}
            S_{\rm{n}}^{A,E}(f) &= 8\Lambda(f)\sin^2(x)\left\{\vphantom{\sin^4\left(\frac{x}{2}\right)} 2\left[3+2\cos(x) \right. \right. \\
            & + \left. \left. \cos(2x)\right]S_{\rm{pm}}(f) + \left[2+\cos(x)\right]S_{\rm{op}}(f) \vphantom{\sin^4\left(\frac{x}{2}\right)}\right\} \\
            & + (3/2)\Lambda(f)S_{\rm{Gal}}(f)\, , \label{eq:SnAE}
        \end{split}\\
        \begin{split}
            S_{\rm{n}}^{T}(f) &= 16\Lambda(f)\sin^2(x)\left\{\vphantom{\sin^4\left(\frac{x}{2}\right)} \left[1-\cos(x)\right]S_{\rm{op}}(f)\right.\\
            & + \left. 8 \sin^4(x/2)S_{\rm{pm}}(f)\vphantom{\sin^4\left(\frac{x}{2}\right)}\right\} + \Lambda(f)S_{\rm{Gal}}(f)\, , \label{eq:SnT}
        \end{split}
    \end{align}
\end{subequations}
where $\Lambda(f)=4\sin^2(2x)$ for 2.0 TDI and $\Lambda(f)=1$ for 1.5 TDI.
See Fig.~\ref{fig:aet_psds} for a visual comparison between PSDs 
with different TDI configurations, both with (upper panel) and without  
(lower panel) the contribution from the galactic foreground $S_{\rm{Gal}}$.

At low frequencies, the power spectral densities given by Eqs.~\eqref{eq:SnAE} and~\eqref{eq:SnT}, 
and shown in Fig.~\ref{fig:aet_psds},
exhibit distinct power-law behavior for different generations of TDI.
For the 2.0-generation, \mbox{$S_{\rm{n}}^{A_{2.0},E_{2.0}}\sim c_1f^0 + c_2f^{2}$}, 
whereas for the 1.5-generation, the trend is approximately 
\mbox{$S_{\rm{n}}^{A_{1.5},E_{1.5}}\sim c_3f^{-2} + c_4f^{0}$}, with $c_i\in\mathbb{R}$.
The difference between the two, which scales as $f^2$, provides
insights on the nature of the TDI mechanism, 
as it is discussed in App.~\ref{sec:app:first_vs_second_tdi}, 
where we present a detailed comparison between the two TDI generations.

In contrast to the PSDs from space-based detectors, where the low-frequency sensitivity band is dominated by acceleration noise, the PSDs from ground-based detectors exhibit significantly different power-law behavior. 
This difference arises from the distinct nature of the noise sources.
At low frequencies, the dominant noise contributions come from the seismic and suspension thermal noise~\cite{seismic_noise, PhysRevD.30.732, PhysRevD.58.122002}, which make the PSD from e.g. LIGO detectors follow \mbox{$S_{\rm{n}}^{\rm{LIGO}}\sim c_5f^{-20} + c_6f^{-5}$}.

While briefly introduced here, the implications of the different low-frequency power laws between the PSDs from ground- and space-based detectors will become clearer in App.~\ref{sec:app:lisa_vs_groundbased}. 
\section{Low-frequency response}\label{sec:lwa_response}

The long-wavelength approximation (LWA) holds for frequencies
$f\ll cL^{-1}=0.12\, \rm{Hz}$, where the GW wavelength $\lambda_{GW}$ matches the LISA armlength.
However, as pointed out in \cite{PhysRevD.103.083011}, deviations from this approximation start to be significant at much lower frequencies. This is because evaluating quantities at delayed light-travel times results in terms with phase factors $e^{-2i\pi f L/c}$ in the Fourier domain, which can only be ignored if \mbox{$2\pi f L/c \ll 1$}, that is, for frequencies \mbox{$f\ll c/(2\pi L)=0.019\,\rm{Hz}$} \cite{Marsat:2018oam} (dotted vertical line in Fig.~\ref{fig:aet_psds}). This corresponds to $\lambda_{GW}=2\pi L$ and timescales smaller than \mbox{$1/0.019 \,\rm{s} \sim50\,\rm{s}$} can be safely ignored.
Therefore, retarded time quantities can be approximated as non-retarded quantities through expansions in powers of $L_l/c\sim 8\,\rm{s}$; for example, $H_l(t-L_l/c) \approx H_l(t) - (L_l/c)\dot{H}_l(t)$, where the dot indicates a time derivative.

From Eq.~\eqref{eq:yslr}, calculating $y_{slr}$ requires the evaluation of $H_l$ at the retarded times $\delta_s$ and $\delta_r$. 
The magnitude of these delays is however larger than $\sim 50\,\rm{s}$, since they take into account the light travel time between the SSB and the constellation ($\sim 8 \, \rm{min}$). 
To be able to expand $H_l$ in Eq.~\eqref{eq:yslr} we focus on lower frequencies
such that $f\ll c/(2\pi \|\mathbf{R}_{s,r}\|)= 3.2\times 10^{-4}\,\rm{Hz}$ (dash-dotted vertical line in Fig.~\ref{fig:aet_psds}), where $\|\mathbf{R}_{s,r}\|\approx 1 \, \rm{AU}$. 
Within this frequency range, timescales smaller than $1/(3.2\times 10^{-4})\,\rm{s}\sim 52\,\rm{min}$ can be neglected and retarded time quantities can be expanded in powers of $L_l/c$, $\|\mathbf{R}_{s,r}\|/c$ or a combination of both.

Motivated by the accuracy of central finite differences we expand $H_l(t-\delta_s)$ and $H_l(t-\delta_r)$, in the numerator of Eq.~\eqref{eq:yslr}, around the middle point $t - (\delta_s + \delta_r)/2$. Following this procedure, and combining Eq.~\eqref{eq:yslr} with Eq.~\eqref{eq:H}, the links $y_{slr}$ are, at leading order, a combination of strain derivatives:
\begin{equation}\label{eq:yslr_strain_derivatives}
    \begin{aligned}
        y_{slr}(t) &\approx -\frac{L_l(t)}{2c}\left[\xi_l^{+}(t)\dot{h}_{+}^{\rm{SSB}}\left(t-\frac{\delta_s(t)+\delta_r(t)}{2}\right)\right. \\
        &  + \left. \xi_l^{\times}(t)\dot{h}_{\times}^{\rm{SSB}}\left(t-\frac{\delta_s(t)+\delta_r(t)}{2}\right)\right]\, .
\end{aligned} 
\end{equation}

The complete expansions of $H_l$ and $y_{slr}$, along with the associated errors in the approximations, are
presented in App.~\ref{subsec:app:expansion_yslr_central_differences}. 
In App.~\ref{subsec:app:expansion_yslr_other_differences}, we further provide a comparison between central and other finite-difference schemes, which yield different expressions for the single-link observables (compare Eq.~\eqref{eq:yslr_strain_derivatives} with, e.g., Eq.~(51) of \cite{Babak:2021mhe}, where $\dot{h}_{+,\times}^{\rm{SSB}}$ are evaluated at a different time).

In the long-wavelength approximation, the light travel time between spacecraft $i$ and $j$ can be approximated as identical in 
both directions, i.e. $L_l\simeq L_{-l}$ \cite{PhysRevD.103.083011}.
Consequently, it follows from Eq.~\eqref{eq:yslr_strain_derivatives} that $y_{slr}(t)\simeq y_{r-ls}(t)$, 
since the retarded time $(\delta_s + \delta_r)/2$ remains unchanged 
under the transformation $r\leftrightarrow s$. 
Using this symmetry, Eqs.~\eqref{eq:X1.5} and~\eqref{eq:X2.0} are simplified and only a subset of the $y_{slr}$ variables 
are required to compute $X, Y, Z$. 
As an example, in Eqs.~\eqref{eq:X1.5LWAsameL_main} and~\eqref{eq:X2.0LWAsameL_main}
we provide the expressions for the Michelson variable $X$ for the simplest LISA orbit. 
Assuming equal and constant time delays ($L_i(t)=L_j(t)=L$), the $X$ channel for 1.5 and 2.0 TDI can be approximated as (see Sec.~\ref{sec:app:expansion_xyz_1tdi} and Sec.~\ref{sec:app:expansion_xyz_2tdi} for details)
\begin{equation}\label{eq:X1.5LWAsameL_main}
    X_{1.5}(t) \approx \frac{4L}{c} \left[\dot{y}_{123}\left(t-\frac{3L}{2c}\right)-\dot{y}_{231}\left(t-\frac{3L}{2c}\right)\right]\, ,
\end{equation}
and
\begin{equation}\label{eq:X2.0LWAsameL_main}
    X_{2.0}(t) \approx \frac{16L^2}{c^2} \left[\ddot{y}_{123}\left(t-\frac{7L}{2c}\right) -  \ddot{y}_{231}\left(t-\frac{7L}{2c}\right)\right]\, .
\end{equation}
Since $Y$ and $Z$ are obtained by cyclic permutation of the indices, 
in this frequency range, only $y_{123}$, $y_{231}$, and $y_{312}$ need to be computed.
Substituting Eq.~\eqref{eq:yslr_strain_derivatives} in Eqs.~\eqref{eq:X1.5LWAsameL_main} and~\eqref{eq:X2.0LWAsameL_main}, and assuming that
\begin{equation}
    \frac{\hat{\mathbf{k}}\cdot\mathbf{R}_s(t)+\hat{\mathbf{k}}\cdot\mathbf{R}_r(t)}{2c} \approx \frac{\hat{\mathbf{k}}\cdot\mathbf{R}_0(t)}{c}\, ,
\end{equation}
with $\mathbf{R}_0(t)$ the position of the center of the constellation relative to SSB, we obtain
\begin{equation}\label{eq:X15_with_strain_derivatives}
    \begin{aligned}
         X_{1.5}(t) & \approx \frac{2L^2}{c^2}\left[\vphantom{\left(t-\frac{2L}{c}-\frac{\hat{\mathbf{k}}\cdot\mathbf{R}_0(t)}{c}\right)} \right.\\
         & \left(\xi_3^{+}(t)-\xi_2^{+}(t)\right)\ddot{h}_{+}^{\rm{SSB}}\left(t-\frac{2L}{c}-\frac{\hat{\mathbf{k}}\cdot\mathbf{R}_0(t)}{c}\right) \\
         & \left. + \left(\xi_3^{\times}(t)-\xi_2^{\times}(t)\right)\ddot{h}_{\times}^{\rm{SSB}}\left(t-\frac{2L}{c}-\frac{\hat{\mathbf{k}}\cdot\mathbf{R}_0(t)}{c}\right)\right]
\end{aligned}
\end{equation}
and 
\begin{equation}\label{eq:X20_with_strain_derivatives}
    \begin{aligned}
         X_{2.0}(t) & \approx \frac{8L^3}{c^3}\left[\vphantom{\left(t-\frac{4L}{c}-\frac{\hat{\mathbf{k}}\cdot\mathbf{R}_0(t)}{c}\right)} \right.\\
         & \left(\xi_3^{+}(t)-\xi_2^{+}(t)\right)\dddot{h}_{+}^{\rm{SSB}}\left(t-\frac{4L}{c}-\frac{\hat{\mathbf{k}}\cdot\mathbf{R}_0(t)}{c}\right) \\
         & \left. + \left(\xi_3^{\times}(t)-\xi_2^{\times}(t)\right)\dddot{h}_{\times}^{\rm{SSB}}\left(t-\frac{4L}{c}-\frac{\hat{\mathbf{k}}\cdot\mathbf{R}_0(t)}{c}\right)\right]\, .
    \end{aligned}
\end{equation}
The 1.5 TDI Michelson variables are, in this limit, proportional to the second time-derivative of the strain 
(i.e.~the Newman-Penrose scalar, $\psi_4(t)$~\cite{Newman1962, PhysRevD.59.124022}), whereas the 2.0 TDI variables are proportional to the third time derivative ($\dot{\psi}_4(t)$).
In this limit, both TDI generations are related via
\begin{equation}
    X_{2.0}(t) = \frac{4L}{c}\dot{X}_{1.5}\left(t-\frac{2L}{c}\right)\, ,
\end{equation}
which, compared to Eq.~\eqref{eq:X2.0}, instead of 8 extra interpolations to compute $X_{2.0}$ from $X_{1.5}$, we only need 1 interpolation and 1 time derivative.

A key result of this section is the recognition that, 
for both first- and second-generation Michelson combinations, 
the strain derivatives appearing in Eqs.~\eqref{eq:X15_with_strain_derivatives} and~\eqref{eq:X20_with_strain_derivatives} 
are evaluated at retarded times that involve delays of
$2L/c$ and $4L/c$, respectively, rather the nominal $L/c$.
In the Fourier domain, according to Eq.~\eqref{eq:FT_shift_property},  
these time shifts introduce oscillatory factors 
of $e^{-4i\pi f L/c}$ for 1.5 TDI 
and $e^{-8i\pi f L/c}$ for 2.0 TDI, 
rather than the often assumed $e^{-2i\pi f L/c}$. 

This finding provides a theoretical justification for the empirical modifications employed 
(at the time of writing) in~\cite{Deng:2025wgk} and in the first version of~\cite{deng2025fastdetectionreconstructionmerging}\footnote{In the second version of this publication, the authors modified the Fourier-domain response by applying the long-wavelength approximation directly to the single-link measurements. This modification introduced an exponential factor $e^{3i\pi f L/c}$ in the expressions for $\tilde{A}$ and $\tilde{E}$. Within our framework, Eq.~(2) of~\cite{deng2025fastdetectionreconstructionmerging} can be derived to leading order in frequency by substituting the forward-difference expression for $y_{slr}$, given in Eq.~\eqref{eq:yslr_forward}, into Eq.~\eqref{eq:X15_equal_timedelays} and computing $\tilde{A}$ and $\tilde{E}$. If we adopt the same Fourier conventions as in~\cite{deng2025fastdetectionreconstructionmerging}, this derivation shows that the exponential factor $e^{3i\pi f L/c}$ appearing in the frequency-domain response is consistent with the time delay of $3L/(2c)$ in Eq.~\eqref{eq:X15_equal_timedelays}.} 
where the substitution of $e^{-2i\pi f L/c}$ with $e^{-4i\pi f L/c}$ was used in the 1.5-generation TDI
to improve the accuracy of the response at low frequencies. 
The derivation presented here demonstrates that these \textit{ad hoc} delays arise naturally 
from the consistent treatment of the low-frequency LISA response as a central finite difference operator 
acting on the GW polarizations. 

For non-equal and time-dependent delays ($L_i(t)\neq L_j(t)\neq L$), but keeping $L_l(t)\simeq L_{-l}(t)$, the expressions for $X$ are 
given by Eqs.~\eqref{eq:X15_nonequal_timedelays}, for 1.5 TDI, and~\eqref{eq:X20_nonequal_timedelays}, 
for 2.0 TDI.

In our code implementation, to minimize the number of interpolations, we do not directly compute the quantities $y_{slr}(t)$. 
Instead, we evaluate the link measurements at the appropriate retarded time dictated by the Michelson variables. 
For example, for 2.0 TDI and assuming equal and constant delays, we directly compute $y_{slr}(t-7L/(2c))\sim \dot{h}_{+,\times}^{\rm{SSB}}(t-4L/c-\hat{\mathbf{k}}\cdot\mathbf{R}_0(t)/c)$ (see Eq.~\eqref{eq:X2.0LWAsameL_main}), where the interpolation is performed on the first-time derivative of the GW polarizations. Temporal derivatives are evaluated numerically using \texttt{numpy.gradient} from the \textsc{NumPy} Python package~\cite{numpy}, for the CPU implementation; and \texttt{cupy.gradient} from the \textsc{CuPy} Python package~\cite{cupy}, for the GPU implementation.

\section{Hybrid response}\label{sec:hybrid_response}

The low-frequency approximation (LFA), described in Sec.~\ref{sec:lwa_response}, holds for frequencies $f\ll 3.2\times 10^{-4}\, \rm{Hz}$. 
Therefore, depending on the system's total mass, the approximation may lose accuracy during the last stages of the binary evolution. In such cases, 
relying on this approximation could introduce significant inaccuracies in the late inspiral and merger-ringdown 
phase, where the majority of the signal-to-noise ratio (SNR) is accumulated for MBHBs~\cite{PhysRevD.103.083011}. While using the full response for the entire signal would avoid the loss of accuracy, it can be computationally expensive, particularly for low-mass systems.

In this section, we introduce a novel hybrid time-domain approach that adapts the response computation based on the binary's 
evolution stage. We apply the LFA during the inspiral phase, leveraging its 
computational efficiency, while using the full LISA response for the merger-ringdown. This strategy 
preserves the computational speed-up of the LFA during the early inspiral, where most binaries spend most of the time in the sensitive frequency band of LISA, while restricting the more costly full-response calculations to just the merger-ringdown part.

The hybridization is performed at the level of the individual links $y_{slr}$.
As described in Sec.~\ref{sec:lwa_response}, in the LFA only three of the six links are computed, 
while the rest are set to zero.
During the last orbits, the full response is applied, where all the links are required.
We then create the hybrid across a previously defined hybridization window, where the links
computed with the low-frequency response transition to the ones from the full response using a weighted
sum. We compute the weights with a sigmoid function given by
\begin{equation}\label{eq:sigmoid}
    \Sigma(t; t_{\rm{Hyb}}, \sigma, \varepsilon) = \left[1+e^{\varepsilon(t-t_{\rm{Hyb}})/\sigma}\right]^{-1}\, ,
\end{equation}
where $t_{\rm{Hyb}}$ and $\sigma$ refer to the typical starting time and timescale of the sigmoid, and 
$\varepsilon=\pm 1$ accounts for its orientation. 

The hybridization time $t_{\rm{Hyb}}$ is determined at leading 
post-Newtonian (PN) order from (for a textbook reference see e.g.~\cite{Maggiore})
\begin{equation}\label{eq:thyb}
    t_{\rm{Hyb}} = t_c - f_{\rm{Hyb}}^{-8/3}\times\frac{5}{(8\pi)^{8/3}}\left(\frac{G\mathcal{M}_c}{c^3}\right)^{-5/3}\, ,
\end{equation}
where $t_c$ is the coalescence time, the chirp mass of the binary is \mbox{$\mathcal{M}_c=(m_1 m_2)^{3/5}/(m_1+m_2)^{1/5}$}, with $m_{1,2}$ the individual mass components, and $f_{\rm{Hyb}}$ is the hybridization frequency, which is free to be chosen by the user. 
Although Eq.~\eqref{eq:thyb} is strictly valid only within the inspiral regime--where PN approximations are applicable--it nonetheless provides a reasonable estimate of the relative portion of the 
waveform that will be projected using the LFA. This is given by the ratio 
\begin{equation}\label{eq:hybrid_ratio}
   \mathcal{R}(f_{\rm{Hyb}}) = \frac{f_{\rm{min}}^{-8/3} - f_{\rm{Hyb}}^{-8/3}}{f_{\rm{min}}^{-8/3} - f_{\rm{max}}^{-8/3}}\, ,
\end{equation}
where $f_{\rm{Hyb}}\in(f_{\rm{min}},f_{\rm{max}})$, with $f_{\rm{min}}$ and $f_{\rm{max}}$ the minimum and 
maximum frequencies of the waveform. 
For the standard LISA sensitivity settings $f_{\rm{min}}=10^{-4} \,\rm{Hz}$~\cite{redbook}, and $f_{\rm{max}}\sim 0.05\,\rm{Hz}$ for MBHB signals. With these values, $\mathcal{R}(f_{\rm{Hyb}})$ does not 
strongly depend on $f_{\rm{max}}$ since $f_{\rm{min}}\ll f_{\rm{max}}$, and Eq.~\eqref{eq:hybrid_ratio} can be simplified to 
\begin{equation}\label{eq:hybrid_ratio_simplified}
   \mathcal{R}(f_{\rm{Hyb}})\approx 1 - \left(\frac{f_{\rm{Hyb}}}{f_{\rm{min}}}\right)^{-8/3}\, .
\end{equation}
For example, choosing $f_{\rm{Hyb}} = 3\times 10^{-4}\,\rm{Hz}$, and setting $f_{\rm{min}}=10^{-4}\,\rm{Hz}$, 
we obtain \mbox{$\mathcal{R}(f_{\rm{Hyb}})\approx 0.95$}. 
This means that the LFA will be applied to 95\% of the waveform, while only the remaining 5\%, which 
corresponds to the final stages of the binary evolution, will be projected by the full response.
Note that $f_{\rm{Hyb}}$ can be freely chosen. For other values of $f_{\rm{Hyb}}$ and their implications on accuracy and efficiency, we point the reader to Sec.~\ref{subsec:impact_of_hybridization_parameters}.

The hybridization procedure corresponds to
\begin{equation}\label{eq:hybridized_yslr}
    \begin{aligned}
         y_{slr}(t) & = \Sigma(t_{\rm{LFA}}; t_{\rm{Hyb}}, \sigma, +1) \times y_{slr}^{\rm{LFA}}(t_{\rm{LFA}}) \\
         & + \Sigma(t_{\rm{Full}}; t_{\rm{Hyb}}, \sigma, -1) \times y_{slr}^{\rm{Full}}(t_{\rm{Full}})\, ,
    \end{aligned}
\end{equation}
where $t_{\rm{LFA}} \le t_{\rm{Hyb}}+T/2$ and $t_{\rm{Full}} \ge t_{\rm{Hyb}}-T/2$, being $T$ the 
duration of the hybridization window and $t=t_{\rm{LFA}}\cup t_{\rm{Full}}$. 
The Michelson observables are also computed separately as
\begin{equation}\label{eq:hybridized_xs}
    \begin{aligned}
         X(t) & = \Sigma(t_{\rm{LFA}}; t_{\rm{Hyb}}, \sigma, +1) \times X^{\rm{LFA}}(t_{\rm{LFA}}) \\
         & + \Sigma(t_{\rm{Full}}; t_{\rm{Hyb}}, \sigma, -1) \times X^{\rm{Full}}(t_{\rm{Full}})\, . 
    \end{aligned}
\end{equation}
In the low-frequency regime, our implementation allows the freedom to choose either 
equal and constant or unequal time-dependent delays.
Thus, depending on the LISA configuration and the chosen TDI generation, $X^{\rm{LFA}}$ is
computed from the expressions given in in Sec.~\ref{subsubsec:app:lwa_1.5}, for 1.5 TDI, 
or in Sec.~\ref{subsubsec:app:lwa_2.0}, for 2.0 TDI. 
For the rest of the frequency spectrum, the full response is applied, which is always computed for the unequal 
time-dependent delay configuration.
Hence, $X^{\rm{Full}}$ follows Eqs.~\eqref{eq:X1.5} or Eq.~\eqref{eq:X2.0} for 1.5 or 2.0 TDI, 
respectively.

It is important to note that multiplying the 
quantities in the response by sigmoids, as in Eqs.~\eqref{eq:hybridized_yslr} and~\eqref{eq:hybridized_xs}, introduces undesired modulations in the low-frequency regime of the 
Fourier transform of the projected signal. 
These modulations persist up to a characteristic 
frequency of approximately $f\lesssim \sigma^{-1}$ (see~\cite{PhysRevD.110.124026} 
for a detailed frequency-domain analysis of steplike functions). 
Ideally, selecting $\sigma \lesssim f_{\rm{min}}^{-1}=10^{4}\, \rm{s}$ 
would shift the frequency modulations outside the LISA sensitivity band. 
However, such a wide sigmoid requires a very large hybridization window, which leads to an
unphysical loss of spectral amplitude near $f_{\rm{Hyb}}$ and, consequently, a 
decrease of the SNR.

We discuss in detail the impact of the hybridization parameters, $f_{\rm{Hyb}}$, $T$, and $\sigma$, in Sec.~\ref{subsec:impact_of_hybridization_parameters}.
Across the parameter space explored in this study,
we find that $f_{\rm{Hyb}}=3\times 10^{-4}\,\rm{Hz}$, $T=800\, \delta t$, and $\sigma = T/40$, with $\delta t$ the time step of the waveform and the response, 
yields a sufficiently smooth hybrid in the time domain. 
In the Fourier domain, these settings constrain the loss of spectral amplitude (compared to the full response) below 0.1\%. 
Lower values of $\sigma$ produce sharper transitions between the LFA and the full response, 
potentially constructing non-smooth hybrids that can introduce artifacts in the frequency domain.
On the other hand, as discussed above, large values of $\sigma$ tend to increase the loss of spectral amplitude in the hybridization region.
As will be discussed in Sec.~\ref{subsec:impact_of_hybridization_parameters}, higher values of  $f_{\rm{Hyb}}$ pushes the hybridization region toward the late inspiral phase, thus reducing the accuracy of the LFA before transitioning to the full response.
Unless otherwise stated, we will use $f_{\rm{Hyb}}=3\times 10^{-4}\,\rm{Hz}$, $T=800\, \delta t$, and $\sigma = T/40$ as our default choices through the paper.

\section{Response model validation}
\label{sec:response_model_validation}

In this section, we evaluate the validity of the low-frequency and the hybrid responses with a variety 
of tests: 
mismatch calculations, timing benchmarks, memory usage, and power consumption.
These two novel approaches have been implemented in the Python framework 
\phenomxpy~\cite{Garcia-Quiros:2025usi}, while the full response was already implemented by 
Garc\'\i{}a-Quir\'os \textit{et. al.}~\cite{Garcia-Quiros:2025usi}. 

Although realistic LISA noise is expected to be non-stationary and non-Gaussian (see e.g.~\cite{redbook, PhysRevLett.116.231101}), for simplicity we will work under the assumption 
of stationary and Gaussian noise.
Based on this, the noise-weighted inner product is defined as~\cite{PhysRevD.44.3819, PhysRevD.46.5236} (see App.~\ref{subsec:app:scalar_product} for a more detailed derivation):
\begin{equation}\label{eq:scalar_product}
    \langle x(t) | y(t) \rangle = 4\mathrm{Re}\int_{f_{\mathrm{min}}}^{f_{\mathrm{max}}} \mathrm{d} f \frac{\tilde{x}^{*}(f) \tilde{y}(f)}{S_{\mathrm{n}}(f)} \,,
\end{equation}
where $x$ and $y$ represent two arbitrary time series, and $S_{\rm{n}}$ is the single-sided PSD.
Complex conjugation is indicated with $(*)$ and tildes denote Fourier transforms (see App.~\ref{sec:app:fourier_conventions} for the Fourier conventions followed).
In practice, the upper cutoff frequency $f_{\mathrm{max}}$ is often set by the signal, while the
lower cutoff $f_{\mathrm{min}}$ is a result of the detector’s technology.
For the accuracy tests we will compute the \emph{overlap}, which quantifies the agreement 
between two signals and is given by
\begin{equation}\label{eq:normalized_overlap}
    \mathcal{O}(x,y) = \frac{\langle x|y\rangle}{\sqrt{\langle x|x\rangle} \, \sqrt{\langle y|y\rangle}},
\end{equation}
between the $A$, $E$, and $T$ outputs obtained with the full response and with our approximate 
versions. We also define the mismatch $\mathcal{M}$ as $\mathcal{M}=1-|\mathcal{O}|$, which takes 
values between 0 and 1. If two signals are identical the mismatch is 0.
Note that here we do not define the mismatch as is usually done, with an optimization over time and phase of coalescence, because we do not vary the waveform, but only the response.
However, by taking the absolute value of the overlap, we implicitly optimize over a constant phase offset between the signals. 
This simplification is justified for the present analysis, as we are comparing different response models where the projected signals are already phase-aligned, as we will show in Sec.~\ref{subsec:golden_binary}.

In the following sections, we adopt the mismatch 
as a metric to assess the accuracy of the 
low-frequency and hybrid responses in comparison to the full response. 
Additionally, we use the recent Python implementation \cite{Garcia-Quiros:2025usi} of the aligned-spin model 
\phTHM~\cite{PhysRevD.103.124060,PhysRevD.105.084039}, to generate 
$h_{+}(t)$ and $h_{\times}(t)$, for the three LISA response models. 
Unless the contrary is specified, we will generate the GW polarizations including all the spherical harmonic mode content available in the model, \mbox{$(l,m)= \{(2,\pm 2), (2,\pm 1), (3,\pm 3), (4,\pm 4), (5,\pm 5)\}$}.

We begin by performing a visual comparison in Sec.~\ref{subsec:accuracy} of the Michelson variables $(X, Y, Z)$ derived from the 
three different models throughout the binary evolution: inspiral, merger, and ringdown. Following this, we compute 
the mismatch against the complete response across the parameter space for the 1.5 and 2.0 TDI 
generations, for both CPU and GPU implementations.
In Sec.~\ref{subsec:impact_of_hybridization_parameters} we examine the impact of different choices of 
hybridization parameters have on the accuracy.
In Sec.~\ref{subsec:timing} we present benchmark comparisons of CPU and GPU performance. Finally, in 
Sec.~\ref{subsec:memory_and_power} we evaluate the memory usage and power consumption of the three responses.

\subsection{Comparison against the full LISA response}\label{subsec:accuracy}

\begin{figure*}[t]
    \centering
    \begin{subfigure}[t]{0.5\textwidth}
        \centering
        \includegraphics[width=\columnwidth]{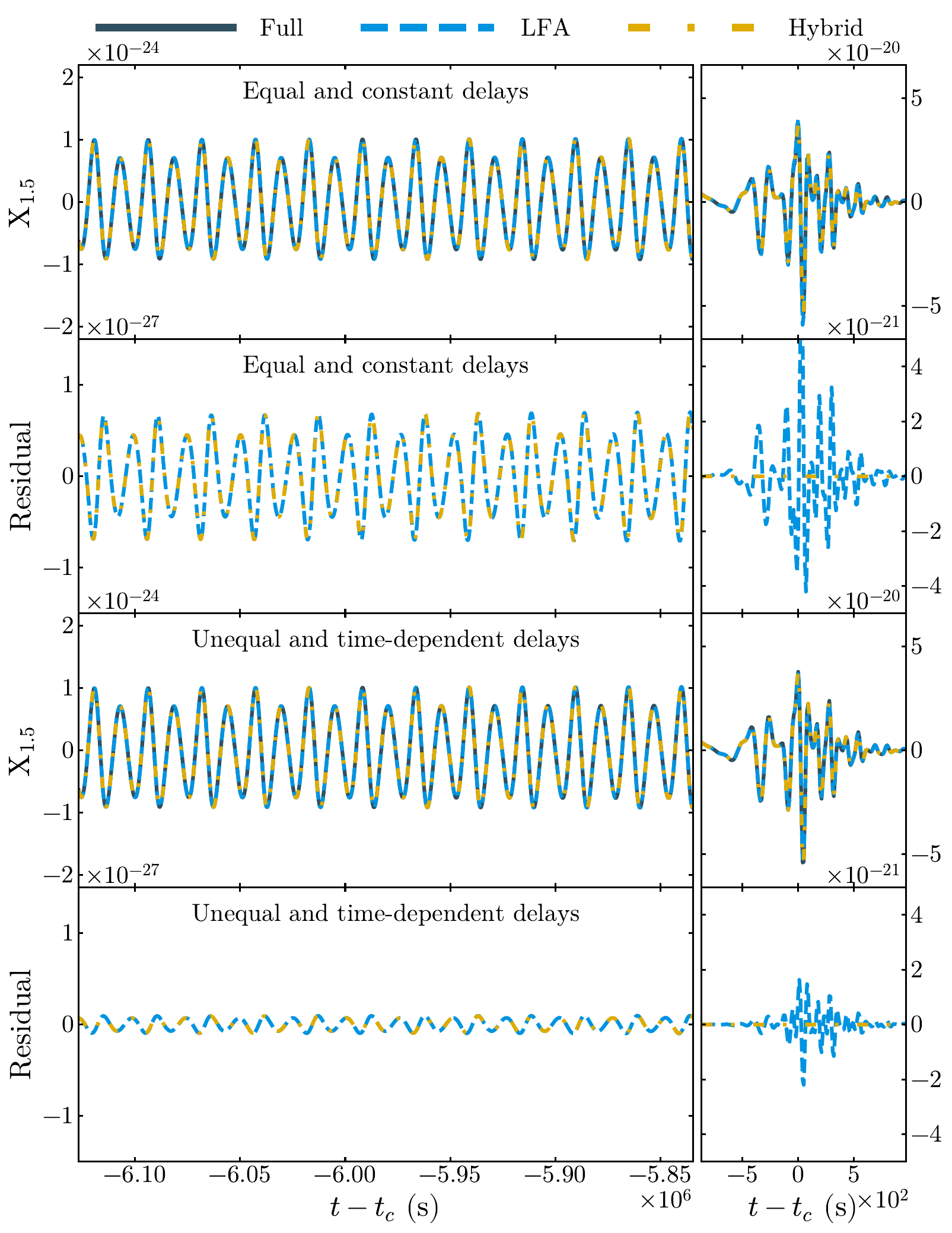}
        \caption{1.5 TDI.}
        \label{fig:cpu_visual_agreement_TDI1}
    \end{subfigure}~
    \begin{subfigure}[t]{0.5\textwidth}
        \centering
        \includegraphics[width=\columnwidth]{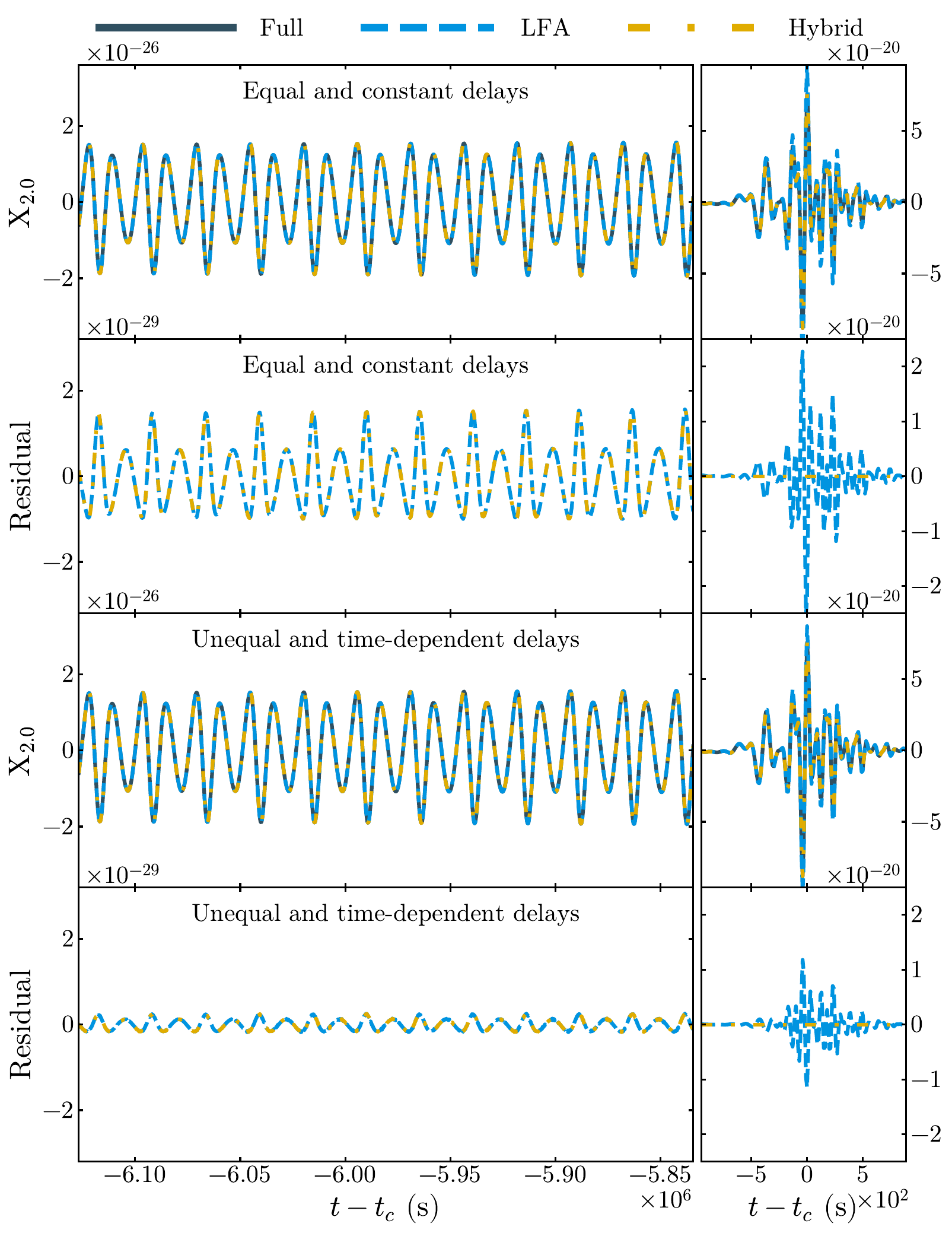}
        \caption{2.0 TDI.}
        \label{fig:cpu_visual_agreement_TDI2}
    \end{subfigure}
    \caption{\textit{Panel (a)}: Michelson combination $X$ and approximation residuals as a function of the time to coalescence $t-t_c$, expressed in seconds, for the 1.5-generation TDI.
    The results are computed with the CPU implementation of the full (black, solid), LFA (blue, dashed), and hybrid (gold, dash-dotted) responses, for an MBHB with $M=3\times 10^6\,\rm{M}_\odot$, $q=4$, $\chi_1=\chi_2=0.8$, $\iota=\pi/3\,\rm{rad}$, $d_L=50\,\rm{Gpc}$, $\varphi=4.0\,\rm{rad}$, $\beta=0.6\,\rm{rad}$, $\lambda=\pi\,\rm{rad}$, and $\psi=0.25\,\rm{rad}$. 
    For the hybrid response we have used $f_{\rm{Hyb}}=3\times10^{-4}\,\rm{Hz}$, $T=800\delta t$, and $\sigma=T/40$, with $\delta t=10\,\rm{s}$.
    In the first two rows, the LFA and hybrid $X$ output is computed using an equal and constant delay configuration (Eq.~\eqref{eq:X15_equal_timedelays}), whereas the last two rows adopt an unequal and time-dependent delay configuration (Eq.~\eqref{eq:X15_nonequal_timedelays}). For all cases, the full response is consistently computed using unequal and time-dependent delay configuration (Eq.~\eqref{eq:X1.5}). The corresponding residuals, defined as $X_{\rm{Full}}-X_{\rm{LFA/Hybrid}}$, are displayed below each panel showing the $X$ observable. \textit{Panel (b)}: The same as in panel (a), but for the 2.0-generation TDI. In this case, the LFA and hybrid $X_{2.0}$ observable is computed from Eq.~\eqref{eq:X20_equal_timedelays} for the equal and constant delay configuration, and from Eq.~\eqref{eq:X20_nonequal_timedelays} for the unequal and time-dependent delay configuration. The full-response $X_{2.0}$ is given by Eq.~\eqref{eq:X2.0}.}
    \label{fig:cpu_visual_agreement_TDI}
\end{figure*} 

The accuracy of the LFA and hybrid approaches is evaluated by computing their deviation from the full response.
As an example, Fig.~\ref{fig:cpu_visual_agreement_TDI} shows the $X$ channel as a function of 
the time to coalescence for the three responses, along with the residuals, defined as 
\mbox{$X_{\text{Full}}-X_{\text{LFA}/\text{Hybrid}}$}, for different LISA configurations and TDI generations. 
The full response assumes non-equal, time-dependent delays, while the LFA and hybrid responses follow the 
configurations specified in the panels. 
During the inspiral part shown in the left panels of Figs.~\ref{fig:cpu_visual_agreement_TDI1} 
and~\ref{fig:cpu_visual_agreement_TDI2}, the two approaches reproduce the 
full response with an error of 0.1\% for equal and constant delays, and 0.01\% for non-equal and time-dependent 
delays, for both TDI generations. 
This error increases as the binary evolves toward the late inspiral, where deviations from the LFA start to become 
significant, and the hybrid 
transitions toward the full response.
During the merger-ringdown phase, the hybrid response precisely matches the full response by construction, resulting 
in zero residuals, while the error for the LFA is maximum, as the approximation breaks down before this phase. 
However, we find that depending on the sky location of the source, adopting non-equal and time-dependent delays 
helps in reducing 
the residual even through the final stages of the binary evolution, as shown in the right panels of 
Figs.~\ref{fig:cpu_visual_agreement_TDI1} and~\ref{fig:cpu_visual_agreement_TDI2}. Additionally, we note that the 
deviations originated by using the LFA in this phase are larger for 2.0 TDI than for 1.5 TDI.  

The example presented in Fig.~\ref{fig:cpu_visual_agreement_TDI} consists of an aligned-spin binary system. 
In these systems, the time-domain GW polarizations exhibit a slowly varying pattern during the early inspiral.
This makes them ideal candidates for applying the LFA response, as higher-order derivatives of the strain components,
$h_{+}$ and $h_{\times}$, can be safely neglected during this phase.
In contrast, more complex systems--such as those involving precession or eccentricity--introduce additional 
features that alter the slowly varying pattern during the inspiral, which might decrease the accuracy of the LFA model. See App.~\ref{sec:app:non_monochromatic inspiral} for a test case involving binaries in eccentric orbits.

\begin{figure*}[t]
    \centering
    \begin{subfigure}[t]{0.5\textwidth}
        \centering
        \includegraphics[width=\textwidth]{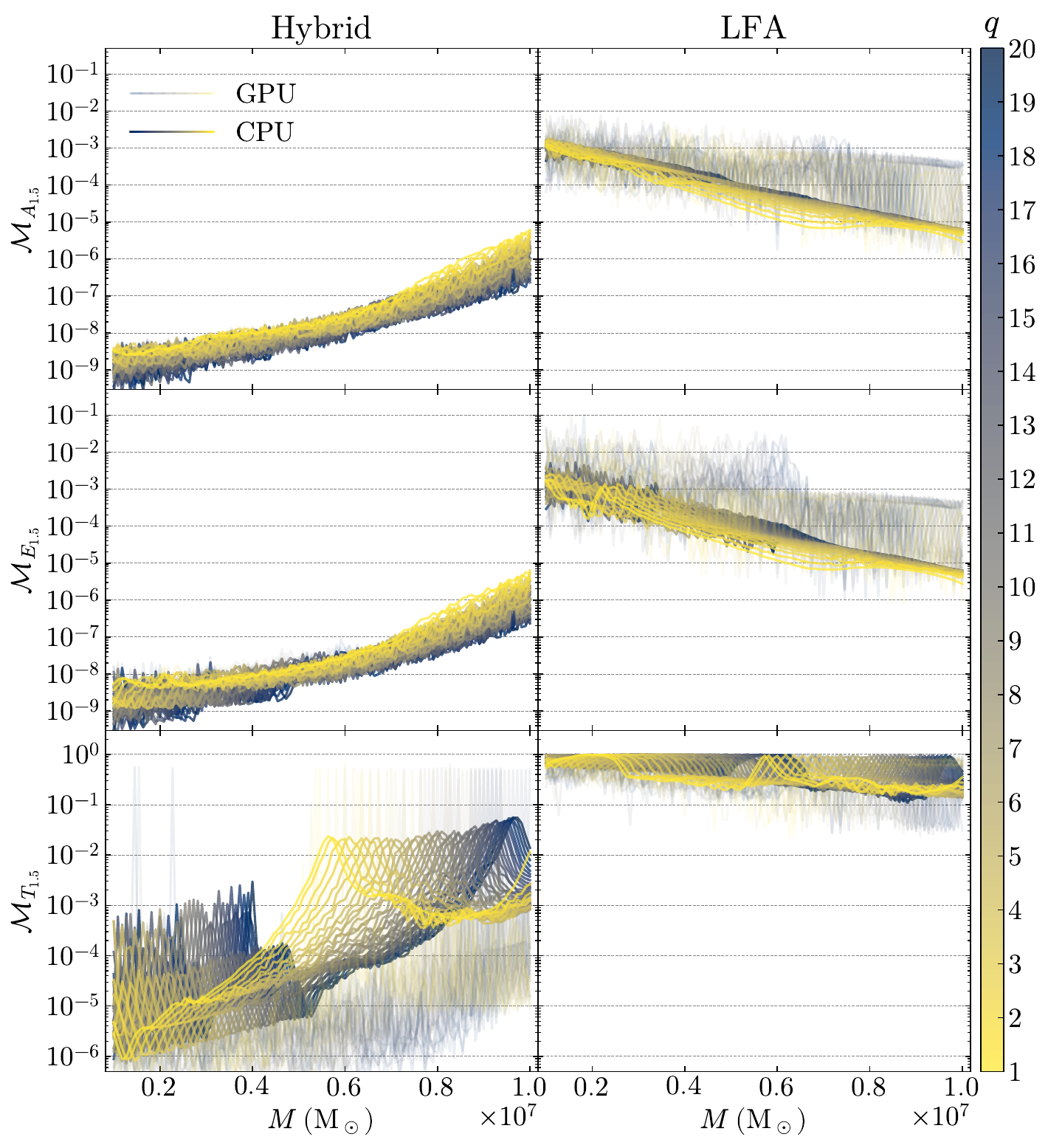}
        \caption{\label{fig:overlap_lwa_hybrid_TDI1} 1.5 TDI.}
    \end{subfigure}~
    \begin{subfigure}[t]{0.5\textwidth}
        \centering
        \includegraphics[width=\textwidth]{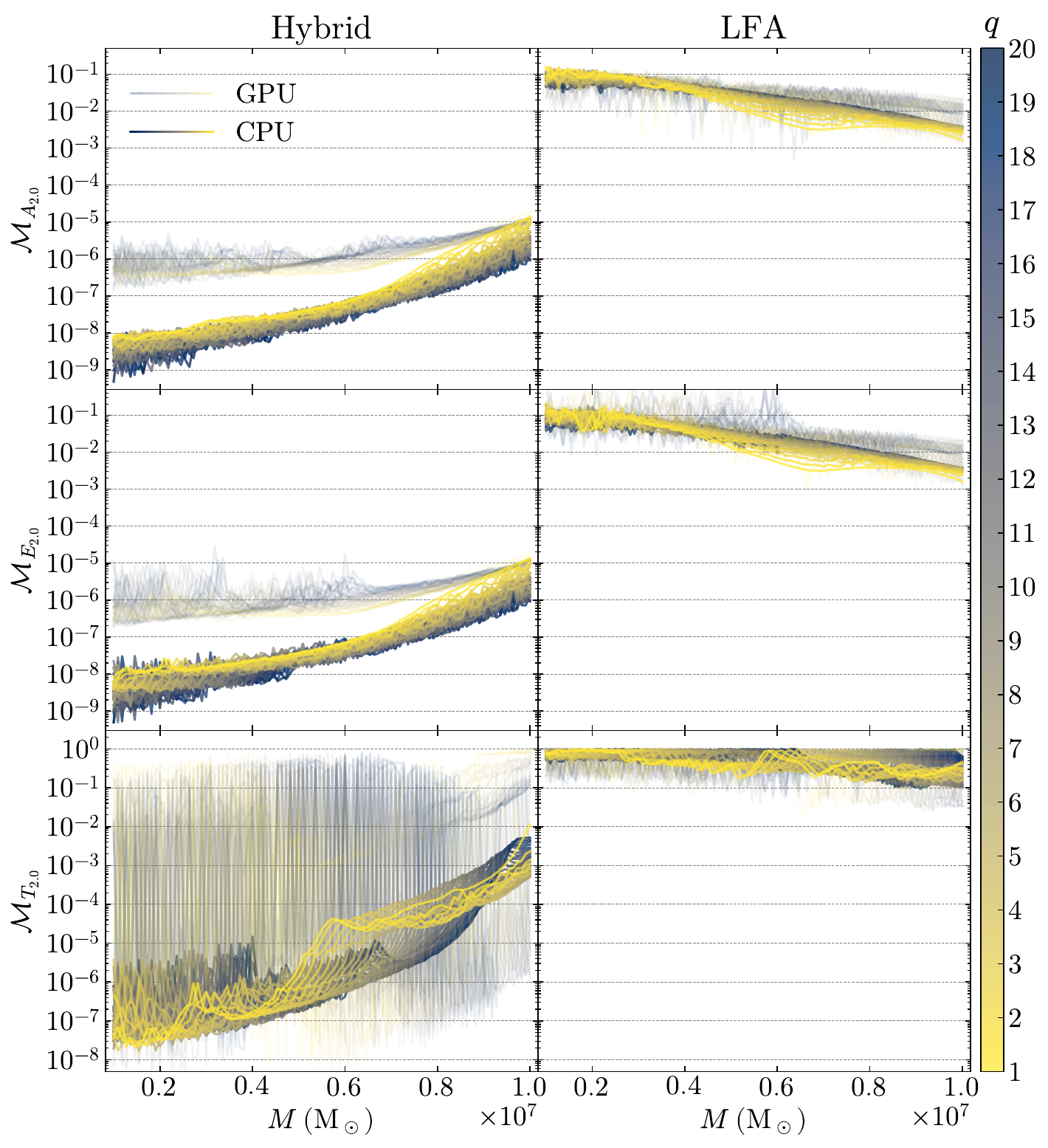}
        \caption{\label{fig:overlap_lwa_hybrid_TDI2} 2.0 TDI.}
    \end{subfigure}
    \caption{\textit{Panel (a)}: Mismatches of the 1.5-generation $A$ (upper row), $E$ (center row), and $T$ (lower row) channels computed with the hybrid (left column) and the LFA (right column) approaches against the same channels computed using the full response, over a range of total masses $M\in[10^{6},10^{7}]\,\rm{M}_\odot$. 
    The color of each curve indicates the value of the mass ratio, $q\in[1,20]$. 
    The opacity of the lines differentiates the CPU (high-opacity) and GPU (low-opacity) implementations of the responses. The mismatch is computed between  $f_{\rm{min}}=10^{-4}\,\rm{Hz}$ and $f_{\rm{max}}=0.05\,\rm{Hz}$. Except for $M$ and $q$, the rest of the parameters are fixed: $\chi_1=0.8$, $\chi_2=0.6$,  $\iota=\pi/3\,\rm{rad}$, $d_L=50\,\rm{Gpc}$, $\varphi=4.0\,\rm{rad}$, $\beta=-0.6\,\rm{rad}$, $\lambda=0.6\,\rm{rad}$, and $\psi=0.4\,\rm{rad}$. \textit{Panel (b)}: The same as in panel (a), but for the 2.0-generation $A$, $E$, and $T$ channels. For this case, the mismatch is computed between $f_{\rm{min}}=10^{-4}\,\rm{Hz}$ and $f_{\rm{max}}=0.02\,\rm{Hz}$.
    }
     \label{fig:overlap_lwa_hybrid}
\end{figure*} 

\begin{figure*}[t]
    \centering
    \includegraphics[width=\textwidth]{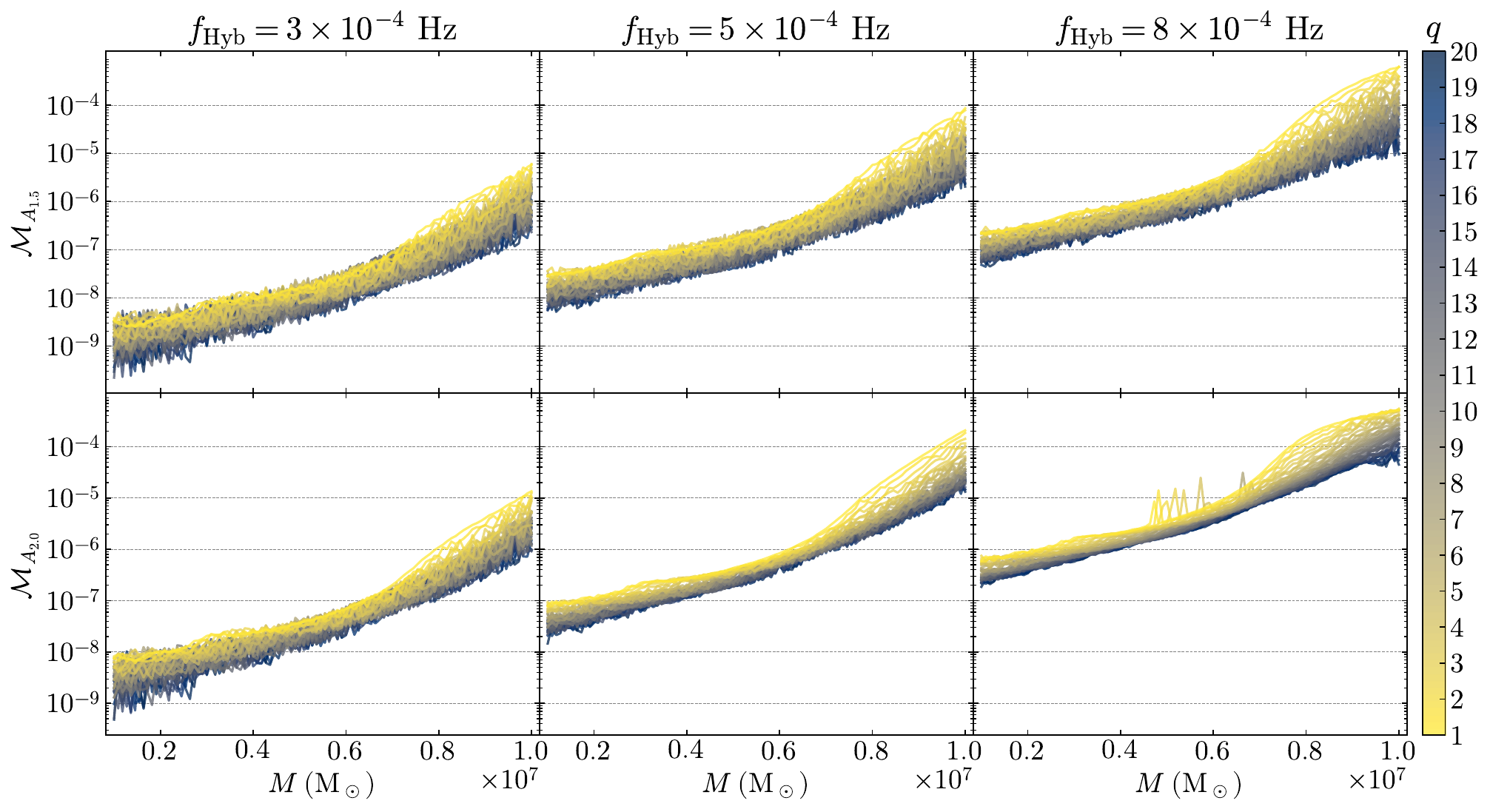}
    \caption{Mismatch of the 1.5-generation (upper row) and 2-0-generation (lower row) $A$ channel computed with the hybrid approach, for different values of $f_{\rm{Hyb}}$, against the full response, over a range of total masses $M\in[10^{6},10^{7}]\,\rm{M}_\odot$. The color of each curve indicates the value of the mass ratio, $q\in[1,20]$. The system's parameters and the frequency limits of the mismatch calculation are the same as in Fig.~\ref{fig:overlap_lwa_hybrid}.}
    \label{fig:overlap_changing_fhyb}
\end{figure*}

To systematically test the robustness across the parameter space, we compute the mismatch
of the hybrid and LFA approaches against the full response as a function of the total mass, $M$, and mass 
ratio, 
\mbox{$q=m_1/m_2 > 1$}. 
In Fig.~\ref{fig:overlap_lwa_hybrid} we show the results for the $A$, $E$, and $T$ channels distributed within 
$M\in[10^{6}, 10^{7}]\,\rm{M}_\odot$ and $q\in[1,20]$, while keeping the rest of the parameters fixed. 
We also set $f_{\rm{min}}=10^{-4}\,\rm{Hz}$ and $f_{\rm{max}}=0.05\,\rm{Hz}$ in 
Eq.~\eqref{eq:scalar_product},
as all binaries within the selected mass range have already merged at that frequency. 
The results are presented separately according to the TDI generation used. 
Additionally, within each generation, we display the results for both the CPU and GPU implementations of the 
models. 
It is important to note that we do not compute mismatches between different TDI generations or hardware 
implementations.
In these plots, each curve quantifies the mismatch between a certain $A,E,T$-channel waveform produced
by the hybrid (left column of the panels) or LFA response (right column of the panels), and the corresponding
waveform produced by the full response.

The results presented in Fig.~\ref{fig:overlap_lwa_hybrid} show that the hybrid response achieves 
mismatches below $10^{-5}$ when compared to the full response for both the $A$ and $E$ channels. 
We find that this behavior is consistent across both CPU and GPU implementations and for both TDI generations. 
In contrast, for the same channels, the LFA produces mismatches below 0.1\% and 10\% for 1.5 TDI and 2.0 
TDI, respectively.
For the $T$ channel, we observe mismatches ranging from 10\% to 100\% for the LFA response, 
while the hybrid approach generally yields values below 10\%. 
This difference arises because, as with the LFA, the hybrid model approximates $T \simeq 0$ during the 
early inspiral, 
but for the merger-ringdown phase,
where the majority of the SNR comes from \cite{PhysRevD.103.083011},
it has already transitioned to the full response description, drastically improving the overlap.

The differences in mismatch between the CPU and GPU implementations arise from the choice of interpolator used to compute the retarded quantities. 
For the CPU, we employ the function 
\texttt{scipy.interpolate.InterpolatedUnivariateSpline} from the \scipy\footnote{We use version \texttt{v1.13.1} of \scipy from the repository \url{https://github.com/scipy/scipy} with git commit \texttt{44e4eba}.} Python package~\cite{scipy}, which implements by default a cubic spline. 
In contrast, for the GPU, we use the linear interpolator function \texttt{cupy.interp}, as more efficient 
interpolators, such as 
\texttt{cupyx.scipy.interpolate.InterpolatedUnivariate\allowbreak Spline}
or \texttt{cupyx.scipy.interpolate.CubicSpline}, have not yet been implemented in the latest release of 
the \cupy Python package~\cite{cupy} at the time of writing.\footnote{The latest release of \cupy at the time of writitng is version \texttt{v13.4.1}, whereas we use version \texttt{v13.3.0} from the repository \url{https://github.com/cupy/cupy} with git 
commit \texttt{118ade4}. Neither of these versions includes the functions \texttt{cupyx.scipy.interpolate.InterpolatedUnivariateSpline}
or \texttt{cupyx.scipy.interpolate.CubicSpline}.} 
As seen in Fig.~\ref{fig:overlap_lwa_hybrid}, the impact is more significant for 2.0 TDI. 
This might be due to the higher number of interpolations required to compute the full-response 2.0-TDI 
Michelson variables (Eq.~\eqref{eq:X2.0}) compared to the 1.5-TDI case (Eq.~\eqref{eq:X1.5}).
For the 2.0-TDI $A$ and $E$ channels, the mismatch difference between the CPU and GPU hybrid responses 
grows with decreasing mass, reaching up to two orders of magnitude difference. 
This is consistent with an increasing number of cycles in the LISA sensitivity band, enabling the mismatch to be more precise, and hence
highlighting small differences arising from the interpolator used.
For the $T$ channel, the use of a linear interpolator, as implemented in the GPU, results in highly 
oscillatory mismatches. 
We have checked that using already implemented, but still slow, spline 
functions like \texttt{cupyx.scipy.interpolate.make_interp_spline} does reduce the GPU mismatches down to the CPU values. An alternative approach would have been to employ other GPU-based cubic-spline packages, e.g.~\cite{jax_cubic_spline, cudakima}.

Additionally, for the same systems analyzed in Fig.~\ref{fig:overlap_lwa_hybrid}, we tested the accuracy of the cubic spline by comparing it with the more accurate 31st-order Lagrange interpolator implemented in \textsc{LISA Instrument}~\cite{LISAinstrument} and \textsc{PyTDI}~\cite{PyTDI}. 
Results showed mismatches increasing from $\mathcal{O}(10^{-7})$ for $M=10^{7}\,\rm{M}_\odot$ to $\mathcal{O}(10^{-3})$ for $M=10^{6}\,\rm{M}_\odot$, highlighting the limitations of cubic splines particularly for low total mass binaries. These limitations are expected to be more pronounced for systems including precession or eccentricity, and motivate the use of more accurate interpolation algorithms. Importantly, note that the hybrid and LFA approaches presented here are compatible with any interpolation scheme and inherit the interpolation accuracy of the chosen interpolation method.

From Fig.~\ref{fig:overlap_lwa_hybrid} we also note that the hybrid and LFA mismatches follow 
different trends when increasing the total mass. As expected, heavier systems merge at lower frequencies,
where the deviations from the LFA are small, and it can accurately reproduce the full response even in the 
merger-ringdown phase. 
For the hybrid response, we observe the opposite: the mismatch grows as the total 
mass increases. 
This is caused by the hybridization procedure. 
The Fourier-domain modulations induced after
multiplying the response by the sigmoid functions of Eq.~\eqref{eq:sigmoid}, 
compromise the accuracy of the Fourier-transformed hybrid approach toward low frequencies.
We will further discuss this in Sec.~\ref{subsec:impact_of_hybridization_parameters}.

\subsection{Impact of hybridization parameters}\label{subsec:impact_of_hybridization_parameters}

In this section, we assess the effect of the hybridization parameters on the 
accuracy of the hybrid response. We will primarily focus on the hybridization frequency, $f_{\rm{Hyb}}$, and the sigmoid's width, $\sigma$. Regarding the duration of the hybridization window, $T$, we only have to ensure it allows the sigmoid to approach its asymptotic values within the window duration.

According to Eq.~\eqref{eq:hybrid_ratio_simplified}, high values for the ratio $f_{\rm{Hyb}}/f_{\rm{min}}$
result in large portions of the waveform being projected by the LFA. 
Depending on the value of $f_{\rm{Hyb}}$, the deviations from the LFA can substantially affect the accuracy 
during the early inspiral, before the hybridization is performed. 

In Fig.~\ref{fig:overlap_changing_fhyb} we present the mismatches between the $A$ channel 
from the CPU-implemented hybrid and full responses, for different values of $f_{\rm{Hyb}}$. 
As expected, the mismatch grows with $f_{\rm{Hyb}}$.
There is an order-of-magnitude increase in the mismatch at each successive sub-panel in the figure (from left to right).
Within each case, we obtain similar mismatches for both TDI generations, which are below $10^{-5}$ for
$f_{\rm{Hyb}}=3\times 10^{-4}\,\rm{Hz}$, $10^{-6}$ for
$f_{\rm{Hyb}}=5\times 10^{-4}\,\rm{Hz}$, and $10^{-3}$ for
$f_{\rm{Hyb}}=8\times 10^{-4}\,\rm{Hz}$. 
This indicates that, depending on the SNR of the system, hybridization frequencies beyond the validity region of the LFA can still lead to sufficiently accurate TDI variables.
However, we do not recommend going beyond that limit, since the computational speedup is minimal compared to the lose in accuracy, as we will discuss in Sec.~\ref{subsec:timing}. 
Additionally, we observe that the mismatch trend toward high total masses does not depend on $f_{\rm{Hyb}}$, which is consistent with the explanation provided in Sec.~\ref{subsec:accuracy}. 

The choice of $\sigma$ in Eq.~\eqref{eq:sigmoid} can also influence the performance of the
response. Ideally, when comparing against the full response, we expect the mismatches from the hybrid response 
to approximate the ones from the LFA during the early inspiral. 
However, as discussed in Secs.~\ref{sec:hybrid_response} and~\ref{subsec:accuracy}, multiplying the 
response variables by sigmoids (see Eqs.~\eqref{eq:hybridized_yslr} and~\eqref{eq:hybridized_xs}) introduces undesired modulations in the low-frequency regime of the 
Fourier-transformed signal, which propagate through the $A$, $E$, and $T$ variables.
This leads to larger mismatch values at low frequencies for the hybrid response as compared to the LFA, which does not exhibit these modulations.

\begin{figure}[t]
    \centering
    \includegraphics[width=\columnwidth]{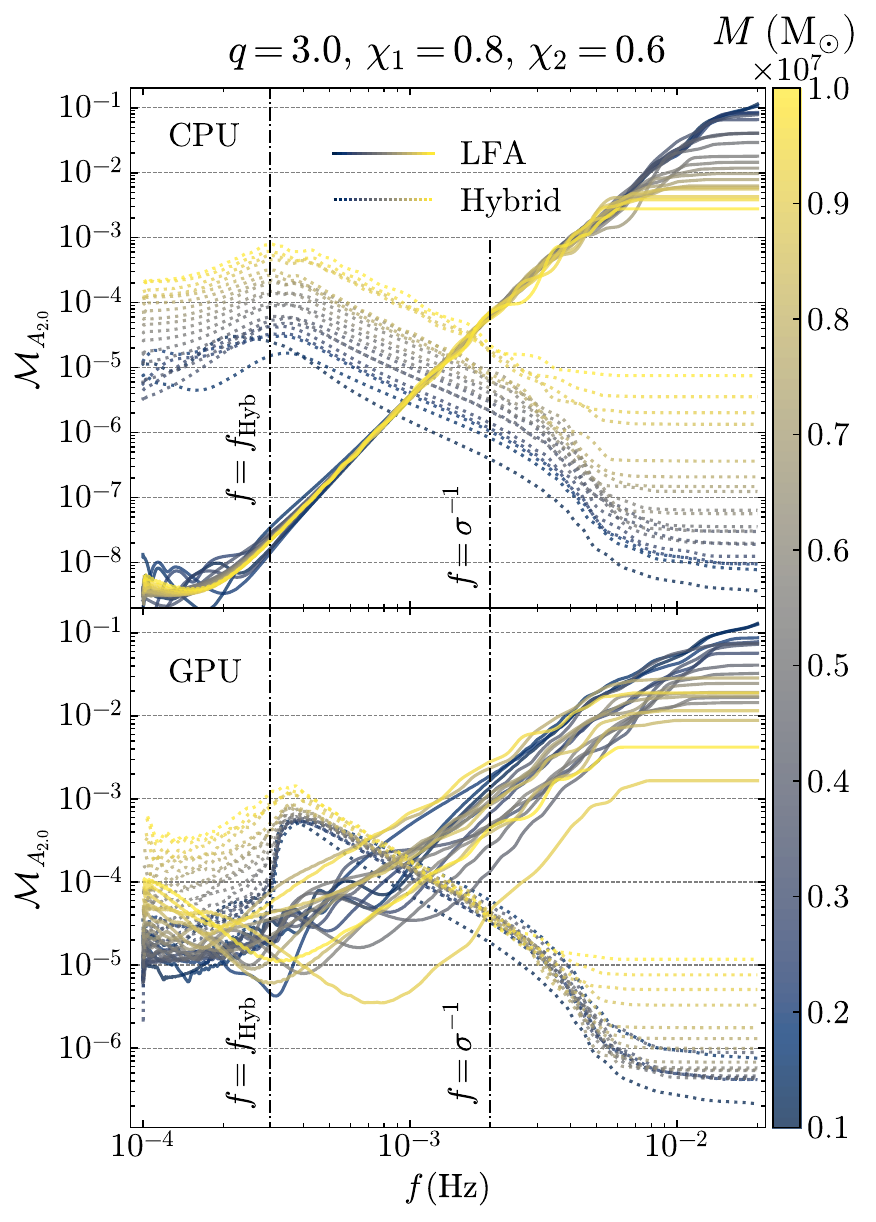}
    \caption{Mismatch of the 2.0-generation $A$ channel computed with the LFA (solid) and hybrid (dotted) approaches against the full response, as a function of the GW frequency $f$, expressed in Hz. The results are shown for both the CPU (upper panel) and GPU (lower panel) implementations of the responses. The color of each line is associated with the value of the system's total mass $M$. 
    The dashed-dotted vertical lines indicate the value of the hybridization frequency $f_{\rm{Hyb}}=3\times 10^{-4}\,\rm{Hz}$, and the inverse of the sigmoid's width of Eq.~\eqref{eq:sigmoid} used in the hybridization procedure, $\sigma^{-1}=0.002\,\rm{Hz}$; which are common to all the hybrid cases.
    The mass ratio is fixed to $q=3$ and the rest of the parameters match the ones of Figs.~\ref{fig:overlap_lwa_hybrid} and~\ref{fig:overlap_changing_fhyb}. The mismatch is computed between $f_{\rm{min}}=10^{-4}\,\rm{Hz}$ and $f_{\rm{max}}=0.02\,\rm{Hz}$.}
    \label{fig:accumulated_overlap_lwa_vs_hybrid_A_TDI2}
\end{figure}

This behavior is illustrated in 
Fig.~\ref{fig:accumulated_overlap_lwa_vs_hybrid_A_TDI2}, where we compute
the mismatch of the 2.0-TDI $A$ channel as a function of the GW frequency.\footnote{Note that the mismatch curves do not increase monotonically with frequency, as we update the normalization of
Eq.~\eqref{eq:normalized_overlap} at each frequency point.}
As an example, we consider aligned-spin systems with $q=3$, $\chi_1=0.8$, $\chi_2=0.6$ and $M\in[10^{6},10^{7}]\,\rm{M}_\odot$. 
As anticipated, the mismatches from the hybrid approach (dotted lines) are larger than those from
the LFA (solid lines) in the lower part of the spectrum, for both CPU and GPU implementations.
The hybrid results show a bump centered around $f_{\rm{Hyb}}$, which then decays towards the mismatch value of considering the entire signal.
This bump, which does not roughly exceed $\mathcal{M}=10^{-3}$, corresponds to the small loss of amplitude in the hybridization procedure.
Also, as discussed in Sec.~\ref{sec:hybrid_response}, the Fourier-domain modulations persist up to a characteristic frequency of approximately $f\lesssim \sigma^{-1}$, from the point where the hybrid response 
approximately starts to be more accurate than the LFA.

The results shown in Fig.~\ref{fig:accumulated_overlap_lwa_vs_hybrid_A_TDI2}
indicate that, consistent with the findings of Fig.~\ref{fig:overlap_lwa_hybrid}, 
the hybrid response provides a more accurate reproduction of the full response projection 
at high frequencies ($f\gtrsim10^{-3}\,\rm{Hz}$) than the LFA approach; thus, making the 
hybrid model suitable for a complete inspiral-merger-ringdown analysis.
However, although the mismatches for the hybrid model are below 0.1\%,
the LFA is preferred for tasks that do not extend beyond \mbox{$f=10^{-3}\, 
\rm{Hz}$}, such as pre-merger analyses. 
We will discuss these two concrete applications in Secs.~\ref{subsec:golden_binary} and~\ref{subsec:deep_alerts}.

\subsection{Timing results}\label{subsec:timing}

\begin{figure*}[htb]
    \centering
    \includegraphics[width=\textwidth]{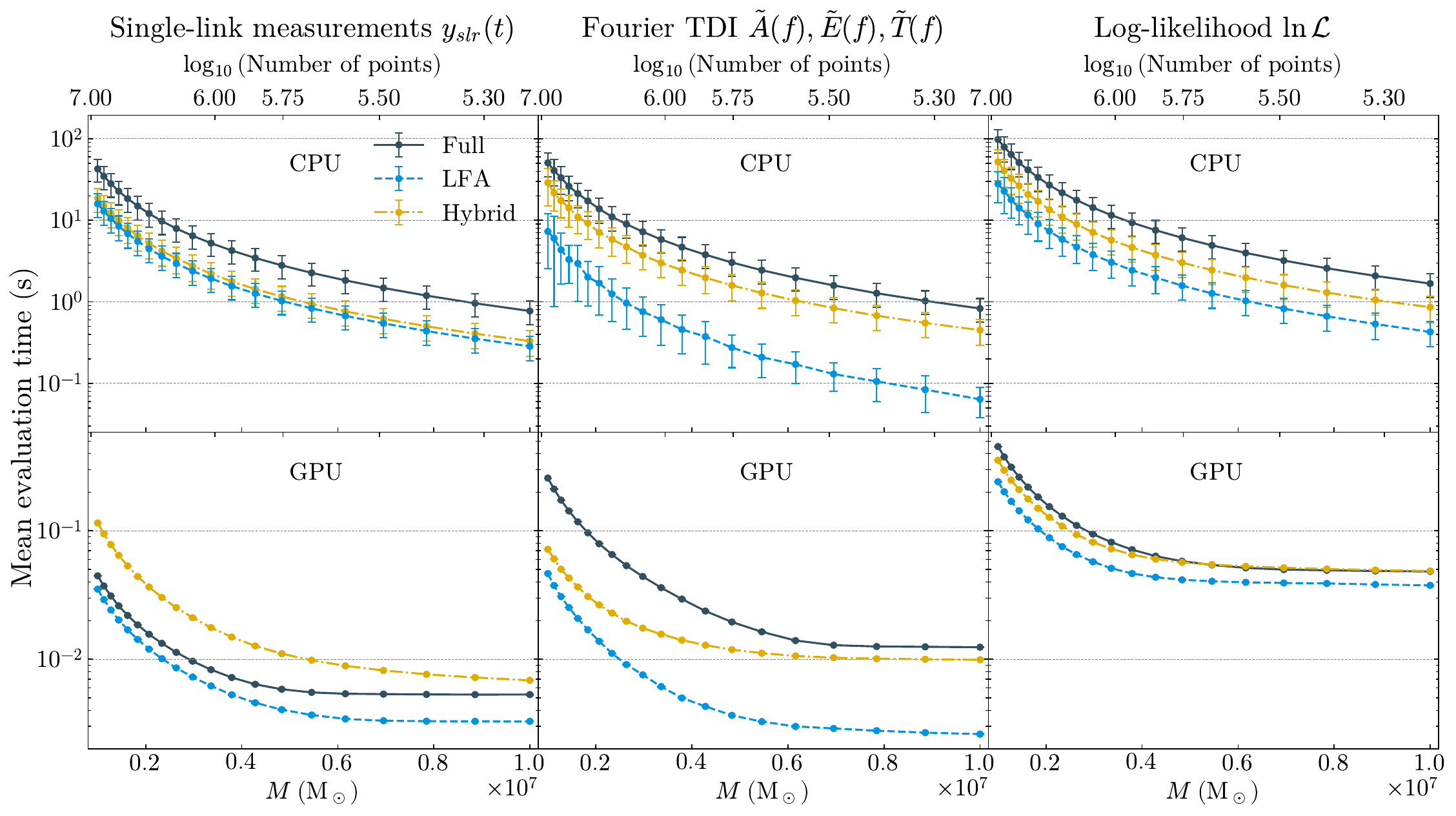}
    \caption{Mean evaluation times, in seconds, for the full (gray, solid), LFA (blue, dashed), and hybrid (gold, dash-dotted) responses using 2.0 TDI, as a function of the total mass $M\in[10^6,10^7]\,\rm{M}_\odot$ (indicated in the lower $x$-axis).
    The results are split into CPU (upper row) and GPU (lower row) benchmarks and into different response quantities: the single-link measurements $y_{slr}$ (left column), the Fourier-transformed $A$, $E$, and $T$ channels (center column), and the full log-likelihood evaluation $\ln \mathcal{L}$ (right column). The systems are characterized by starting at a minimum frequency $f_{\rm{min}}=0.8\times 10^{-4}\,\rm{Hz}$, and being sampled at $\delta t=5\,\rm{s}$. For each value of $M$, the evaluation times are averaged over 100 realizations where the mass ratio is uniformly sampled between 1 and 10, and the dimensionless spin components are uniformly sampled between $-1$ and 1. The error bars represent the $1\sigma$ standard deviation. The top $x$-axis provides a leading-order estimation of the number of points of a representative case with $q=5$.}
    \label{fig:timing_cpu_vs_gpu_tdi2_loglikelihood}
\end{figure*}

To assess the computational speed of the hybrid and LFA responses,
we measure the mean evaluation time required for the algorithms to evaluate the likelihood,
which involves \textit{i)} generating $h_{+}(t)$ and $h_{\times}(t)$, \textit{ii)} computing $y_{slr}(t)$, \textit{iii)} computing $X(t)$, $Y(t)$, and $Z(t)$, and linearly combining them to build $A(t)$, $E(t)$, and $T(t)$, \textit{iv)} Fourier-transforming those quantities to get $\tilde{A}(f)$, $\tilde{E}(f)$, and $\tilde{T}(f)$, and finally \textit{v)} evaluating a scalar product like the one defined in Eq.~\eqref{eq:scalar_product}.

In Fig.~\ref{fig:timing_cpu_vs_gpu_tdi2_loglikelihood},
we show the mean evaluation times and the associated standard deviations
as a function of the binary's total mass,
for different quantities computed with the three 
approaches and using 2.0-generation TDI.
In the left-column panels, we show the timing performance for the single-link measurements 
$y_{slr}(t)$, which correspond to step \textit{ii)} previously defined.
In the center-column panels, we summarize the results for steps \textit{iii)} and
\textit{iv)}, which involve the construction of $\tilde{A}(f)$, $\tilde{E}(f)$, and $\tilde{T}(f)$ from $y_{slr}(t)$.
Finally, the right-column panels report the total computational time required for a full likelihood evaluation—comprising steps \textit{i)} through \textit{v)}—including the generation of  $h_{+}(t)$ and $h_{\times}(t)$.

The mass range in the $x$-axis goes from $M=10^{6}\,\rm{M}_\odot$ to  $M=10^{7}\,\rm{M}_\odot$.
Given a value of $M$, we average the evaluation time
over 100 realizations in which we uniformly distribute the mass-ratio, $q\sim \mathcal{U}(1,10)$, and 
dimensionless spin components, $\chi_{1,2}\sim \mathcal{U}(-1,1)$. 
We generate the time-domain polarizations with \phTHM including 
\mbox{$(l,m)= \{(2,\pm 2), (2,\pm 1), (3,\pm 3), (4,\pm 4), (5,\pm 5)\}$}, 
starting from $f_{\rm{min}}=0.8\times 10^{-4}\,\rm{Hz}$. 
The sampling rate is fixed at $\delta t = 5\,\rm{s}$ for all masses.
Based on these parameters, we additionally show, along the top $x$-axis, an estimate of the discrete array length used in the calculations. 
At leading order, this estimate varies with $M$ and $q$.
However, in this context, we limit ourselves to a representative mass ratio of $q = 5$ for each total mass, which gives an approximate range between $10^7$ and $10^5$ points.

The CPU benchmark results are presented in the upper-row panels of Fig.~\ref{fig:timing_cpu_vs_gpu_tdi2_loglikelihood}. 
The small overhead observed in the hybrid evaluation times for $y_{slr}$ relative to 
the LFA (upper-left panel) arises from the need to compute all six inter-spacecraft 
links when the hybrid model transitions to the full response regime. 
As indicated by Eq.~\eqref{eq:yslr}, this transition requires 12 additional 
interpolations of $H_l$, which are exclusively applied during the late 
inspiral and merger-ringdown phases. 
The computational overhead further increases when computing the Michelson observables 
and the subsequent $A$, $E$, and $T$ variables, as shown in the upper-center panel. 
This is due to a substantially greater number of additional interpolations required 
during the hybrid-to-full transition in this case, as can be inferred by 
comparing Eqs.~\eqref{eq:X2.0} and~\eqref{eq:X20_nonequal_timedelays}.
Overall, for the end-to-end likelihood evaluation (upper-right panel),
the hybrid approach is twice as fast as the full response, whereas the 
LFA achieves speedup factors ranging from 3 to 5 depending on the total mass.

The timing results for the responses implemented on GPUs are shown 
in the lower-row panels of Fig.~\ref{fig:timing_cpu_vs_gpu_tdi2_loglikelihood}.
Here, the evaluation time tends to stabilize toward high masses. 
The reason for this is that the short duration of the waveform makes the GPU utilization 
inefficient and the overhead associated with transferring data from the CPU to the GPU dominates over 
the computational parallelization.

Compared to the CPU case, the hybrid evaluation times exhibit different behavior on GPUs.
In particular, the hybrid timings do not consistently follow the trends observed for the LFA and full models and, in certain cases, do not lie between the corresponding LFA and full response timings. 
This deviation arises from the limited fraction of the waveform processed by the full response in the hybrid approach.
For this example, with $f_{\rm{min}}=0.8\times 10^{-4}\,\rm{Hz}$ 
and $f_{\rm{Hyb}}=3\times 10^{-4}\,\rm{Hz}$, only 3\% of the waveform is 
handled by the full response, as indicated by Eq.~\eqref{eq:hybrid_ratio_simplified}. 
Thus, depending on the value of $M$ and $\delta t$, this small subset of data points 
can result in inefficient GPU performance.
For instance, for $M = 5 \times 10^6,\mathrm{M}_\odot$ 
and $\delta t = 5\,\rm{s}$, the waveform comprises 
approximately $10^{6}$ samples, as can be read from the top $x$-axis of Fig.~\ref{fig:timing_cpu_vs_gpu_tdi2_loglikelihood}, of which only about 
$3 \times 10^4$ (3\%) are projected using the full response.
This introduces an additional data-transfer overhead intrinsic to the hybrid approach, 
which shifts the evaluation times toward larger values and that is more pronounced for short waveforms, i.e. high total masses.
To address these limitations, we are working on different strategies to accelerate and optimally implement the GPU versions of the responses.

Timing tests were also performed for $f_{\rm{Hyb}}=5\times 10^{-4}\,\rm{Hz}$ and 
$f_{\rm{Hyb}}=8\times 10^{-4}\,\rm{Hz}$. The results indicate that the improvement in efficiency is negligible compared to the default case of $f_{\rm{Hyb}}=3\times 10^{-4}\,\rm{Hz}$.
This is due to the fact that, according to Eq.~\eqref{eq:hybrid_ratio_simplified}, $\mathcal{R}(3\times 10^{-4}\,\rm{Hz})\approx 0.97$, while $\mathcal{R}(8\times 10^{-4}\,\rm{Hz})\approx 0.99$,
with the difference being too small to affect the evaluation times significantly.
Therefore, as commented in Sec.~\ref{subsec:impact_of_hybridization_parameters}, 
we do not recommend using hybridization frequencies beyond 
$f_{\rm{Hyb}}=3\times 10^{-4}\,\rm{Hz}$, as the gain in speedup is negligible compared to the 
loss in accuracy, previously shown in Fig.~\ref{fig:overlap_changing_fhyb}.

The benchmarks were conducted using the MareNostrum5 computer cluster at 
Barcelona Supercomputer Center. 
The CPU benchmarks were performed on a single node using one core of an Intel Xeon Platinum 8480+ 56C 2GHz processor\footnote{Each core has an effective usable memory of 2 GiB (where $1\,\rm{GiB}\approx 1.074\,\rm{GiB}$). The benchmark computations were performed on a single core, but we requested the full memory capacity of the node, which is 256 GiB.}, while the GPU 
benchmarks were executed on an NVIDIA Hopper H100 64GB HBM2.

\subsection{Memory usage and power consumption}\label{subsec:memory_and_power}

\begin{figure}[t]
    \centering
    \includegraphics[width=\columnwidth]{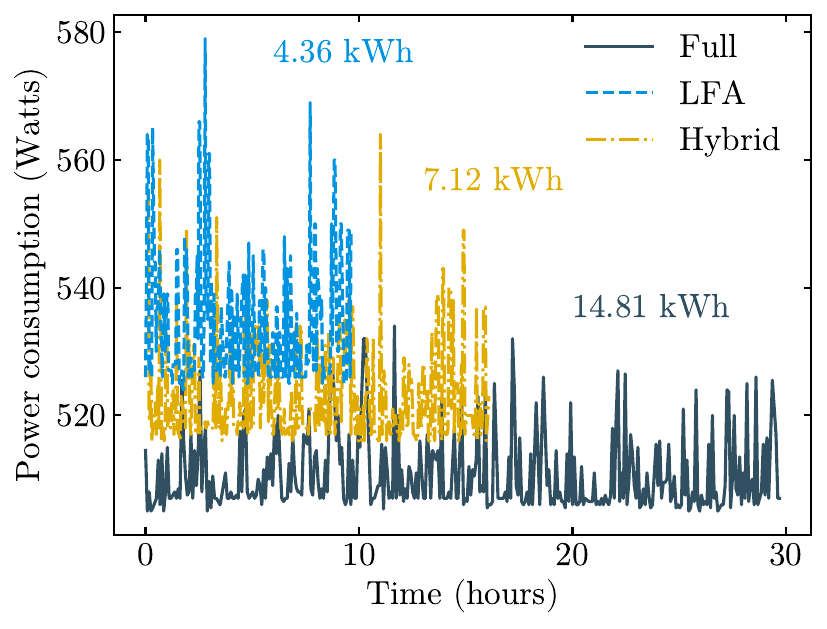}
    \caption{Power consumption, in Watts, for the CPU implementations of the full (gray, solid), LFA (blue, dashed), and hybrid (gold, dash-dotted) responses, as a function of the computational time, in hours. For each approach, a total of 2000 response initializations and likelihood evaluations are performed over a range of total mass $M\in[10^6,10^7]\,\rm{M}_\odot$, mass ratio $q\in[1,10]$, and dimensionless spin components $\chi_{1,2}\in[-1,1]$. The numbers annotated on the axis (4.36, 7.12, and 14.81) represent the total consumed energy in kilowatt-hours associated with each algorithm (LFA, Hybrid, and Full, respectively).}
    \label{fig:power_consumption}
\end{figure}

In this section, we compare the performance of the three response algorithms in terms of resource utilization, focusing on their memory usage and power consumption.\footnote{The power consumption and memory usage values presented in this section have been provided by the High-Performance Computing (HPC) User Portal of the Barcelona Supercomputing Center: \url{https://hpcportal.bsc.es}.} As an example, we choose to profile the same code used to generate the data presented in the upper-row panels of Fig.~\ref{fig:timing_cpu_vs_gpu_tdi2_loglikelihood}. 
It is important to note that, for each likelihood evaluation, the LISA S/C positions are recomputed, a computational step that was not included in the benchmarking results of Fig.\ref{fig:timing_cpu_vs_gpu_tdi2_loglikelihood}, but that is accounted for in the analysis presented here. Nonetheless, the impact of this additional computation is identical across all three response algorithms.

The results for the three approaches are shown in Fig.~\ref{fig:power_consumption}, where 
the power consumption is plotted as a function of runtime. 
The benchmarks were performed on a single node using one CPU core.\footnote{As for the timing benchmarks of Sec.~\ref{subsec:timing}, the power consumption comparisons were performed on a single core, but we requested the full memory capacity of the node.}
We observe that the power usage for the hybrid and LFA approaches increases, reaching 
mean values of approximately 520~W and 530~W, respectively, compared to the full response, which has a mean of around 510~W. 
Consistently with Fig.~\ref{fig:timing_cpu_vs_gpu_tdi2_loglikelihood},
the runtime for the hybrid code is roughly half, and for the LFA approach, about one-third, 
of that of the full response.
To calculate the total consumed energy we integrate the power over time. 
For each of the codes, the results are quoted in the axis and are given in kilowatt-hour 
(kWh).\footnote{Note that $1\,\rm{kWh}=3.6\,\rm{MJ}$.} 
As shown in Fig.~\ref{fig:power_consumption}, the hybrid code consumes 
about half the energy of the full LISA response, while the LFA consumes roughly one-third.

From the measured energy consumption, we estimate the corresponding carbon footprint, expressed as carbon dioxide equivalent (CO$_2$e), for each algorithm. 
As a representative emissions factor, we adopt a value of $210\,\rm{gCO}_2\rm{e}/\mathrm{kWh}$, as the ratio of emitted CO$_2$e (in grams) to total energy generation (in kWh).\footnote{This value, obtained from Fig.~1 of \cite{co2} (accessed on 24 April 2025), reflects the European average for the year 2023.}
Based on this factor, the estimated CO$_2$e emissions are approximately 0.92~kg, 1.5~kg, and 3.1~kg for the LFA, hybrid, and full response algorithms, respectively.

Since this task involves computing the likelihood around 2000 times for different positions of the 
parameter space, the relative values for the energy consumption and CO$_2$ emission between the different codes
can be scaled to a full Bayesian parameter estimation study.

The random-access memory (RAM) peak for the three 
approaches is approximately 42~GB (using double-precision float).

\section{Parameter estimation studies}\label{sec:parameter_estimation_studies}

\begin{table*}[htb]
\renewcommand{\arraystretch}{1.6}  \resizebox{\textwidth}{!}{
\begin{tabular}{cccccccc}
\tableline\tableline
\multirow{3}{*}{\textbf{\renewcommand{\arraystretch}{1}\begin{tabular}[c]{@{}c@{}}Injection\\ (SNR)\end{tabular}}} & \multirow{3}{*}{\textbf{\renewcommand{\arraystretch}{1}\begin{tabular}[c]{@{}c@{}}Response\\ for recovery\end{tabular}}} & \multirow{3}{*}{\renewcommand{\arraystretch}{1.}\begin{tabular}[c]{@{}c@{}}\textbf{Waveform}\\  \textbf{settings}\\ $f_{\rm{min}}$ (Hz)\end{tabular}} & \multirow{3}{*}{\renewcommand{\arraystretch}{1.}\begin{tabular}[c]{@{}c@{}}\textbf{Likelihood}\\  \textbf{integration}\\ $[f_{\rm{min}},f_{\rm{max}}]$ (Hz)\end{tabular}} & \multirow{3}{*}{\renewcommand{\arraystretch}{1.}\begin{tabular}[c]{@{}c@{}}\textbf{Sampling}\\  \textbf{frequency}\\ $f_{\rm{s}}$ (Hz)\end{tabular}} & \multicolumn{2}{c}{\textbf{Computing resources}} & \multirow{3}{*}{\textbf{Runtime}} \\ \cline{6-7}
&  &  & & & \multirow{2}{*}{\renewcommand{\arraystretch}{1}\begin{tabular}[c]{@{}c@{}}\textbf{CPU}\\ \textbf{cores} $\times$ \textbf{nodes} \end{tabular}} & \multirow{2}{*}{\renewcommand{\arraystretch}{1}\begin{tabular}[c]{@{}c@{}}\textbf{GPU}\\ \textbf{NVIDIA H100} \end{tabular}} & \\
& & & & & & & \\\tableline
\multirow{3}{*}{\textbf{\renewcommand{\arraystretch}{1}\begin{tabular}[c]{@{}c@{}}Golden binary\\ (1876)\end{tabular}}} & Full & $8\times 10^{-5}$ & $[10^{-4}, 0.02]$ & 0.04 & 112 $\times$ 1 & $-$ & 3h 18min \\
& Hybrid & $8\times 10^{-5}$ & $[10^{-4}, 0.02]$ & 0.04 & 112 $\times$ 1 & $-$ & 1h 31min \\
& LFA & $8\times 10^{-5}$ & $[10^{-4}, 0.02]$ & 0.04 & $-$ & \checkmark & 9d 8h 25min \\\tableline
\multirow{3}{*}{\textbf{\renewcommand{\arraystretch}{1}\begin{tabular}[c]{@{}c@{}}Deep alert\\ 10 days\\ (24)\end{tabular}}}  & Full & $8\times 10^{-5}$ & $[10^{-4}, 3\times10^{-4}]$ & 0.1 & $-$ & \checkmark & 1d 22h 30min \\
& Hybrid & $8\times 10^{-5}$ & $[10^{-4}, 3\times10^{-4}]$ & 0.1 & $-$ & \checkmark & 1d 10h 4min \\
& LFA & $8\times 10^{-5}$ & $[10^{-4}, 3\times10^{-4}]$ & 0.1 & $-$ & \checkmark & 1d 3h 27min \\\tableline
\multirow{3}{*}{\textbf{\renewcommand{\arraystretch}{1}\begin{tabular}[c]{@{}c@{}}Deep alert\\ 2.5 days\\ (46)\end{tabular}}} & Full & $8\times 10^{-5}$ & $[10^{-4}, 5\times10^{-4}]$ & 0.1 & $-$ & \checkmark & 2d 21h 58min \\
& Hybrid & $8\times 10^{-5}$ & $[10^{-4}, 5\times10^{-4}]$ & 0.1 & $-$ & \checkmark & 1d 7h 50min \\
& LFA & $8\times 10^{-5}$ & $[10^{-4}, 5\times10^{-4}]$ & 0.1 & $-$ & \checkmark & 20h 57min \\\tableline
\multirow{3}{*}{\textbf{\renewcommand{\arraystretch}{1}\begin{tabular}[c]{@{}c@{}}Deep alert\\ 10 hours\\ (90)\end{tabular}}} & Full & $8\times 10^{-5}$ & $[10^{-4},10^{-3}]$ & 0.1 & $-$ & \checkmark & 1d 20h 30min \\
& Hybrid & $8\times 10^{-5}$ & $[10^{-4},10^{-3}]$ & 0.1 & $-$ & \checkmark & 1d 5h 25min \\
& LFA & $8\times 10^{-5}$ & $[10^{-4},10^{-3}]$ & 0.1 & $-$ & \checkmark & 18h 41min \\\tableline\tableline  
\end{tabular}}
\caption{Summary of the settings for the PE runs performed in this work. The first column indicates the different injections along with the corresponding SNR: golden binary (Sec.~\ref{subsec:golden_binary}), and the deep alert cases (Sec.~\ref{subsec:deep_alerts}). While the injected waveform is always projected with the full response, the second column specifies the response model employed for the Bayesian inference process. The minimum frequency of the waveform generation is provided in the third column. The likelihood integration limits, i.e. the frequency lower and upper cutoffs in the scalar product of Eq.~\eqref{eq:scalar_product}, are given in the fourth column, and the sampling frequency for both the waveform and the response is indicated in the fifth column. We also list the computing resources and the runtime. For the CPU cases, the total computational time, in core-hours, can be estimated as the product of the computational resources and the runtime.}\label{tab:pe_settings}
\end{table*}

We analyze the effect of using different LISA responses for Bayesian inference of source parameters.
We assess the performance of the hybrid and LFA approaches by performing signal injections into zero 
noise\footnote{For zero-noise injections, the data measured by the detector, 
generally $d(t)=s(t)+n(t)$ with $s(t)$ the GW signal and $n(t)$ the noise, comprises only the GW signal, i.e. $n(t)=0$. 
This is a simplified case that enables direct analysis of the topology of the likelihood surface without any biases caused by the random noise process.} 
and comparing the results with those of the full response.
For all the cases, we assume 2.0-generation TDI, equal-arm-length orbits, and
unequal and time-dependent delays. 
The injected waveform is always projected with the full response, whereas the recovery is 
performed with the full, hybrid, or LFA algorithms, maintaining the same TDI generation, 
orbits, and time-delay configuration that is used for the injection.

We adopt the methodology outlined in Ref.~\cite{Garcia-Quiros:2025usi} for the parameter estimation runs, which is briefly summarized below. 
We use the function \texttt{bilby.core.sampler.run_sampler} from the \bilby\footnote{
We use the version \texttt{v2.3.0} of \bilby from the repository \url{https://github.com/bilby-dev/bilby} with git commit \texttt{7d45e5c}.
}~\cite{bilby} Python package to sample the likelihood, and
for the sampler, we utilize \ptemcee\footnote{
We use the version \texttt{1.0.0} of \ptemcee from the repository \url{https://github.com/willvousden/ptemcee} with git commit \texttt{c06ffef}.
}~\cite{ptemcee} with Fisher initialization (see Sec.~II~C of~\cite{Garcia-Quiros:2025usi} for details).
For our analyses, we set the number of walkers $\texttt{nwalkers} = 28$ and the number of 
temperatures $\texttt{ntemps} = 8$.

Since the LISA response is implemented in the time-domain, to compute the likelihood we follow the steps \textit{i)-v)} outlined in Sec.~\ref{subsec:timing}: we first evaluate the GW polarizations and the response in the time domain and then we Fourier transform the projected signal to compute the likelihood in the frequency domain (see App.~\ref{subsec:app:scalar_product} for details on the likelihood function).

For simplicity, we adopt uniform priors in all parameters.
The priors on the inclination $\iota$, binary's reference orbital phase $\varphi$, ecliptic latitude $\beta$, 
ecliptic longitude $\lambda$, and polarization angle $\psi$ are uniformly distributed, with
$\iota\in[0,\pi]$, $\varphi\in[0,2\pi]$, $\beta\in[-\pi,\pi]$, $\lambda\in[0,2\pi]$, and 
$\psi\in[-\pi,\pi]$. The limits of the luminosity distance $d_{L}$ and the mass components $m_{1,2}$ 
are specified depending on the injection. 
The prior on the dimensionless spin components $\chi_{1,2}$ are 
uniformly distributed across $\chi_{1,2}\in[-1,1]$.
For the runs that also sample the coalescence time $t_c$, we choose a uniform prior
with limits of 0.75 and 1.25 times the injected value.

In the following, we present the results of Bayesian inference performed on a golden binary with SNR 1876. Additionally, we demonstrate the applicability of our approaches to GW deep alert pipelines by analyzing an injected signal at multiple time intervals before the merger. The specific settings employed for each case, along with the runtime, are summarized in Tab.~\ref{tab:pe_settings}.\footnote{The reported runtimes correspond to the wall-clock sampling time. 
For the CPU-based configurations, the use of 112 computational cores implies that a wall-clock runtime of one hour corresponds to a total of 112 core-hours. 
The computational cost of the runs results, in the case of the golden binary, from employing different LISA response models, at high SNR, for the injected and recovered signals (this high computational demand was previously noted in~\cite{Garcia-Quiros:2025usi}). In contrast, the computational expense of the deep alert runs arises primarily from the faintness of the injected signals, which exhibit low SNRs (24, 46, and 90), thereby increasing the required sampling time. For reference, the align-spin injections considered in~\cite{Garcia-Quiros:2025usi} correspond to an SNR of approximately 2800, and were recovered keeping the same LISA settings within $\mathcal{O}(6-10)$ hours on a GPU. Here, the use of different response codes for the inspiral-merger-ringdown recovery at high SNR and the lower injected SNRs for the deep alert cases significantly increase the computational cost.}

\subsection{Golden binary example}\label{subsec:golden_binary}

\begin{figure*}[t]
    \centering
    \includegraphics[width=\textwidth]{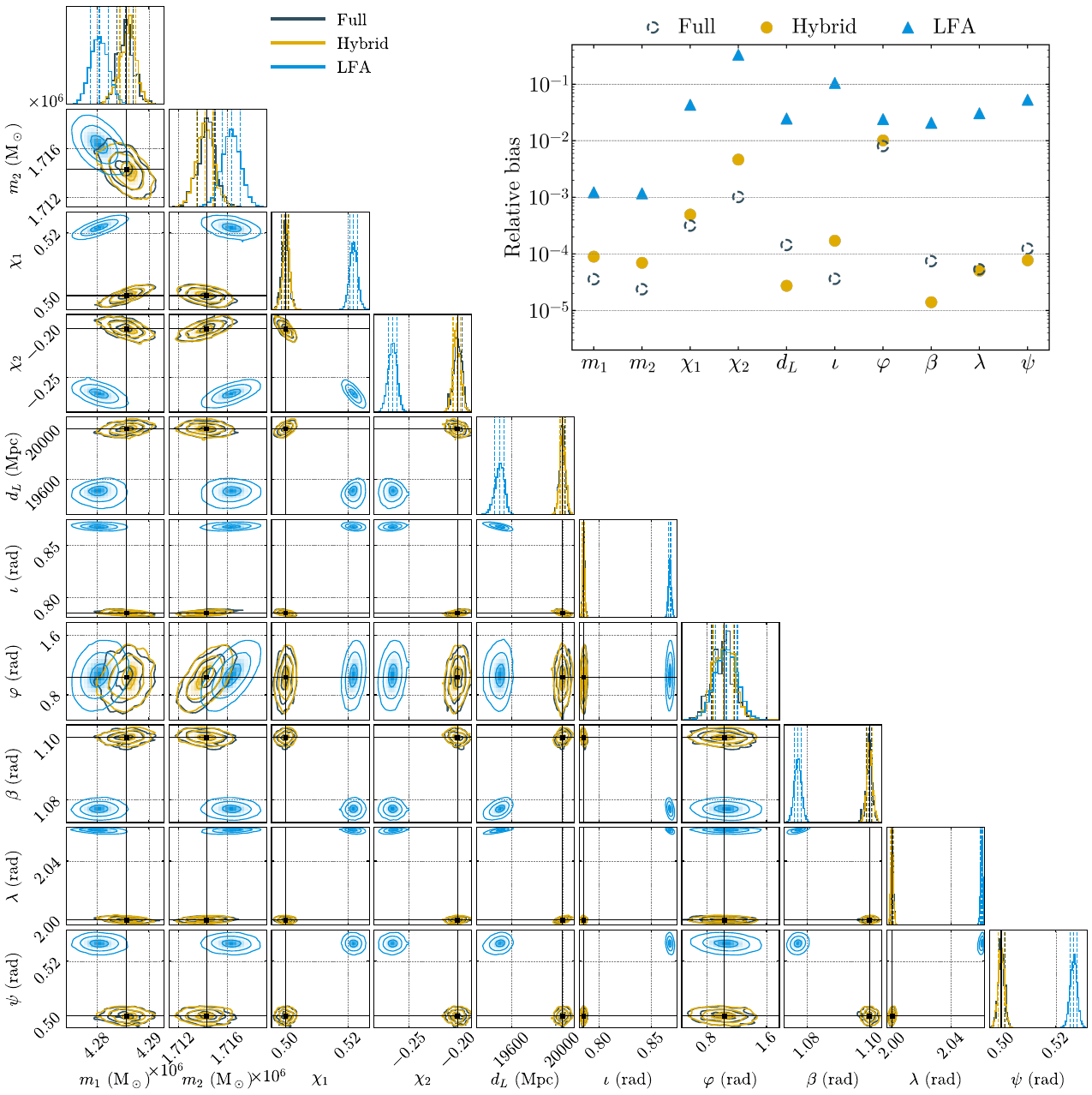}
    \caption{\textit{Corner panel}: Marginalized one- and two-dimensional parameter posterior distributions recovered with the full (gray), hybrid (gold), and LFA (blue) responses, for the MBHB injection described in Sec.~\ref{subsec:golden_binary}. The black cross indicates the true injected parameter. \textit{Upper-right panel}: Relative bias in the recovery of the injected parameters for the full (gray, dashed circumference), hybrid (gold, circle), and LFA (blue, triangle) approaches.}
    \label{fig:pe_golden_binary}
\end{figure*}

As a first example, we perform a full parameter estimation run for a quasi-circular non-precessing MBHB with
$M=6.6\times10^6\,\rm{M}_\odot$ and $q=2.5$ (see Tab.~\ref{tab:golden_binary} for the rest of the parameters). The SNR, computed as
\begin{equation}
    \rho = \sqrt{
    \langle A|A \rangle 
    + \langle E|E \rangle
    + \langle T|T \rangle
    },
\end{equation}
of the injected signal projected with the full LISA response and adopting 2.0 TDI is 1876.
We set the upper-frequency limit at 0.02~Hz to avoid ``0/0'' numerical instabilities
(see \cite{deng2025fastdetectionreconstructionmerging} for a proper 1.5-TDI upper-frequency limit,
and \cite{PhysRevD.103.083011} for a redefinition of the TDI variables that factor these numerical instabilities out).

The frequency at the innermost stable circular orbit (ISCO)~\cite{isco} for this source
is approximately \mbox{$f_{\rm{ISCO}}\approx 3.3\times 10^{-4}\,\rm{Hz}$}.
It will therefore merge at frequencies beyond the limit of applicability of the LFA 
and we expect to see biases in the recovery of the source parameters when applying this response.

\begin{table*}[t]
\setlength{\tabcolsep}{12pt}  \renewcommand{\arraystretch}{1.6}  \begin{tabular}{lrrrr}
\tableline\tableline
\multicolumn{1}{c}{\multirow{2}{*}{\textbf{Parameter}}} & \multicolumn{1}{c}{\multirow{2}{*}{\textbf{\renewcommand{\arraystretch}{1}\begin{tabular}[c]{@{}c@{}}Injected\\ value\end{tabular}}}}& \multicolumn{3}{c}{\textbf{Response for the recovery}}                                                                     \\ \cline{3-5} 
\multicolumn{1}{c}{}                                    & \multicolumn{1}{c}{}                                                                                   & \multicolumn{1}{c}{Full (CPU)}      & \multicolumn{1}{c}{Hybrid (CPU)}    & \multicolumn{1}{c}{LFA (GPU)}     \\ \hline
$m_1$ ($10^6$ M$_\odot$)                                & 4.285714                                                                                               & $4.2856^{+0.0014}_{-0.0017}$ & $4.2861^{+0.0013}_{-0.0017}$ & $4.2805^{+0.0017}_{-0.0017}$ \\
$m_2$ ($10^6$ M$_\odot$)                                & 1.714286                                                                                               & $1.7143^{+0.0006}_{-0.0008}$ & $1.7142^{+0.0007}_{-0.0007}$  & $1.7163^{+0.0008}_{-0.0008}$  \\
$\chi_1$                                                & 0.5                                                                                                    & $0.4998^{+0.0011}_{-0.0011}$   & $0.5002^{+0.0010}_{-0.0011}$   & $0.5217^{+0.0013}_{-0.0012}$    \\
$\chi_2$                                                & $-0.2$                                                                                                 & $-0.200^{+0.004}_{-0.004}$     & $-0.201^{+0.004}_{-0.004}$      & $-0.266^{+0.004}_{-0.005}$     \\
$d_L$ (Mpc)                                             & 20000                                                                                                  & $19997^{+22}_{-19}$           & $20001^{+24}_{-21}$           & $19504^{+33}_{-39}$         \\
$\iota$ (rad)                                           & 0.785398                                                                                               & $0.7854^{+0.0010}_{-0.0009}$ & $0.7853^{+0.0011}_{-0.0010}$   & $0.8679^{+0.0014}_{-0.0013}$   \\
$\varphi$ (rad)                                         & 1.035                                                                                                  & $1.027^{+0.138}_{-0.161}$       & $1.024^{+0.137}_{-0.136}$       & $1.06^{+0.15}_{-0.15}$    \\
$\beta$ (rad)                                         & 1.1                                                                                                    & $1.0999^{+0.0009}_{-0.0009}$  & $1.1000^{+0.0010}_{-0.0008}$  & $1.0770^{+0.0011}_{-0.0011}$  \\
$\lambda$ (rad)                                           & 2.0                                                                                                    & $2.0001^{+0.0007}_{-0.0008}$  & $1.9999^{+0.0010}_{-0.0009}$  & $2.0611^{+0.0006}_{-0.0009}$  \\
$\psi$ (rad)                                            & 0.5                                                                                                    & $0.5000^{+0.0012}_{-0.0011}$   & $0.5000^{+0.0011}_{-0.0011}$  & $0.5265^{+0.0012}_{-0.0013}$  \\ \hline
\multicolumn{2}{l}{$\ln\mathcal{L}$}                                                                                                                            & $1759645.5^{+2.2}_{-2.8}$              & $1759644.2^{+2.0}_{-2.6}$              & $1407784.6^{+1.8}_{-2.5}$            \\
\multicolumn{2}{l}{Runtime}                                                                                                                                & 3h 18min                               & 1h 31min                              & 9d 8h 25min                         \\
\multicolumn{2}{l}{Number of samples}                                                                                                                                   & 11592                                  & 13300                                  & 11732                                \\
\tableline\tableline
\end{tabular}
\caption{Injected, median values, and 90\% credible intervals for the posterior distributions shown in Fig.~\ref{fig:pe_golden_binary}, recovered with the three response models: full, hybrid, and LFA. The displayed parameters are the primary and secondary mass components $m_1$ and $m_2$, the dimensionless spin components $\chi_1$ and $\chi_2$, the luminosity distance $d_L$, the inclination $\iota$, the reference orbital phase $\varphi$, the ecliptic latitude $\beta$, the ecliptic longitude $\lambda$, and the polarization angle $\psi$. Additionally, we show the recovered log-likelihood values, along with the runtime and number of samples. In brackets, we also indicate the hardware used for each response, see Tab.~\ref{tab:pe_settings} for more details.}\label{tab:golden_binary}
\end{table*}

In Fig.~\ref{fig:pe_golden_binary} we show the posterior distributions recovered by the
three algorithms. 
The two-dimensional projections include the $1\sigma$, $2\sigma$, and
$3\sigma$ contour levels, whereas for the one-dimensional marginal distributions we only show the $1\sigma$
level and the median, marked by dashed vertical lines.
The true (injected) parameter values are highlighted by 
solid black lines for reference. 
Additionally, we show a plot to quantify the relative biases in the recovery of the parameters.
From the median of the single-parameter 1D distributions, we subtract the injected value and then we divide 
by the same quantity.
For example, a relative bias of $10^{-3}$ indicates that the parameter estimate deviates from the true value by 0.1\% of the injected value.

In this analysis, we employed the CPU implementations of both the full and hybrid responses, while utilizing the GPU implementation for the LFA.
This choice was motivated by the results shown in Fig.~\ref{fig:overlap_lwa_hybrid_TDI2},
which, for $M=6.6\times10^6\,\rm{M}_\odot$, demonstrate a notable improvement in accuracy for the CPU-based full and hybrid responses.
In contrast, the accuracy gain for the LFA between CPU and GPU implementations 
was comparatively minor, prompting the choice of the GPU version to 
benefit from an accelerated waveform projection.

From Fig.~\ref{fig:pe_golden_binary} we observe the hybrid response achieves 
results consistent with those of the full response.
Concretely, the statistical errors, illustrated by the contour lines in the corner panel, 
and the systematic biases, shown in the upper right panel, are robust between both algorithms.
Overall, the relative biases obtained by the hybrid response are of the same order of 
magnitude as those achieved by the full response.
These values are below $10^{-3}$ for all the recovered parameters except for the 
secondary spin $\chi_2$ and the reference orbital phase $\varphi$, which are below 1\%.

As previously noted, the LFA exhibits systematic biases across all parameters, 
except for $\varphi$.
This is because the approximations that are made for this response, and also for the hybrid, 
do not misalign the waveform when projecting it into the detector frame.
The relative biases obtained with this algorithm exceed 1\% for all the parameters except
for the component masses, yielding values of 0.1\%.

A summary of the results for the three response algorithms, including the corresponding runtimes, is presented in Tab.~\ref{tab:golden_binary}. 

In agreement with the CPU timing benchmarks of Fig.~\ref{fig:timing_cpu_vs_gpu_tdi2_loglikelihood}, 
the inference using the hybrid response required approximately half the computational 
time compared to the full response.
In contrast, the LFA exhibited a longer total sampling time ($\sim 224$ hours) than the total computational time, in core-hours, required for the hybrid ($\sim 168$ hours). 
This discrepancy comes from the fact that the speedup factors reported 
in Fig.~\ref{fig:timing_cpu_vs_gpu_tdi2_loglikelihood} can only be directly extrapolated to 
a full parameter estimation run provided the underlying likelihood surfaces are similar 
across responses.
As evidenced by the posterior distributions in Fig.~\ref{fig:pe_golden_binary}, 
the hybrid and full responses share highly similar likelihood surfaces;
hence preserving the observed factor-of-two improvement in efficiency.
However, in the case of the LFA, the morphology of the likelihood surface 
deviates substantially from that of the full and hybrid responses.
This divergence hinders the sampler to convergence 
and consequently leads to a significant increase in the sampling time.

In summary, the hybrid response yields comparable parameter estimation results 
with the full response at half the CPU computational time even at an SNR of 1876.

\subsection{Deep alert analysis example}\label{subsec:deep_alerts}

\begin{figure*}[t]
    \centering
    \includegraphics[width=\textwidth]{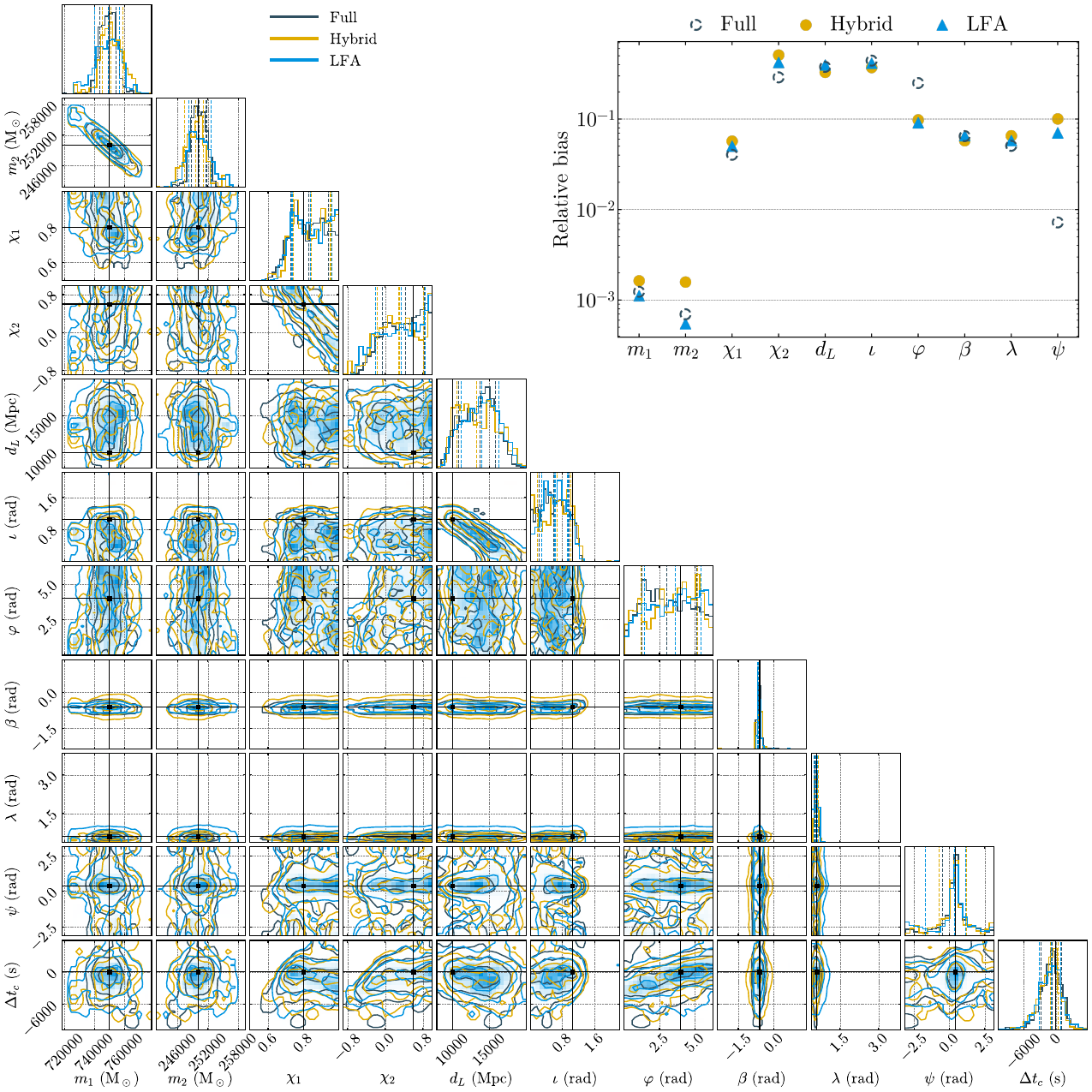}
    \caption{\textit{Corner panel}: Marginalized one- and two-dimensional parameter posterior distributions recovered with the full (gray), hybrid (gold), and LFA (blue) responses, for the pre-merger MBHB injection described in Sec.~\ref{subsec:deep_alerts}.
    We set an upper-frequency limit in the likelihood integration of $f_{\rm{max}}=3\times10^{-4}\,\rm{Hz}$, which is associated with a time to coalescence of approximately 10 days and an SNR of 24. The coalescence time parameter $t_c$ is centered so that $\Delta t_c=0\,\rm{s}$ corresponds to the injection.
    The black cross indicates the true injected parameter. \textit{Upper-right panel}: Relative bias in the recovery of the injected parameters for the full (gray, dashed circumference), hybrid (gold, circle), and LFA (blue, triangle) approaches.}
    \label{fig:pe_early_warining_0dot0003Hz}
\end{figure*}

\begin{figure*}[t]
    \centering
    \includegraphics[width=\textwidth]{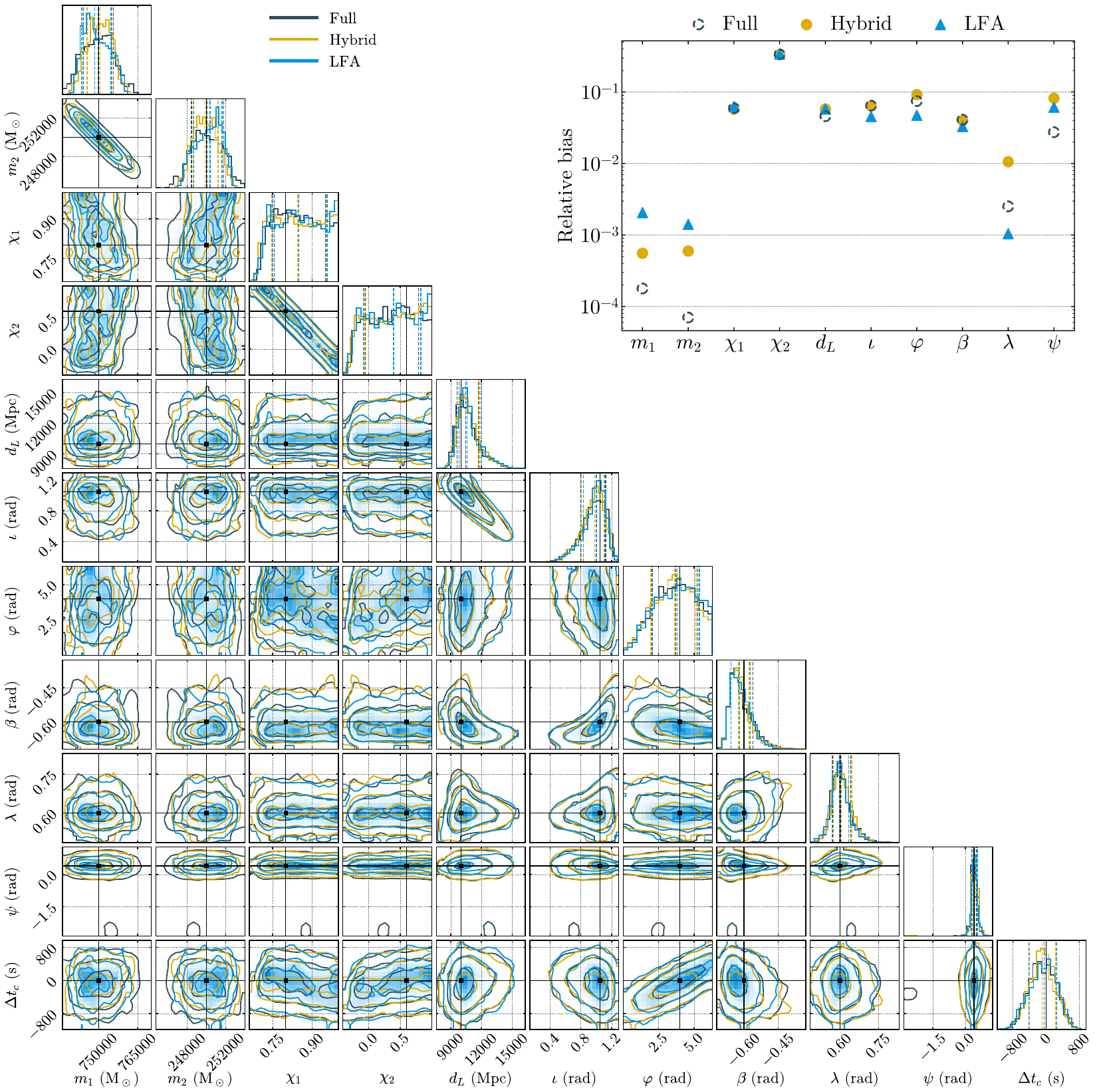}
    \caption{Same as Fig.~\ref{fig:pe_early_warining_0dot0003Hz}, but setting an upper-frequency limit in the likelihood integration of $f_{\rm{max}}=5\times10^{-4}\,\rm{Hz}$, which maps to a time to coalescence of approximately 2.5 days and an SNR of 46.}
    \label{fig:pe_early_warining_0dot0005Hz}
\end{figure*}

\begin{figure*}[t]
    \centering
    \includegraphics[width=\textwidth]{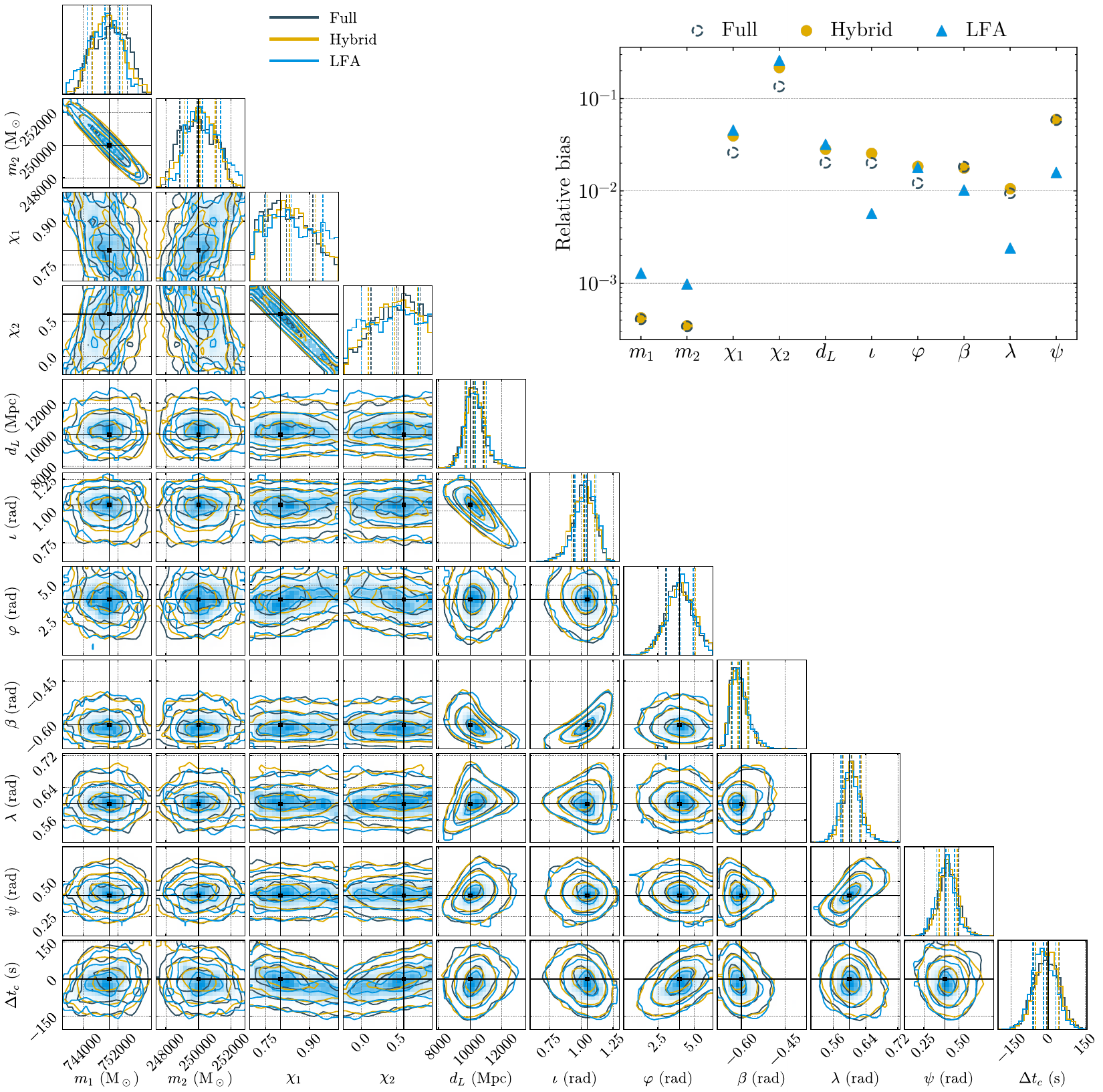}
    \caption{Same as Fig.~\ref{fig:pe_early_warining_0dot0005Hz}, but setting an upper-frequency limit in the likelihood integration of $f_{\rm{max}}=10^{-3}\,\rm{Hz}$, which corresponds to a time to coalescence of approximately 10 hours and an SNR of 90.}
    \label{fig:pe_early_warining_0dot001Hz}
\end{figure*}

\begin{table*}[htb]
\renewcommand{\arraystretch}{1.6}  \resizebox{\textwidth}{!}{
\begin{tabular}{lrrrrrrrrrr}
\tableline\tableline
\multicolumn{1}{c}{\multirow{2}{*}{\textbf{Parameter}}} & \multicolumn{1}{c}{\multirow{2}{*}{\textbf{\renewcommand{\arraystretch}{1}\begin{tabular}[c]{@{}c@{}}Injected\\ value\end{tabular}}}}  & \multicolumn{3}{c}{\textbf{10 days before merger}}                                   & \multicolumn{3}{c}{\textbf{2.5 days before merger}}                                  & \multicolumn{3}{c}{\textbf{10 hours before merger}}                                    \\ \cline{3-11} 
\multicolumn{1}{c}{}                                    & \multicolumn{1}{c}{}                                                                                   & \multicolumn{1}{c}{Full}   & \multicolumn{1}{c}{Hybrid} & \multicolumn{1}{c}{LFA}    & \multicolumn{1}{c}{Full}   & \multicolumn{1}{c}{Hybrid} & \multicolumn{1}{c}{LFA}    & \multicolumn{1}{c}{Full}   & \multicolumn{1}{c}{Hybrid}   & \multicolumn{1}{c}{LFA}    \\ \hline 
$m_1$ ($10^5$ M$_\odot$)                                & 7.5000                                                                                                 & $7.491^{+0.051}_{-0.054}$  & $7.512^{+0.079}_{-0.062}$  & $7.492^{+0.077}_{-0.090}$  & $7.499^{+0.058}_{-0.066}$  & $7.504^{+0.047}_{-0.050}$  & $7.485^{+0.063}_{-0.045}$  & $7.503^{+0.039}_{-0.043}$  & $7.497^{+0.034}_{-0.035}$    & $7.490^{+0.034}_{-0.041}$  \\
$m_2$ ($10^5$ M$_\odot$)                                & 2.5000                                                                                                 & $2.502^{+0.015}_{-0.014}$  & $2.496^{+0.015}_{-0.022}$  & $2.501^{+0.025}_{-0.020}$  & $2.500^{+0.018}_{-0.016}$  & $2.499^{+0.013}_{-0.012}$  & $2.504^{+0.013}_{-0.017}$  & $2.499^{+0.012}_{-0.011}$  & $2.5009^{+0.0095}_{-0.0093}$ & $2.502^{+0.011}_{-0.009}$  \\
$\chi_1$                                                & 0.80                                                                                                   & $0.83^{+0.11}_{-0.10}$     & $0.85^{+0.11}_{-0.10}$     & $0.84^{+0.12}_{-0.10}$     & $0.85^{+0.10}_{-0.09}$     & $0.85^{+0.11}_{-0.10}$     & $0.85^{+0.11}_{-0.10}$     & $0.821^{+0.091}_{-0.076}$  & $0.832^{+0.090}_{-0.077}$    & $0.84^{+0.11}_{-0.09}$     \\
$\chi_2$                                                & 0.60                                                                                                   & $0.43^{+0.41}_{-0.48}$     & $0.29^{+0.52}_{-0.46}$     & $0.33^{+0.51}_{-0.53}$     & $0.40^{+0.39}_{-0.45}$     & $0.40^{+0.43}_{-0.47}$     & $0.40^{+0.43}_{-0.48}$     & $0.52^{+0.31}_{-0.38}$     & $0.47^{+0.33}_{-0.37}$       & $0.44^{+0.36}_{-0.44}$     \\
$d_L$ (Mpc)                                             & 10000                                                                                                  & $13770^{+2070}_{-2910}$    & $13280^{+3070}_{-2690}$    & $13960^{+2330}_{-2810}$    & $10460^{+1300}_{-840}$     & $10580^{+1200}_{-770}$     & $10580^{+1400}_{-780}$     & $10200^{+620}_{-550}$      & $10280^{+640}_{-560}$        & $10320^{+700}_{-550}$      \\
$\iota$ (rad)                                           & 1.0472                                                                                                 & $0.58^{+0.38}_{-0.37}$     & $0.66^{+0.36}_{-0.49}$     & $0.61^{+0.32}_{-0.32}$     & $0.98^{+0.14}_{-0.18}$     & $0.98^{+0.12}_{-0.17}$     & $1.00^{+0.11}_{-0.17}$     & $1.026^{+0.080}_{-0.085}$  & $1.020^{+0.085}_{-0.093}$    & $1.041^{+0.075}_{-0.091}$  \\
$\varphi$ (rad)                                         & 4.0                                                                                                    & $3.0^{+2.1}_{-1.7}$        & $3.6^{+1.6}_{-2.4}$        & $3.6^{+1.8}_{-2.2}$        & $3.7^{+1.5}_{-1.7}$        & $3.6^{+1.5}_{-1.7}$        & $3.8^{+1.6}_{-1.7}$        & $4.0^{+1.0}_{-0.9}$        & $4.1^{+1.0}_{-1.0}$          & $4.07^{+0.85}_{-0.95}$     \\
$\lambda$ (rad)                                         & 0.6                                                                                                    & $0.570^{+0.063}_{-0.052}$  & $0.561^{+0.080}_{-0.056}$  & $0.565^{+0.054}_{-0.068}$  & $0.602^{+0.041}_{-0.030}$  & $0.606^{+0.030}_{-0.032}$  & $0.599^{+0.043}_{-0.030}$  & $0.606^{+0.022}_{-0.023}$  & $0.606^{+0.023}_{-0.021}$    & $0.601^{+0.023}_{-0.023}$  \\
$\beta$ (rad)                                           & $-0.6$                                                                                                   & $-0.639^{+0.057}_{-0.045}$ & $-0.635^{+0.059}_{-0.051}$ & $-0.639^{+0.057}_{-0.044}$ & $-0.625^{+0.063}_{-0.033}$ & $-0.624^{+0.048}_{-0.034}$ & $-0.620^{+0.048}_{-0.037}$ & $-0.611^{+0.033}_{-0.023}$ & $-0.611^{+0.034}_{-0.024}$   & $-0.606^{+0.033}_{-0.024}$ \\
$\psi$ (rad)                                            & 0.4                                                                                                    & $0.40^{+0.79}_{-0.90}$     & $0.4^{+0.8}_{-1.1}$        & $0.4^{+0.8}_{-2.1}$        & $0.41^{+0.12}_{-0.10}$     & $0.43^{+0.11}_{-0.11}$     & $0.42^{+0.13}_{-0.11}$     & $0.423^{+0.071}_{-0.068}$  & $0.424^{+0.065}_{-0.063}$    & $0.406^{+0.063}_{-0.063}$  \\
$\Delta t_c$ (s)                                        & 0                                                                                                      & $-760^{+1750}_{-1960}$     & $-930^{+1400}_{-2140}$     & $-890^{+1720}_{-2210}$     & $-98^{+340}_{-310}$         & $-97^{+330}_{-310}$        & $-46^{+300}_{-370}$        & $-7^{+51}_{-51}$           & $-4^{+46}_{-48}$             & $-19^{+49}_{-43}$          \\ \hline
\multicolumn{2}{l}{$\ln\mathcal{L}$}                                                                                                                            & $267.1^{+2.3}_{2.5}$      & $267.4^{+2.0}_{-2.4}$     & $267.2^{+2.2}_{-2.6}$     & $1067.6^{+2.2}_{-2.8}$    & $1067.6^{+1.9}_{-2.4}$    & $1067.8^{+1.6}_{-2.4}$    & $4129.9^{+1.8}_{-2.3}$    & $4129.8^{+1.8}_{-2.3}$      & $4130.1^{+1.7}_{-2.4}$    \\
\multicolumn{2}{l}{Runtime}                                                                                                                                & 1d 22h 30min               & 1d 10h 4min                & 1d 3h 27min                & 2d 21h 58min               & 1d 7h 50min                & 20h 57min                  & 1d 20h 30min               & 1d 5h 25min                  & 18h 41min                  \\
\multicolumn{2}{l}{Number of samples}                                                                                                                                   & 10024                      & 10444                      & 10164                      & 10724                      & 11564                      & 11228                      & 10836                      & 10248                        & 10780                      \\ \tableline\tableline
\end{tabular}}
\caption{Injected, median values, and 90\% credible intervals for the posterior distributions shown in Fig.~\ref{fig:pe_early_warining_0dot0003Hz} (10 days before the merger), in Fig.~\ref{fig:pe_early_warining_0dot0005Hz} (2.5 days before the merger), and in Fig.~\ref{fig:pe_early_warining_0dot001Hz} (10 hours before the merger), recovered with the three response models: full, hybrid, and LFA. The displayed parameters are the primary and secondary mass components $m_1$ and $m_2$, the dimensionless spin components $\chi_1$ and $\chi_2$, the luminosity distance $d_L$, the inclination $\iota$, the reference orbital phase $\varphi$, the ecliptic latitude $\beta$, the ecliptic longitude $\lambda$, the polarization angle $\psi$, and the bias in the coalescence time $\Delta t_c$. Additionally, we show the recovered log-likelihood values, along with the runtime and number of samples.}
\label{tab:early_alert}
\end{table*}

During LISA's operational phase, pre-merger detection alerts will be issued by 
the low-latency alert pipeline (LLAP).
These alerts will provide a first rough estimate of the parameters of the emitting source, that
will be forwarded to the deep alert analysis pipeline (DAAP).
This pipeline will assess the validity of the detection and, in case of detection, will 
perform a full parameter estimation study to refine the parameter values.

In this section, we test the validity of the hybrid and LFA responses 
in the context of a simplified DAAP.
We assume the detection has already been confirmed and we perform a zero-noise 
injection-recovery analysis of an MBHB signal at different times 
before the merger. (See e.g.~\cite{deng2025fastdetectionreconstructionmerging, PhysRevD.111.043045} for MBHB detection strategies with LISA). 
As in Sec.~\ref{subsec:golden_binary}, we use Fisher initialization for the sampler,
which can be considered as the preliminary parameter estimations provided by the LLAP.
We employ the GPU-accelerated implementation of the response algorithms, as 
the primary objective is to rapidly refine parameter estimates from LAAP to 
send alerts to electromagnetic (EM) telescopes for EM counterpart searches.
In addition, we aim to deliver estimates of the time to coalescence to enable the 
activation of the designated ``protection periods''.
These periods correspond to predefined time windows during which all 
scheduled maintenance operations of the detector constellation are suspended 
to avoid potential overlap with the binary coalescence.
Hence, the scenarios analyzed here correspond to the deep inspiral phase of the binary, 
where low-frequency approximations are valid and the LFA, 
as opposed to Sec.~\ref{subsec:golden_binary}, 
is expected to yield results comparable to that of the full response.

For an MBHB with $M=10^{6}\,\rm{M}_\odot$ and $q=3$, we show in 
Figs.~\ref{fig:pe_early_warining_0dot0003Hz},~\ref{fig:pe_early_warining_0dot0005Hz}, 
and~\ref{fig:pe_early_warining_0dot001Hz} 
the evolution of the posterior distributions and relative biases as the system approaches coalescence.
In the corner plots, we have defined $\Delta t_c$ as the difference between the injected and recovered values for the coalescence time.
The likelihood integration is performed with a fixed lower frequency cut-off of 0.0001 Hz, 
and upper cut-offs of 0.0003 Hz, 0.0005 Hz, and 0.001 Hz, 
corresponding to the approximate times to coalescence $t_c$ 
of 10 days, 2.5 days, and 10 hours, and
SNRs of 24, 46, and 90, respectively.
To ensure consistent comparisons between the response models,
we generate the GW polarizations and the LISA response at the same sampling frequency of 0.1 
Hz (oversampling) in all cases.
We keep the same waveform model and mode content for the waveforms, and identical priors and 
sampling settings, detailed in Sec.~\ref{subsec:golden_binary}, for the PE runs.
The remaining prior distributions are chosen as follows: 
for the component masses, we adopt a uniform prior spanning $\pm30\%$ around the injected
value; for the coalescence time $t_c$, a uniform prior within $\pm25\%$ of the injected
value; and for the luminosity distance, a uniform prior in the range $[1000,20000]$ Mpc.
The injected waveform is always projected using the full LISA response with unequal and 
time-dependent delays, as described in Sec.~\ref{sec:full_response}, while
the recovery is performed with each of the response models keeping the same time-delay configuration. 
The injected and recovered parameter values for each scenario are summarized in Tab.~\ref{tab:early_alert}.

Figs.~\ref{fig:pe_early_warining_0dot0003Hz}-\ref{fig:pe_early_warining_0dot001Hz}
demonstrate good agreement among the posterior distributions obtained using the three response 
models for all the parameters. 
The statistical errors consistently decrease as the system approaches merger, 
with a comparable rate across the three responses.
For instance, the relative statistical error\footnote{The relative statistical error is computed by dividing the width of the 90\% credible interval in the posterior distribution of a parameter by its true injected value} in the parameter $\lambda$
decreases from 19\% to 12\% to 7\% for the full response;
from 23\% to 12\% to 7\% for the hybrid response; and
from 20\% to 11\% to 8\% for the LFA response, 
as the time to coalescence decreases from 10 days to 2.5 days and then to 10 hours.

The relative systematic errors,\footnote{The relative systematic error or relative bias, as commented in Sec.~\ref{subsec:golden_binary}, is computed as the difference between the median of the posterior distribution and the injected value of a given parameter, normalized by the injected value} 
or relative biases with respect to the injected value,
are shown in the upper right panel of \mbox{Figs.~\ref{fig:pe_early_warining_0dot0003Hz}-\ref{fig:pe_early_warining_0dot001Hz}}. 
For the intrinsic parameters, these values remain approximately 
constant from 10 days to 10 hours before the merger.
The individual masses, $m_1$ and $m_2$, show relative biases
on the order of 0.1\%;
whereas the dimensionless spin components have larger relative biases,
ranging from 1\% to 10\% for $\chi_1$, and exceeding 10\% for $\chi_2$.
For the extrinsic parameters, a reduction in the relative systematic errors 
is observed over the analyzed time window. 
Specifically, these parameters show relative biases on the order of 10\% or 
greater approximately 10 days before coalescence. 
As the system approaches the merger, the relative biases decrease
and they fall below 5\% for nearly all extrinsic parameters 
10 hours before the merger.

At very low frequencies, below 0.0003 Hz, the relative biases produced
by the three responses are nearly identical practically for all the parameters, 
as shown in the upper right panel of Fig.~\ref{fig:pe_early_warining_0dot0003Hz}.
However, as the analysis extends toward higher frequencies, 
the LFA starts to deviate from the full and hybrid results, 
as illustrated in the relative bias plot of Figs.~\ref{fig:pe_early_warining_0dot0005Hz}
and~\ref{fig:pe_early_warining_0dot001Hz}.
In general, the LFA is expected to exhibit larger biases than the hybrid response, which itself should show larger biases than the full response.
This hierarchical ordering of systematic errors becomes more pronounced as the binary approaches merger and is first observed in the intrinsic parameters: initially in the individual component masses 
(Fig.~\ref{fig:pe_early_warining_0dot0005Hz}), followed by the spin components
(Fig.~\ref{fig:pe_early_warining_0dot001Hz}).
For the remaining parameters, this bias hierarchy may only become 
apparent later in the binary evolution, or in the post-merger phase, 
as previously demonstrated in Fig.~\ref{fig:pe_golden_binary}.
The cases where the LFA yields lower relative systematic errors than the hybrid or full responses are attributed to the randomness of the sampling process.

The 90\% credible interval of posterior distributions 
for $t_c$ is constrained to approximately 
1 hour 10 days before the merger, 25 minutes 2.5 days before 
the merger, and 1.5 minutes 10 hours before the merger.
In all cases and for the three responses, the statistical error is dominant compared to the 
systematic error, which ranges from approximately 13 minutes 
10 days before the merger, to 1.7 minutes 2.5 days before the merger, 
and 20 seconds 10 hours before the merger.
Note that the precision in the estimation of $t_c$ can be reduced in the 
presence of noise, see \cite{deng2025fastdetectionreconstructionmerging}.

According to the lower panel of Fig.~\ref{fig:timing_cpu_vs_gpu_tdi2_loglikelihood}, for a total mass $M=10^{6}\,\rm{M}_\odot$ the GPU-based LFA 
implementation is 1.8 and 1.5 faster than the full and hybrid responses, 
respectively.
The sampling times reported in Tab.~\ref{tab:early_alert} approximately 
follow these speedup factors, 
except for the full response 2.5 days before the merger, 
where the PE run was significantly slower.
As discussed in Sec.\ref{subsec:golden_binary}, this is 
caused by the substantial differences in the multidimensional 
likelihood surface between the three responses.
In particular, Fig.~\ref{fig:pe_early_warining_0dot0005Hz} shows that
the full response identifies a secondary peak in $\psi$,\footnote{This secondary peak in the polarization angle is an exact $\pi$-degeneracy \cite{PhysRevD.103.083011}. It is centered at $\psi^*\approx\psi-\pi$, which leaves the trigonometric functions invariant in Eqs.~\eqref{eq:hplus_ssb} and~\eqref{eq:hcross_ssb}.} which is not
captured by the hybrid or LFA approaches. 
This additional structure slows the convergence of the sampler relative to the other approaches. 
As a result, the speedup factor does not provide a meaningful comparison of sampling times under these conditions.

In short, the LFA response yields statistically the same results as the full
response for low-frequency pre-merger signals at a significantly reduced computational time.

\section{Discussion}\label{sec:discussion}

In this study, we have developed a hybrid formulation of the LISA response function that employs distinct modeling prescriptions for the low- and high-frequency regimes of the LISA GW spectrum. 
To assess the performance of this approach, we have conducted Bayesian parameter estimation analyses using full inspiral–merger–ringdown waveforms that include higher-order harmonics and aligned-spin configurations. Additionally, we have evaluated the low-frequency component of the response in the context of inspiral-only analyses, relevant to deep alert scenarios.

We improved the speed of likelihood computations by implementing a low-frequency approximation (LFA) to the LISA response which consists of Taylor expanding the projection of the GW polarizations into the constellation links $H_l$ (Eq.~\eqref{eq:H}) in the expression for the single-link measurements $y_{slr}$ (Eq.~\eqref{eq:yslr}) around a chosen evaluation time, such that the resulting expression for $y_{slr}$ corresponds to a central finite difference scheme. 
We used the same technique to rewrite the expressions for the Michelson variables $X$, $Y$, and $Z$, which resulted in simpler expressions proportional to the second and third-time derivative of the GW polarizations for the 1.5- and 2.0-TDI generation, respectively.
With these algebraic expressions, we have not only exploited the connection between TDI and finite differences but have also clarified the origin of the empirical substitution $e^{2i\pi f L/c}\rightarrow e^{4i\pi f L/c}$ in the LISA response that was adopted (at the time of writing) in~\cite{Deng:2025wgk} and in the first version of~\cite{deng2025fastdetectionreconstructionmerging}.
Additionally, in the absence of noise, we have shown that the whitened waveforms computed with 1.5- and 2.0-generation TDI are equivalent in the low-frequency limit, which reinforces previous observations reported in~\cite{Garcia-Quiros:2025usi, PhysRevD.111.044039}.

The LFA loses accuracy as the binary evolves toward the late inspiral, merger, and ringdown phases. 
To prevent that, we hybridize this approach with the full LISA response at a hybridization frequency specified by the user. 
This hybrid response benefits from the speedup of the LFA at early times while reserving the more computationally costly full-response projection just for the latest stages of the evolution, 
where most of the SNR accumulates for MBHBs~\cite{PhysRevD.103.083011}.

Comparisons against the full LISA response for \mbox{$M\in[10^6,10^7]\,\rm{M}_\odot$} and \mbox{$q\in[1,20]$}, both on CPUs and GPUs, yield mismatches below $10^{-5}$ for the hybrid approach, 
while the LFA, characterized by a poor description of the merger-ringdown phase, only reached those low values during the early inspiral. 
The impact on the accuracy of different hybridization parameters confirmed 
that errors grow as the hybridization frequency is pushed toward the late inspiral.

Timing benchmarks were performed in Sec.~\ref{subsec:timing}. 
Results have shown a relative CPU speedup of a factor of 2 for the hybrid and around 3-5 for the LFA with respect to the full response when computing the likelihood. 
On the GPU, the hybrid approach was penalized by data-transfer overhead between the CPU and GPU, due to the small number of data points projected by the full response—an effect that was especially pronounced for high total mass systems. 
This causes the likelihood evaluation times for the hybrid to be similar to those of the full response. 
The LFA, which is not affected by this issue, showed relative speedups of roughly a factor of 2.
In terms of memory usage, both the LFA and the hybrid approach exhibit no significant overhead compared to the full response.
Additionally, our new response models demonstrated reduced energy consumption. 
Specifically, the LFA consumed approximately one-third, and the hybrid about half, of the total energy consumed by the full response, performing the same task. 
Because these benchmarks were conducted on a single CPU core, these results cannot yet be considered representative of a production Bayesian inference or global fit analysis, where several nodes will be busy. Nevertheless, these are promising results that are aligned with the environmental goals of the Distributed Data Processing Center (DDPC) and can potentially contribute to 
reducing the overall carbon footprint of the mission.

We have used these algorithms to simulate parameter recovery for a ``golden binary" and for deep alert pipelines.
In Sec.~\ref{subsec:golden_binary} we have tested the responses by performing a high SNR (1876), $M=6.6\times 10^{6}\,\rm{M}_\odot$, inspiral-merger-ringdown, zero-noise injection with the full response and employing the hybrid and LFA in the inference process, allowing to quantify the potential biases introduced by the new approaches.
The posterior probability distribution of the parameters, shown in Fig.~\ref{fig:pe_golden_binary}, highlights the accuracy of the hybrid response, which yielded statistical errors and relative biases of the order of those achieved by the full response, while requiring only half the computational time.
As expected for injections including the merger-ringdown phase, the LFA leads to a biased recovery of the parameters, except for the phase since there is no misalignment of the projected waveform between responses. Concretely, the relative biases are of the order of $0.1\%$ for the individual mass components and between 2\% and 30\% for the rest of the parameters.

In Sec.~\ref{subsec:deep_alerts} we performed an inspiral-only Bayesian inference study of a single MBHB 10 days, 2.5 days and 10 hours before the coalescence. 
Both the hybrid and LFA approaches were able to constraint the coalescence time of the MBHB merger with a similar accuracy as the full response (with a $1\sigma$ error of about $\sim 30$ min, 10 days before the merger; $\sim 6$ min, 2.5 days before the merger; and $\sim 1$ min, 10 hours before the merger) with lower runtimes.
Note these results are based on zero-noise injections and cannot be directly compared to other more realistic analyses performed in the presence of noise (see e.g.~\cite{deng2025fastdetectionreconstructionmerging}).
Nonetheless, the results presented in Sec.~\ref{subsec:deep_alerts} further support the suitability of a hybrid LISA response, wherein the low-frequency component can be accurately described using an approximate model (as shown in Figs.~\ref{fig:pe_early_warining_0dot0003Hz}-\ref{fig:pe_early_warining_0dot001Hz}), while the high-frequency portion of the spectrum is modeled by a more sophisticated method.

Recently, the implementation of waveform models using \textsc{JAX}~\cite{jax} or \textsc{Julia}~\cite{julia} has gained popularity within the community (see e.g.~\cite{ripple, tenorio_sfts, ann-sur, jaxnrsur, GWjulia}). These frameworks support automatic differentiation, offering the potential for seamless integration with the framework developed in this work.

A key challenge toward developing data analysis algorithms for the actual LISA mission will be to optimize tradeoffs between accuracy and speed in the response evaluation (and also in the waveform model evaluation) in the context of global fit pipelines, e.g. it may not only be useful to treat different MBHB events with different algorithms, but also to tune the tradeoff between accuracy and speed dynamically as the global fit proceeds.

\section*{Acknowledgements}

The a   uthors gratefully thank Cecilio García-Quirós
for providing early access to \phenomxpy and for useful discussions,
and Rodrigo Tenorio for his help during the first stages of the project.
We also thank Senwen Deng, Sylvain Marsat, and Stanislav Babak for clarifications on the
Fourier-domain LISA response; Juan Calderón Bustillo for discussions about data analysis with the Newman-Penrose scalar; and Lorenzo Speri for useful comments on the draft.

We thankfully acknowledge the computer resources at 
MareNostrum and the technical support provided 
by Barcelona Supercomputing Center (BSC) through Grants No. 
AECT-2024-3-0033 and No. AECT-2025-1-0043 from the Red Española Supercomputación (RES).

J. V. is supported by the Spanish Ministry of Universities Grant No. FPU22/02211.
This work was supported by the
Universitat de les Illes Balears (UIB); the Spanish Agencia
Estatal de Investigación Grants No. PID2022-138626NB-
I00, No. RED2022-134204-E, and No. RED2022-134411-T,
funded by Ministerio de Ciencia, Innovación y Universidades
(MICIU)/Agencia Estatal de Investigación (AEI)/10.13039/
501100011033, by the European Social Fund Plus (ESF+)
and European Regional Development Fund (ERDF)/EU; the Comunitat
Autònoma de les Illes Balears through the Conselleria d'Educació
i Universitats with funds from the the European Union
NextGenerationEU/Plan de Recuperación, Transformación
y Resiliencia (PRTR) PRTR-C17.I1 (SINCO2022/6719) and from the 
European Union - ERDF (SINCO2022/18146).

\appendix

\begin{widetext}

\section{Fourier conventions}\label{sec:app:fourier_conventions}

We define the Fourier transform (FT) operator, denoted by $\mathcal{F}$, acting on a time-domain function $x(t)$, and its inverse
to be consistent with the conventions adopted in the 
LIGO Algorithms Library~\cite{lalsuite}:
\begin{subequations}
    \begin{align}
        \tilde{x}(f) = \mathcal{F}[x(t)](f) &= \int_{-\infty}^{\infty} \mathrm{d}t\, e^{-i2\pi f t}x(t)\,, \label{eq:defFT}\\
        x(t) = \mathcal{F}^{-1}[\tilde{x}(f)](t) &= \int_{-\infty}^{\infty} \mathrm{d}f\, e^{+i2\pi f t}\tilde{x}(f)\,. \label{eq:defIFT}
    \end{align}
\end{subequations}

With this convention:
\begin{enumerate}
  \renewcommand{\labelenumi}{\textit{\roman{enumi})}}
  \item The FT of a time-shifted function $x(t-t_0)$ gains an additional, frequency-dependent phase in the Fourier domain with respect to Eq.~\eqref{eq:defFT}: 
  \begin{align}\label{eq:FT_shift_property}
    \mathcal{F}[x(t-t_0)](f)
&= e^{-i  2 \pi f t_0} \tilde{x}(f) \,.
  \end{align}
  \item The FT of the $n^{\rm{th}}$-order time derivative of a function $\mathrm{d}^nx(t)/\mathrm{d}t^n$ adds a frequency-dependent pre-factor in the Fourier domain
  with respect to Eq.~\eqref{eq:defFT}: 
  \begin{align}\label{eq:FT_derivative_property}
    \mathcal{F}\left[\frac{\mathrm{d}^nx(t)}{\mathrm{d}t^n}\right](f)
&= (i2\pi f)^n \tilde{x}(f) \,.
  \end{align}
\end{enumerate}

\section{Taylor-expanded TDI variables} \label{sec:app:expansions}

In this section, we expand the variables $y_{slr}$ and $X,Y,Z$ in powers of a combination of 
$L_l/c$ and $\|\mathbf{R}_{s,r}\|/c$ for both 1.5 and 2.0 TDI generations. 
We then simplify those expressions by assuming the low-frequency approximation described in Sec.~\ref{sec:lwa_response} for two time-delay configurations: equal and constant, and nonequal time-dependent. 
Within this limit, dynamics with timescales much smaller than $\sim 50 \, \rm{min}$ can be ignored, and Taylor expansions of the TDI variables are expected to provide an accurate approximation of the full response. 

Previous studies have investigated Taylor expansions of TDI variables in realistic LISA configurations--accounting for time-varying arm lengths, nested delays, and on-board antialiasing filters--for the purpose of modeling and simulating the residual laser frequency noise in the Michelson variables with remarkable precision (see e.g.~\cite{PhysRevD.99.084023}). In the present work, as an initial study, we consider instead a simplified configuration (as described in Sec.~\ref{sec:full_response}) that enables sufficiently fast response evaluations to support parameter estimation. Specifically, we assess both the accuracy and computational efficiency of a newly derived set of Taylor expansions (presented in this appendix) within an end-to-end parameter estimation pipeline (see Sec.~\ref{sec:parameter_estimation_studies}). Furthermore, as we show in the next sections, this analysis demonstrates that the use of finite-difference stencils offers valuable intuitive insights into the behavior of the TDI mechanism in the low-frequency regime of the LISA sensitivity band.

\subsection{Single-link measurements}\label{sec:app:expansion_yslr}

\subsubsection{Central finite differences}\label{subsec:app:expansion_yslr_central_differences}

Similar to a finite difference algorithm, Eq.~\eqref{eq:yslr} involves the evaluation of $H_l$ at 
different times: $H_l(t-\delta_s(t))$ and $H_l(t-\delta_r(t))$. 
We can rewrite the single-link observables $y_{slr}$ in terms of derivatives of $H_l$ evaluated at 
some retarded time. Where to evaluate the derivatives is crucial to reduce the error of the 
approximation. Motivated by the central finite difference technique, we expect to minimize the 
error by evaluating the derivatives at the center of the domain, i.e. at $\tau$ given by
\begin{equation}\label{eq:tau_t_dependence}
    \tau = t-\frac{\delta_s(t)+\delta_r(t)}{2} = t-\frac{L_l(t)}{2c}-\frac{\hat{\mathbf{k}}\cdot[\mathbf{R}_s(t) + \mathbf{R}_r(t)]}{2c}.
\end{equation}
We expand $H_l(t-\delta_s(t))$ and $H_l(t-\delta_r(t))$ in the numerator of Eq.~\eqref{eq:yslr} around $\tau$:
\begin{subequations}
    \begin{align}
        H_l(t-\delta_s(t)) &= H_l\left(\tau - \frac{\delta_s(t)-\delta_r(t)}{2}\right) = \sum\limits_{m=0}^{\infty}\frac{(-1)^m}{m!}\left[\frac{\delta_s(t)-\delta_r(t)}{2}\right]^m\frac{\mathrm{d}^m}{\mathrm{d}\tau^m}H_l(\tau),\label{eq:central_expand_Hlds}\\
        H_l(t-\delta_r(t)) &= H_l\left(\tau + \frac{\delta_s(t)-\delta_r(t)}{2}\right) = \sum\limits_{m=0}^{\infty}\frac{1}{m!}\left[\frac{\delta_s(t)-\delta_r(t)}{2}\right]^m\frac{\mathrm{d}^m}{\mathrm{d}\tau^m}H_l(\tau). \label{eq:central_expand_Hldr}
    \end{align}
\end{subequations}
Substituting Eqs.~\eqref{eq:central_expand_Hlds} and~\eqref{eq:central_expand_Hldr} in Eq.~\eqref{eq:yslr} we obtain 
\begin{equation}\label{eq:central_series_yslr}
    y_{slr}(t) = \frac{-1}{1-\mathbf{\hat{k}}\cdot\mathbf{\hat{n}}_l(t)}\sum\limits_{m=0}^{\infty}\frac{1}{(2m+1)!}\left[\frac{\delta_s(t)-\delta_r(t)}{2}\right]^{2m+1}\frac{\mathrm{d}^{2m+1}}{\mathrm{d}\tau^{2m+1}}H_l(\tau),
\end{equation}
where the even powers of $[\delta_s(t)-\delta_r(t)]/2$ have canceled out. 
This means that the error after truncating the series would 
not be of the order of the next-leading term, but rather of the next-to-next leading term, which is 
consistent with central finite difference algorithms. Using the definitions of $\delta_s(t)$ and 
$\delta_r(t)$ in Eqs.~\eqref{eq:delta_s} and~\eqref{eq:delta_r}, and approximating 
$L_l(t)\hat{\mathbf{n}}_l(t) \approx \mathbf{R}_r(t) - \mathbf{R}_s(t)$~\cite{Babak:2021mhe,PhysRevD.71.022001}, we can rewrite Eq.~\eqref{eq:central_series_yslr} as
\begin{equation}\label{eq:yslr_with_tau_derivatives}
    y_{slr}(t) = -\frac{L_l(t)}{2c}\sum\limits_{m=0}^{\infty}\frac{1}{(2m+1)!}\left[\frac{L_l(t)}{2c}\left(1-\hat{\mathbf{k}}\cdot\hat{\mathbf{n}}_l(t)\right)\right]^{2m}\frac{\mathrm{d}^{2m+1}}{\mathrm{d}\tau^{2m+1}}H_l(\tau).
\end{equation}
At leading order, the single-link measurements are given by
\begin{equation}\label{eq:yslr_first_order_tau_derivatives}
    y_{slr}(t) = -\frac{L_l(t)}{2c}\frac{\mathrm{d}}{\mathrm{d}\tau}H_l(\tau) + \mathcal{O}\left[\frac{L_l^3}{c^3}\frac{\mathrm{d}^3}{\mathrm{d}\tau^3}H_l\right].
\end{equation}
Note that $\tau$ depends on $t$ through Eq.~\eqref{eq:tau_t_dependence}. 
We can convert $\tau$-derivatives into $t$-derivatives using the chain rule:
\begin{equation}\label{eq:Htauderivative_as_Htderivative}
    \frac{\mathrm{d}H}{\mathrm{d}\tau} = \left(\frac{\mathrm{d}\tau}{\mathrm{d}t}\right)^{-1}\frac{\mathrm{d}H_l}{\mathrm{d}t} = \frac{1}{1-\frac{1}{2c}[\dot{L}_l(t)+\hat{\mathbf{k}}\cdot\dot{\mathbf{R}}_s(t)+\hat{\mathbf{k}}\cdot\dot{\mathbf{R}}_r(t)]}\frac{\mathrm{d}H_l}{\mathrm{d}t},
\end{equation}
where the dot indicates derivation with respect to $t$. Here, $\dot{L}_l < 12 \, \rm{m}\,\rm{s}^{-1}$ is the armlength rate of change as quoted in \cite{redbook}, and $\dot{\mathbf{R}}_{r,s}$ is the spacecraft's linear velocity. We can safely approximate Eq.~\eqref{eq:Htauderivative_as_Htderivative} as
\begin{equation}\label{eq:zero_order_Htauderivative_as_Htderivative}
    \frac{\mathrm{d}H}{\mathrm{d}\tau} = \left[1+\mathcal{O}\left(\frac{\dot{L}_l}{c}\right) + \mathcal{O}\left(\frac{\|\dot{\mathbf{R}}_{r,s}\|}{c}\right)\right]\frac{\mathrm{d}H_l}{\mathrm{d}t},
\end{equation}
since $\dot{L}_l/c\sim\mathcal{O}(10^{-8})$ and 
\mbox{$\hat{\mathbf{k}}\cdot\dot{\mathbf{R}}_{r,s}/c\lesssim\mathcal{O}(R\Omega/c)\sim\mathcal{O}
(10^{-4})$} are much smaller than 1. We have used \mbox{$R=1\,\rm{AU}=1.496\times 10^{11}\,\rm{m}$} 
as the 
magnitude of $\mathbf{R}_{r,s}$, and $\Omega=1.992\times 10^{-7} \, \rm{rad}\, \rm{s}^{-1}$ as the 
magnitude of the constellation's angular velocity. If we substitute 
Eq.~\eqref{eq:zero_order_Htauderivative_as_Htderivative} into Eq.~\eqref{eq:yslr_first_order_tau_derivatives} we 
obtain 
the expression for the single-link observables in terms of time derivatives of $H_l$:
\begin{equation}\label{eq:yslr_without_tau_derivatives}
    \begin{aligned}
        y_{slr}(t) & = -\frac{L_l(t)}{2c}\dot{H}_l\left(t-\frac{L_l(t)}{2c}-\frac{\hat{\mathbf{k}}\cdot(\mathbf{R}_s(t) + \mathbf{R}_r(t))}{2c}\right) \\
        & + \mathcal{O}\left[\frac{L_l\dot{L_l}}{c^2}\dot{H}_l\right]
        + \mathcal{O}\left[\frac{L_l\|\dot{\mathbf{R}}_{r,s}\|}{c^2}\dot{H}_l\right]
        \hspace{0.5cm} \text{error from $\frac{\mathrm{d}}{\mathrm{d}\tau}\approx \frac{\mathrm{d}}{\mathrm{d}t}$ in Eq.~\eqref{eq:zero_order_Htauderivative_as_Htderivative}} \\
        & + \mathcal{O}\left[\frac{L_l^3}{c^3}\dddot{H}_l\right]
        \hspace{0.5cm} \text{error from Taylor series truncation after $\frac{\mathrm{d}^3}{\mathrm{d}\tau^3}\approx \frac{\mathrm{d}^3}{\mathrm{d}t^3}$}.
    \end{aligned}
\end{equation}
The error term proportional to $\dot{L}_l/c$ in Eq.~\eqref{eq:yslr_without_tau_derivatives} gives a 
relative error of $\mathcal{O}(10^{-8})$, whereas the one containing $\|\dot{\mathbf{R}}_{r,s}\|/c$ 
yields a relative error of $\mathcal{O}(10^{-4})$. The series truncation error involves third-time 
derivatives of the strain which are negligible during the deep inspiral for slowly varying signals. 
For non-slowly varying signals such as eccentric binaries, the contribution of higher derivatives of the
strain are larger and more terms in the expansion of Eq.~\eqref{eq:yslr_with_tau_derivatives} may be needed 
depending on the specific accuracy requirements (see App.~\ref{sec:app:non_monochromatic inspiral} for an eccentric example case).
For simplicity, in Eq.~\eqref{eq:yslr_without_tau_derivatives} we omit 
additional error terms arising from the approximation $\mathrm{d}^3/\mathrm{d}\tau^3 \approx 
\mathrm{d}^3/\mathrm{d}t^3$, as these involve, for instance, terms such as 
$(L_l^3\ddot{L}_l^2/c^5)\dot{H}_l$ and $(L_l^3\|\ddot{\mathbf{R}}_{r,s}\|^2/c^5)\dot{H}_l$, which 
contribute with relative errors of $\mathcal{O}(10^{-33})$ and $\mathcal{O}(10^{-17})$, 
respectively\footnote{To compute the relative error of terms involving the spacecraft acceleration 
$\ddot{L}_l$ we have used $\ddot{L}_l\sim 2\times 10^{-9}\,\rm{m}\,\rm{s}^{-2}$ from~\cite{redbook}. 
}. 
Other terms, also involving $\ddot{H}_l$, contribute with even smaller relative error.

In order to obtain $y_{slr}$ in terms of the GW polarizations, we have to substitute $H_l$ in 
Eq.~\eqref{eq:yslr_without_tau_derivatives} by its expression in Eq.~\eqref{eq:H}. 
To do that, we make two approximations on the antenna response functions $\xi_l^{+,\times}$, given 
by Eqs.~\eqref{eq:xi_plus} and~\eqref{eq:xi_cross}. 
First, we ignore any retarded evaluation of $\xi_l^{+,\times}$ assuming $\dot{\xi}_l^{+,\times}$ is 
negligible (this is consistent with other codes, e.g.~\cite{PhysRevD.106.103001}). 
For example, 
\begin{equation}\label{eq:xi_evaluation_error}
    \xi_l^{+,\times}\left(t-\frac{L_l(t)}{2c}-\frac{\hat{\mathbf{k}}\cdot(\mathbf{R}_s(t) + \mathbf{R}_r(t))}{2c}\right) = \xi_l^{+,\times}(t) 
    + \mathcal{O}\left[\frac{L_l}{c}\dot{\xi}_l^{+,\times}\right]
    + \mathcal{O}\left[\frac{\|\mathbf{R}_{r,s}\|}{c}\dot{\xi}_l^{+,\times}\right].
\end{equation}
Second, we neglect the contribution coming from $\dot{\xi}_l^{+}h^{\rm{SSB}}_{+}$ and 
$\dot{\xi}_l^{\times}h^{\rm{SSB}}_{\times}$ when computing $\dot{H}_l$. 
Under these approximations, we obtain
\begin{equation}
    \begin{aligned}
        y_{slr}(t) & = -\frac{L_l(t)}{2c}\left[
        \xi_l^+(t)\dot{h}^{\rm{SSB}}_+\left(t-\frac{L_l(t)}{2c}-\frac{\hat{\mathbf{k}}\cdot(\mathbf{R}_s(t) + \mathbf{R}_r(t))}{2c}\right) 
        + \xi_l^\times(t)\dot{h}^{\rm{SSB}}_\times\left(t-\frac{L_l(t)}{2c}-\frac{\hat{\mathbf{k}}\cdot(\mathbf{R}_s(t) + \mathbf{R}_r(t))}{2c}\right)
        \right] \\
        & + \mathcal{O}\left[\frac{L_l\dot{\xi}_l^{+,\times}}{c}h_{+,\times}^{\rm{SSB}}\right]
        \hspace{0.5cm} \text{error from neglecting $\dot{\xi}_l^{+}h^{\rm{SSB}}_{+}$ and $\dot{\xi}_l^{\times}h^{\rm{SSB}}_{\times}$ when computing $\dot{H}_l$} \\
        & + \mathcal{O}\left[\frac{L_l^2\dot{\xi}_l^{+,\times}}{c^2}\dot{h}_{+,\times}^{\rm{SSB}}\right]
        + \mathcal{O}\left[\frac{L_l\|\mathbf{R}_{r,s}\|\dot{\xi}_l^{+,\times}}{c^2}\dot{h}_{+,\times}^{\rm{SSB}}\right]
        \hspace{0.5cm} \text{error from Eq.~\eqref{eq:xi_evaluation_error}} \\
        & 
        \left.
        \begin{aligned}
            +& \, \mathcal{O}\left[\frac{L_l\dot{L}_l\dot{\xi}_l^{+,\times}}{c^2}h_{+,\times}^{\rm{SSB}}\right]
            + \mathcal{O}\left[\frac{L_l\dot{L}_l}{c^2}\dot{h}_{+,\times}^{\rm{SSB}}\right]\\
            +& \, \mathcal{O}\left[\frac{L_l\|\dot{\mathbf{R}}_{r,s}\|\dot{\xi}_l^{+,\times}}{c^2}h_{+,\times}^{\rm{SSB}}\right]
            + \mathcal{O}\left[\frac{L_l\|\dot{\mathbf{R}}_{r,s}\|}{c^2}\dot{h}_{+,\times}^{\rm{SSB}}\right]
        \end{aligned}
        \hspace{0.5cm} \right\}
        \text{error from $\frac{\mathrm{d}}{\mathrm{d}\tau}\approx \frac{\mathrm{d}}{\mathrm{d}t}$ in Eq.~\eqref{eq:yslr_without_tau_derivatives} after substituting $\dot{H}_l$} \\
        & + \mathcal{O}\left[\frac{L_l^3}{c^3}\dddot{h}_{+,\times}^{\rm{SSB}}\right]
        \hspace{0.5cm} \text{error from Taylor series truncation in Eq.~\eqref{eq:yslr_without_tau_derivatives} after substituting $\dot{H}_l$.}
    \end{aligned}
\end{equation}
In the inspiral regime, where the GW signal is slowly varying, the error term 
proportional to $\dddot{h}^{\rm{SSB}}_{+,\times}$ is negligible compared to the ones 
proportional to $\dot{\xi}_l^{+,\times}$, $\|\dot{\mathbf{R}}_{r,s}\|$, and $\dot{L}_l$, that 
account for the changes in the motion of the constellation. 
The opposite behavior occurs during the merger-ringdown phase, where the fast chirp 
of the signal results in large strain derivatives while LISA is approximately frozen in its orbit.

\subsubsection{Other finite differences}\label{subsec:app:expansion_yslr_other_differences}

Instead of expanding around the middle point of the domain, we could expand the numerator of 
Eq.~\eqref{eq:yslr} around any other point between $t-\delta_s(t)$ and $t-\delta_r(t)$. 
We parametrize this point by $\tau_\epsilon = t + [\delta_s(t)-\delta_r(t)]\epsilon - \delta_s(t)$, 
where $\epsilon\in[0,1]\subset\mathbb{R}$, and $\tau_0=t-\delta_s(t)$ and $\tau_1=t-\delta_r(t)$. 
Note that the results from Sec.~\ref{subsec:app:expansion_yslr_central_differences} correspond to 
the case $\epsilon=1/2$. 
In general, if we expand $H_l(t-\delta_s(t))$ and $H_l(t-\delta_r(t))$ 
around $\tau_\epsilon$ we obtain:
\begin{subequations}
    \begin{align}
        H_l(t - \delta_s(t)) &= H_l(\tau_\epsilon - \epsilon(\delta_s(t)-\delta_r(t))) = \sum\limits_{m=0}^{\infty}\frac{(-1)^m\epsilon^m}{m!}[\delta_s(t)-\delta_r(t)]^m\frac{\mathrm{d}^m}{\mathrm{d}\tau_\epsilon^m}H_l(\tau_\epsilon),\\
        H_l(t - \delta_r(t)) &= H_l(\tau_\epsilon - (\epsilon-1)(\delta_s(t)-\delta_r(t))) = \sum\limits_{m=0}^{\infty}\frac{(-1)^m(\epsilon-1)^m}{m!}[\delta_s(t)-\delta_r(t)]^m\frac{\mathrm{d}^m}{\mathrm{d}\tau_\epsilon^m}H_l(\tau_\epsilon).
    \end{align}
\end{subequations}
For any value of $\epsilon$, the single-link measurements $y_{slr}$ are given by
\begin{equation}
    y_{slr}(t) = -\frac{L_l(t)}{2c}\sum\limits_{m=0}^\infty \frac{(-1)^m[(\epsilon-1)^m-\epsilon^m]}{m!}\left[\frac{L_l(t)}{c}\left(1-\hat{\mathbf{k}}\cdot \hat{\mathbf{n}}_l(t)\right)\right]^{m-1}\frac{\mathrm{d}^m}{\mathrm{d}\tau_\epsilon^m}H_l(\tau_\epsilon).
\end{equation}
The $m=0$ term vanishes for any value of $\epsilon$, while the even $m>0$ terms only do it for 
$\epsilon=1/2$. 
Therefore, if $\epsilon\neq1/2$ the error originated by the Taylor expansion would 
be of the order of the next-leading term, instead of the next-to-next leading term, as described in 
Sec.~\ref{subsec:app:expansion_yslr_central_differences}. 

For $\epsilon=0$ (backward difference), the single-link measurements are given by
\begin{equation}\label{eq:yslr_backward}
    \begin{aligned}
        y_{slr}(t) & = -\frac{L_l(t)}{2c}\dot{H}_l\left(t-\frac{L_l(t)}{c}-\frac{\hat{\mathbf{k}}\cdot\mathbf{R}_s(t)}{c}\right) \\
        & + \mathcal{O}\left[\frac{L_l\dot{L_l}}{c^2}\dot{H}_l\right]
        + \mathcal{O}\left[\frac{L_l\|\dot{\mathbf{R}}_{r,s}\|}{c^2}\dot{H}_l\right]
        \hspace{0.5cm} \text{error from $\frac{\mathrm{d}}{\mathrm{d}\tau}\approx \frac{\mathrm{d}}{\mathrm{d}t}$} \\
        & + \mathcal{O}\left[\frac{L_l^2}{c^2}\ddot{H}_l\right]
        \hspace{0.5cm} \text{error from Taylor series truncation after $\frac{\mathrm{d}^2}{\mathrm{d}\tau^2}\approx \frac{\mathrm{d}^2}{\mathrm{d}t^2}$,}
    \end{aligned}
\end{equation}
and for $\epsilon=1$ (forward difference) by 
\begin{equation}\label{eq:yslr_forward}
    \begin{aligned}
        y_{slr}(t) & = -\frac{L_l(t)}{2c}\dot{H}_l\left(t-\frac{\hat{\mathbf{k}}\cdot\mathbf{R}_r(t)}{c}\right) \\
        & + \mathcal{O}\left[\frac{L_l\|\dot{\mathbf{R}}_{r,s}\|}{c^2}\dot{H}_l\right]
        \hspace{0.5cm} \text{error from $\frac{\mathrm{d}}{\mathrm{d}\tau}\approx \frac{\mathrm{d}}{\mathrm{d}t}$} \\
        & + \mathcal{O}\left[\frac{L_l^2}{c^2}\ddot{H}_l\right]
        \hspace{0.5cm} \text{error from Taylor series truncation after $\frac{\mathrm{d}^2}{\mathrm{d}\tau^2}\approx \frac{\mathrm{d}^2}{\mathrm{d}t^2}$,}
    \end{aligned}
\end{equation}
which is equivalent to Eq.~(51) of~\cite{Babak:2021mhe}.

\subsection{Michelson 1.5 TDI variables}\label{sec:app:expansion_xyz_1tdi}

In this section, we apply the central finite difference procedure described in 
Sec.~\ref{subsec:app:expansion_yslr_central_differences} to the 1.5 TDI Michelson variable 
$X_{1.5}$ ($Y_{1.5}$ and $Z_{1.5}$ are obtained by cyclic permutations of the indexes).
For this section and also for Sec.~\ref{sec:app:expansion_xyz_2tdi} we will keep the dominant error
terms proportional to $\dot{L}_l$ and $\ddot{L}_l$. 

First, we group the terms in Eq.~\eqref{eq:X1.5} that share the same value of $|l|$, 
i.e.\footnote{This 
split was motivated by the footnote~4 of~\cite{PhysRevD.71.022001}.}
\begin{equation}\label{eq:X1.5_l_split}
    X_{1.5}(t) = X_{1.5}(t;l=\pm 2) + X_{1.5}(t;l=\pm 3),
\end{equation}
where
\begin{equation}\label{eq:X1.5l2}
    \begin{aligned}
        X_{1.5}(t;l=\pm 2) & = y_{123}\left(t-\frac{L_{-2}(t)}{c}\right) - y_{123}\left(t-\frac{L_{-2}(t)}{c}-\frac{L_{-3}(t)}{c}-\frac{L_{3}(t)}{c}\right)\\
        & + y_{3-21}(t) - y_{3-21}\left(t-\frac{L_{-3}(t)}{c}-\frac{L_{3}(t)}{c}\right),\\
    \end{aligned}
\end{equation}
and
\begin{equation}\label{eq:X1.5l3}
    \begin{aligned}
         X_{1.5}(t;l=\pm 3) & = y_{1-32}\left(t-\frac{L_{-2}(t)}{c}-\frac{L_2(t)}{c}-\frac{L_3(t)}{c}\right) - y_{1-32}\left(t-\frac{L_3(t)}{c}\right)\\
         & + y_{231}\left(t-\frac{L_2(t)}{c}-\frac{L_{-2}(t)}{c}\right) - y_{231}(t).
    \end{aligned}
\end{equation}
Eqs.~\eqref{eq:X1.5l2} and~\eqref{eq:X1.5l3} involve the evaluation of $y_{slr}$ and  $y_{r-ls}$ at 
different retarded times. As in Sec.~\ref{subsec:app:expansion_yslr_central_differences}, we 
expand them around the middle points of the corresponding $|l|$-domain which are
\begin{subequations}
    \begin{align}
        \tau_{2} &= t-\frac{L_{-3}(t)+L_{-2}(t)+L_{3}(t)}{2c}, \\
        \tau_{3} &= t-\frac{L_{-2}(t)+L_{2}(t)+L_{3}(t)}{2c},
    \end{align}       
\end{subequations}
for $l=\pm2$ and $l=\pm 3$, respectively.
For example:
\begin{equation}
    \begin{aligned}
         y_{1-32}\left(t-\frac{L_{-2}(t)}{c}-\frac{L_2(t)}{c}-\frac{L_3(t)}{c}\right) &= y_{1-32}\left(\tau_{3} - \frac{L_{-2}(t)+L_2(t)+L_3(t)}{2c}\right) \\
         & = \sum\limits_{m=0}^{\infty}\frac{(-1)^m}{m!}\left[\frac{L_{-2}(t)+L_2(t)+L_3(t)}{2c}\right]^m\frac{\mathrm{d}^m}{\mathrm{d}\tau_{3}^m}y_{1-32}(\tau_{3}).
\end{aligned}
\end{equation}

We repeat this for each of the 8 terms in Eqs.~\eqref{eq:X1.5l2} and~\eqref{eq:X1.5l3}. We then 
compute $X_{1.5}$ from Eq.~\eqref{eq:X1.5_l_split}, which at leading order reads
\begin{equation}
    \begin{aligned}
        X_{1.5}(t) & = \left[\frac{L_{-3}(t)+L_3(t)}{c}\right] \left[\frac{\mathrm{d}}{\mathrm{d}\tau_{2}}y_{123}(\tau_{2})-\frac{L_{-2}(t)}{2c} \frac{\mathrm{d}^2}{\mathrm{d}\tau^2_{2}}y_{123}(\tau_{2})\right] \\
        & + \left[\frac{L_{-3}(t)+L_3(t)}{c}\right] \left[\frac{\mathrm{d}}{\mathrm{d}\tau_{2}}y_{3-21}(\tau_{2})+\frac{L_{-2}(t)}{2c} \frac{\mathrm{d}^2}{\mathrm{d}\tau^2_{2}}y_{3-21}(\tau_{2})\right] \\
        & - \left[\frac{L_{-2}(t)+L_2(t)}{c}\right] \left[\frac{\mathrm{d}}{\mathrm{d}\tau_{3}}y_{1-32}(\tau_{3})-\frac{L_3(t)}{2c} \frac{\mathrm{d}^2}{\mathrm{d}\tau^2_{3}}y_{1-32}(\tau_{3})\right] \\
        & - \left[\frac{L_{-2}(t)+L_2(t)}{c}\right] \left[\frac{\mathrm{d}}{\mathrm{d}\tau_{3}}y_{231}(\tau_{3})+\frac{L_3(t)}{2c} \frac{\mathrm{d}^2}{\mathrm{d}\tau^2_{3}}y_{231}(\tau_{3})\right] \\
        & + \mathcal{O}\left[\frac{L_l^3}{c^3}\frac{\mathrm{d}^3}{\mathrm{d}\tau^3_{2,3}}y_{slr}(\tau_{2,3})\right],
    \end{aligned}
\end{equation}
in terms of derivatives with respect to $\tau_2$ and $\tau_3$, and 
\begin{equation}\label{eq:X1.5_time_derivatives}
    \begin{aligned}
        X_{1.5}(t) & = \left[\frac{L_{-3}(t)+L_3(t)}{c}\right] \left[\dot{y}_{123}\left(t-\frac{L_{-3}(t)+L_{-2}(t)+L_{2}(t)}{2c}\right)-\frac{L_{-2}(t)}{2c} \ddot{y}_{123}\left(t-\frac{L_{-3}(t)+L_{-2}(t)+L_{2}(t)}{2c}\right)\right] \\
        & + \left[\frac{L_{-3}(t)+L_3(t)}{c}\right] \left[\dot{y}_{3-21}\left(t-\frac{L_{-3}(t)+L_{-2}(t)+L_{2}(t)}{2c}\right)+\frac{L_{-2}(t)}{2c} \ddot{y}_{3-21}\left(t-\frac{L_{-3}(t)+L_{-2}(t)+L_{2}(t)}{2c}\right)\right] \\
        & - \left[\frac{L_{-2}(t)+L_2(t)}{c}\right] \left[\dot{y}_{1-32}\left(t-\frac{L_{-2}(t)+L_{2}(t)+L_{3}(t)}{2c}\right)-\frac{L_3(t)}{2c} \ddot{y}_{1-32}\left(t-\frac{L_{-2}(t)+L_{2}(t)+L_{3}(t)}{2c}\right)\right] \\
        & - \left[\frac{L_{-2}(t)+L_2(t)}{c}\right] \left[\dot{y}_{231}\left(t-\frac{L_{-2}(t)+L_{2}(t)+L_{3}(t)}{2c}\right)+\frac{L_3(t)}{2c} \ddot{y}_{231}\left(t-\frac{L_{-2}(t)+L_{2}(t)+L_{3}(t)}{2c}\right)\right] \\
        & + \mathcal{O}\left[\frac{L_l\dot{L}_l}{c^2}\dot{y}_{slr}\right] 
        \hspace{0.5cm}\text{error from $\frac{\mathrm{d}}{\mathrm{d}\tau_{2,3}}\approx \frac{\mathrm{d}}{\mathrm{d}t}$} \\
        & 
        \left.
        \begin{aligned}
            +& \, \mathcal{O}\left[\frac{L_l^2\ddot{L}_l}{c^3}\dot{y}_{slr}\right]
            + \mathcal{O}\left[\frac{L_l^2\dot{L}\ddot{L}_l}{c^4}\dot{y}_{slr}\right]\\
            +& \, \mathcal{O}\left[\frac{L_l^2\dot{L}_l}{c^3}\ddot{y}_{slr}\right]
            + \mathcal{O}\left[\frac{L_l^2\dot{L}_l^2}{c^4}\ddot{y}_{slr}\right]
        \end{aligned}
        \hspace{0.5cm} \right\}
        \text{error from $\frac{\mathrm{d}^2}{\mathrm{d}\tau^2_{2,3}}\approx \frac{\mathrm{d}^2}{\mathrm{d}t^2}$} \\
        & + \mathcal{O}\left[\frac{L_l^3\ddot{L}_l}{c^4}\ddot{y}_{slr}\right]
        + \mathcal{O}\left[\frac{L_l^3\dot{L}_l\ddot{L}_l}{c^5}\ddot{y}_{slr}\right]
        + \mathcal{O}\left[\frac{L_l^3}{c^3}\dddot{y}_{slr}\right]
        \hspace{0.5cm} \text{error from Taylor series truncation after $\frac{\mathrm{d}^3}{\mathrm{d}\tau^3}\approx \frac{\mathrm{d}^3}{\mathrm{d}t^3}$,}
    \end{aligned}
\end{equation}
in terms of time derivatives.

\subsubsection{Low-frequency approximation}\label{subsubsec:app:lwa_1.5}

In the low-frequency regime, a series of approximations can be applied to the LISA response. In this 
section, we simplify Eq.~\eqref{eq:X1.5_time_derivatives} by assuming link reversal symmetry ($L_l\simeq L_{-l}$) in two 
time-delay configurations: one with unequal and time-dependent delays, and the other with equal and 
constant delays.

The effective armlengths experienced by light propagating between spacecraft, up to second order in 
eccentricity, are given by \cite{PhysRevD.67.022001, PhysRevD.71.022001} 
\begin{equation}\label{eq:L_of_t}
    L_l(t) = L + \frac{1}{32}(eL)\sin(3\Omega t - 3\xi_0) + \left[\left(\frac{R\Omega L}{c}\right)\text{sgn}(l) - \frac{15}{32}(eL)\right]\sin(\Omega t - \delta_{|l|}),
\end{equation}
where $\Omega=2\pi \, \rm{yr}^{-1}$ is the LISA orbital angular velocity, $e=0.004811$ is the orbital eccentricity, and $\xi_0$ and 
$\delta_{|l|}$ are parameters defined in~\cite{PhysRevD.71.022001, PhysRevD.67.022001}. 

As described in Sec.~\ref{sec:lwa_response}, at low frequencies we can approximate $L_l\simeq L_{-
l}$~\cite{PhysRevD.103.083011} and consequently, $y_{slr}\simeq y_{r-ls}$. According to 
Eq.~\eqref{eq:L_of_t}
\begin{equation}\label{eq:link_reversal_sym_error_L}
     L_{-l}(t) = L_{l}(t) + \mathcal{O}\left[\frac{R\Omega}{c}L\right],
\end{equation}
which combined with Eq.~\eqref{eq:yslr_without_tau_derivatives} yields
\begin{equation}\label{eq:link_reversal_sym_error_yslr}
    y_{r-ls}(t) = y_{slr}(t) + \mathcal{O}\left[\frac{R\Omega}{c}y_{slr}\right].
\end{equation}
From this, we estimate that assuming the link reversal symmetry amounts to relative errors of $\mathcal{O}
(R\Omega/c)\sim 10^{-4}$ (0.01\%) for the single-link measurements $y_{slr}$. Similarly, $X_{1.5}$ 
simplifies to
\begin{equation}\label{eq:X15_nonequal_timedelays}
    \begin{aligned}
        X_{1.5}(t) & = \frac{4}{c} \left[ L_3(t) \dot{y}_{123}\left(t-\frac{L_3(t)}{c}-\frac{L_2(t)}{2c}\right)- L_2(t) \dot{y}_{231}\left(t-\frac{L_2(t)}{c}-\frac{L_3(t)}{2c}\right)\right] \\
        & + \mathcal{O}\left[\frac{R\Omega L}{c^2}\dot{y}_{slr}\right]
        + \mathcal{O}\left[\frac{R\Omega L^2}{c^3}\ddot{y}_{slr}\right]
        \hspace{0.5cm}\text{error from $L_l\simeq L_{-l}$ and $y_{slr}\simeq y_{r-ls}$} \\
        & + \mathcal{O}\left[\frac{L_l\dot{L}_l}{c^2}\dot{y}_{slr}\right] 
        \hspace{0.5cm}\text{error from $\frac{\mathrm{d}}{\mathrm{d}\tau_{2,3}}\approx \frac{\mathrm{d}}{\mathrm{d}t}$} \\
        & 
        \left.
        \begin{aligned}
            +& \, \mathcal{O}\left[\frac{L_l^2\ddot{L}_l}{c^3}\dot{y}_{slr}\right]
            + \mathcal{O}\left[\frac{L_l^2\dot{L}\ddot{L}_l}{c^4}\dot{y}_{slr}\right]\\
            +& \, \mathcal{O}\left[\frac{L_l^2\dot{L}_l}{c^3}\ddot{y}_{slr}\right]
            + \mathcal{O}\left[\frac{L_l^2\dot{L}_l^2}{c^4}\ddot{y}_{slr}\right]
        \end{aligned}
        \hspace{0.5cm} \right\}
        \text{error from $\frac{\mathrm{d}^2}{\mathrm{d}\tau^2_{2,3}}\approx \frac{\mathrm{d}^2}{\mathrm{d}t^2}$} \\
        & + \mathcal{O}\left[\frac{L_l^3\ddot{L}_l}{c^4}\ddot{y}_{slr}\right]
        + \mathcal{O}\left[\frac{L_l^3\dot{L}_l\ddot{L}_l}{c^5}\ddot{y}_{slr}\right]
        + \mathcal{O}\left[\frac{L_l^3}{c^3}\dddot{y}_{slr}\right]
        \hspace{0.5cm} \text{error from Taylor series truncation after $\frac{\mathrm{d}^3}{\mathrm{d}\tau^3}\approx \frac{\mathrm{d}^3}{\mathrm{d}t^3}$,}
    \end{aligned}
\end{equation}
where we have inserted Eqs.~\eqref{eq:link_reversal_sym_error_L} and~\eqref{eq:link_reversal_sym_error_yslr} in Eq.~\eqref{eq:X1.5_time_derivatives}. 
The larger relative error contribution of adopting this approximation in the response is 
$\mathcal{O}(R\Omega/c)\sim 10^{-4}$ (0.01\%), which is consistent with 
Fig.~\ref{fig:cpu_visual_agreement_TDI1}. 
This contribution grows as the binary approaches the 
latest stages of coalescence, and error terms proportional to higher-order derivatives of $y_{slr}$ 
dominate during the merger-ringdown phase.

A further simplification of the response consists of assuming equal and constant delays. According 
to Eq.~\eqref{eq:L_of_t} this implies
\begin{equation}
    L_l(t) = L + \mathcal{O}\left[eL\right] + \mathcal{O}\left[\frac{\Omega R L}{c}\right],
\end{equation}
which originates errors of 0.5\% of the nominal armlength $L$. In this case, $X_{1.5}$ is computed from
\begin{equation}\label{eq:X15_equal_timedelays}
    \begin{aligned}
        X_{1.5}(t) & = \frac{4L}{c} \left[\dot{y}_{123}\left(t-\frac{3L}{2c}\right) - \dot{y}_{231}\left(t-\frac{3L}{2c}\right)\right]\\
        & + \mathcal{O}\left[\frac{e L}{c}\dot{y}_{slr}\right]
        \hspace{0.5cm}\text{error from $L_l(t)=L$} \\
        & + \mathcal{O}\left[\frac{R\Omega L}{c^2}\dot{y}_{slr}\right]
        + \mathcal{O}\left[\frac{R\Omega L^2}{c^3}\ddot{y}_{slr}\right]
        \hspace{0.5cm}\text{error from $L_l\simeq L_{-l}$ and $y_{slr}\simeq y_{r-ls}$} \\
        & + \mathcal{O}\left[\frac{L_l\dot{L}_l}{c^2}\dot{y}_{slr}\right] 
        \hspace{0.5cm}\text{error from $\frac{\mathrm{d}}{\mathrm{d}\tau_{2,3}}\approx \frac{\mathrm{d}}{\mathrm{d}t}$} \\
        & 
        \left.
        \begin{aligned}
            +& \, \mathcal{O}\left[\frac{L_l^2\ddot{L}_l}{c^3}\dot{y}_{slr}\right]
            + \mathcal{O}\left[\frac{L_l^2\dot{L}\ddot{L}_l}{c^4}\dot{y}_{slr}\right]\\
            +& \, \mathcal{O}\left[\frac{L_l^2\dot{L}_l}{c^3}\ddot{y}_{slr}\right]
            + \mathcal{O}\left[\frac{L_l^2\dot{L}_l^2}{c^4}\ddot{y}_{slr}\right]
        \end{aligned}
        \hspace{0.5cm} \right\}
        \text{error from $\frac{\mathrm{d}^2}{\mathrm{d}\tau^2_{2,3}}\approx \frac{\mathrm{d}^2}{\mathrm{d}t^2}$} \\
        & + \mathcal{O}\left[\frac{L_l^3\ddot{L}_l}{c^4}\ddot{y}_{slr}\right]
        + \mathcal{O}\left[\frac{L_l^3\dot{L}_l\ddot{L}_l}{c^5}\ddot{y}_{slr}\right]
        + \mathcal{O}\left[\frac{L_l^3}{c^3}\dddot{y}_{slr}\right]
        \hspace{0.5cm} \text{error from Taylor series truncation after $\frac{\mathrm{d}^3}{\mathrm{d}\tau^3}\approx \frac{\mathrm{d}^3}{\mathrm{d}t^3}$,}
    \end{aligned}
\end{equation}
where the relative error is of order $\mathcal{O}(e)\sim 10^{-3}$, which agrees with 
Fig.~\ref{fig:cpu_visual_agreement_TDI1}. 

In this approximation, the Michelson variables are 
proportional to second-time derivatives of the GW polarizations ($\psi_4$):
\begin{equation}
    X_{1.5}(t) \sim \ddot{h}_{+,\times}\left(t-\frac{2L}{c}-\frac{\hat{\mathbf{k}}\cdot\mathbf{R}_0(t)}{c}\right),
\end{equation}
where we have used that in this limit
\begin{equation}\label{eq:Rs_Rr_R0}
    \frac{\hat{\mathbf{k}}\cdot\mathbf{R}_s(t)+\hat{\mathbf{k}}\cdot\mathbf{R}_r(t)}{2c} \approx \frac{\hat{\mathbf{k}}\cdot\mathbf{R}_0(t)}{c}.
\end{equation}
This behavior is shared not only with $Y_{1.5}$ and $Z_{1.5}$ but also with 
$A_{1.5}$, $E_{1.5}$, and $T_{1.5}$, since these are linear outputs of the former ones.

\subsection{Michelson 2.0 TDI variables}\label{sec:app:expansion_xyz_2tdi}

Following the same procedure described in Sec.~\ref{sec:app:expansion_xyz_1tdi}, here we derive the 
expression for $X_{2.0}$.
We first group the $y_{slr}$ with the same $|l|$ in Eq.~\eqref{eq:X2.0} and, for each subset, we 
expand the terms around the center of the corresponding finite difference domain. After recombining 
them, we obtain
\begin{equation}\label{eq:X2.0_with_tau_derivatives}
    \begin{aligned}
        X_{2.0}(t) & = \left[\frac{L_{-3}(t)+L_{-2}(t)+L_{2}(t)+L_{3}(t)}{c}\right] \left\{\vphantom{\frac{\left(\frac{1}{2}\right)}{2}}\right. \\
        & -\left[\frac{L_{-2}(t)+L_2(t)}{c}\right] \left[\frac{\mathrm{d}^2}{\mathrm{d}\tau_{3}^2}y_{1-32}(\tau_{3}) - \frac{L_3(t)}{2c}\frac{\mathrm{d}^3}{\mathrm{d}\tau_{3}^3}y_{1-32}(\tau_{3}) \right] \\
        & -\left[\frac{L_{-2}(t)+L_2(t)}{c}\right] \left[\frac{\mathrm{d}^2}{\mathrm{d}\tau_{3}^2}y_{231}(\tau_{3}) + \frac{L_3(t)}{2c}\frac{\mathrm{d}^3}{\mathrm{d}\tau_{3}^3}y_{231}(\tau_{3}) \right] \\
        & +\left[\frac{L_{-3}(t)+L_3(t)}{c}\right] \left[\frac{\mathrm{d}^2}{\mathrm{d}\tau_{2}^2}y_{123}(\tau_{2}) - \frac{L_{-2}(t)}{2c}\frac{\mathrm{d}^3}{\mathrm{d}\tau_{2}^3}y_{123}(\tau_{2}) \right] \\
        & + \left.\left[\frac{L_{-3}(t)+L_3(t)}{c}\right] \left[\frac{\mathrm{d}^2}{\mathrm{d}\tau_{2}^2}y_{3-21}(\tau_{2}) + \frac{L_{-2}(t)}{2c}\frac{\mathrm{d}^3}{\mathrm{d}\tau_{2}^3}y_{3-21}(\tau_{2}) \right]\vphantom{\frac{\left(\frac{1}{2}\right)}{2}}\right\} \\
        & + \mathcal{O}\left[\frac{L_l^4}{c^4}\frac{\mathrm{d}^4}{\mathrm{d}\tau_{2,3}^4}y_{slr}(\tau_{2,3})\right],\\
    \end{aligned}
\end{equation}
where we have defined
\begin{subequations}
    \begin{align}
        \tau_{2} &= t-\frac{L_{-3}(t)}{c}-\frac{L_{-2}(t)}{c}-\frac{L_3(t)}{c}-\frac{L_2(t)}{2c}, \\
        \tau_{3} &= t-\frac{L_{-2}(t)}{c}-\frac{L_2(t)}{c}-\frac{L_3(t)}{c}-\frac{L_{-3}(t)}{2c}.
    \end{align}       
\end{subequations}
According to Eq.~\eqref{eq:X2.0_with_tau_derivatives}, unlike for 1.5 TDI, the 2.0 TDI variables are, at leading-order, proportional to second derivatives 
 of the single-link measurements. In terms of time derivatives, 
 Eq.~\eqref{eq:X2.0_with_tau_derivatives} becomes
\begin{equation}
    \begin{aligned}
        X_{2.0}(t) & = \left[\frac{L_{-3}(t)+L_{-2}(t)+L_{2}(t)+L_{3}(t)}{c}\right] \left\{\vphantom{\frac{\left(\frac{1}{2}\right)}{2}}\right. \\
        & -\left[\frac{L_{-2}(t)+L_2(t)}{c}\right] 
        \left[ \ddot{y}_{1-32}\left(t-\frac{L_{-2}(t)}{c}-\frac{L_2(t)}{c}-\frac{L_3(t)}{c}-\frac{L_{-3}(t)}{2c}\right) \right.\\
        & \hspace{9em} \left. - \frac{L_3(t)}{2c}\dddot{y}_{1-32}\left(t-\frac{L_{-2}(t)}{c}-\frac{L_2(t)}{c}-\frac{L_3(t)}{c}-\frac{L_{-3}(t)}{2c}\right) \right] \\
        & -\left[\frac{L_{-2}(t)+L_2(t)}{c}\right] \left[\ddot{y}_{231}\left(t-\frac{L_{-2}(t)}{c}-\frac{L_2(t)}{c}-\frac{L_3(t)}{c}-\frac{L_{-3}(t)}{2c}\right)\right. \\
        & \hspace{9em} \left. + \frac{L_3(t)}{2c}\dddot{y}_{231}\left(t-\frac{L_{-2}(t)}{c}-\frac{L_2(t)}{c}-\frac{L_3(t)}{c}-\frac{L_{-3}(t)}{2c}\right) \right] \\
        & +\left[\frac{L_{-3}(t)+L_3(t)}{c}\right] \left[\ddot{y}_{123}\left(t-\frac{L_{-3}(t)}{c}-\frac{L_{-2}(t)}{c}-\frac{L_3(t)}{c}-\frac{L_2(t)}{2c}\right) \right. \\
        & \hspace{9em} \left. - \frac{L_{-2}(t)}{2c}\dddot{y}_{123}\left(t-\frac{L_{-3}(t)}{c}-\frac{L_{-2}(t)}{c}-\frac{L_3(t)}{c}-\frac{L_2(t)}{2c}\right) \right] \\
        & + \left[\frac{L_{-3}(t)+L_3(t)}{c}\right] \left[\ddot{y}_{3-21}\left(t-\frac{L_{-3}(t)}{c}-\frac{L_{-2}(t)}{c}-\frac{L_3(t)}{c}-\frac{L_2(t)}{2c}\right) \right. \\
        & \hspace{9em} \left. \left. + \frac{L_{-2}(t)}{2c}\dddot{y}_{3-21}\left(t-\frac{L_{-3}(t)}{c}-\frac{L_{-2}(t)}{c}-\frac{L_3(t)}{c}-\frac{L_2(t)}{2c}\right) \right] \vphantom{\frac{\left(\frac{1}{2}\right)}{2}}\right\} \\
        & 
        \left.
        \begin{aligned}
            +& \, \mathcal{O}\left[\frac{L_l^2\ddot{L}_l}{c^3}\dot{y}_{slr}\right]
            + \mathcal{O}\left[\frac{L_l^2\dot{L}\ddot{L}_l}{c^4}\dot{y}_{slr}\right]\\
            +& \, \mathcal{O}\left[\frac{L_l^2\dot{L}_l}{c^3}\ddot{y}_{slr}\right]
            + \mathcal{O}\left[\frac{L_l^2\dot{L}_l^2}{c^4}\ddot{y}_{slr}\right]
        \end{aligned}
        \hspace{0.5cm} \right\}
        \text{error from $\frac{\mathrm{d}^2}{\mathrm{d}\tau^2_{2,3}}\approx \frac{\mathrm{d}^2}{\mathrm{d}t^2}$} \\
        & 
        \left.
        \begin{aligned}
            +& \, \mathcal{O}\left[\frac{L_l^3\ddot{L}_l}{c^4}\ddot{y}_{slr}\right]
            + \mathcal{O}\left[\frac{L_l^3\dot{L}_l\ddot{L}_l}{c^5}\ddot{y}_{slr}\right]\\
            +& \, \mathcal{O}\left[\frac{L_l^3\dot{L}_l}{c^4}\dddot{y}_{slr}\right]
            + \mathcal{O}\left[\frac{L_l^3\dot{L}_l^2}{c^5}\dddot{y}_{slr}\right]
        \end{aligned}
        \hspace{0.5cm} \right\}
        \text{error from $\frac{\mathrm{d}^2}{\mathrm{d}\tau^3_{2,3}}\approx \frac{\mathrm{d}^3}{\mathrm{d}t^3}$} \\        
        & + \mathcal{O}\left[\frac{L_l^4\ddot{L}_l}{c^5}\dddot{y}_{slr}\right]
        + \mathcal{O}\left[\frac{L_l^4\dot{L}_l\ddot{L}_l}{c^6}\dddot{y}_{slr}\right]
        + \mathcal{O}\left[\frac{L_l^4}{c^4}\ddddot{y}_{slr}\right]
        \hspace{0.3cm} \text{error from Taylor series truncation after $\frac{\mathrm{d}^4}{\mathrm{d}\tau_{2,3}^4}\approx \frac{\mathrm{d}^4}{\mathrm{d}t^4}$.}
    \end{aligned}
\end{equation}

\subsubsection{Low-frequency approximation}\label{subsubsec:app:lwa_2.0}

We provide the approximate expressions for $X_{2.0}$ in the low-frequency regime with the same two 
configurations of Sec.~\ref{subsubsec:app:lwa_1.5}.

First, assuming $L_l(t)=L_{-l}(t)$:
\begin{equation}\label{eq:X20_nonequal_timedelays}
    \begin{aligned}
        X_{2.0}(t) & = \frac{8}{c}\left[\frac{L_2(t)+L_3(t)}{c}\right] \left[ L_3(t) \ddot{y}_{123}\left(t-\frac{2L_3(t)}{c}-\frac{3L_2(t)}{2c}\right) - L_2(t) \ddot{y}_{231}\left(t-\frac{2L_2(t)}{c}-\frac{3L_3(t)}{2c}\right)\right] \\
        & + \mathcal{O}\left[\frac{R\Omega L^2}{c^3}\ddot{y}_{slr}\right] + \mathcal{O}\left[\frac{R\Omega L^3}{c^4}\dddot{y}_{slr}\right]
        \hspace{0.5cm}\text{error from $L_l\simeq L_{-l}$ and $y_{slr}\simeq y_{r-ls}$} \\
        & 
        \left.
        \begin{aligned}
            +& \, \mathcal{O}\left[\frac{L_l^2\ddot{L}_l}{c^3}\dot{y}_{slr}\right]
            + \mathcal{O}\left[\frac{L_l^2\dot{L}\ddot{L}_l}{c^4}\dot{y}_{slr}\right]\\
            +& \, \mathcal{O}\left[\frac{L_l^2\dot{L}_l}{c^3}\ddot{y}_{slr}\right]
            + \mathcal{O}\left[\frac{L_l^2\dot{L}_l^2}{c^4}\ddot{y}_{slr}\right]
        \end{aligned}
        \hspace{0.5cm} \right\}
        \text{error from $\frac{\mathrm{d}^2}{\mathrm{d}\tau^2_{2,3}}\approx \frac{\mathrm{d}^2}{\mathrm{d}t^2}$} \\
        & 
        \left.
        \begin{aligned}
            +& \, \mathcal{O}\left[\frac{L_l^3\ddot{L}_l}{c^4}\ddot{y}_{slr}\right]
            + \mathcal{O}\left[\frac{L_l^3\dot{L}_l\ddot{L}_l}{c^5}\ddot{y}_{slr}\right]\\
            +& \, \mathcal{O}\left[\frac{L_l^3\dot{L}_l}{c^4}\dddot{y}_{slr}\right]
            + \mathcal{O}\left[\frac{L_l^3\dot{L}_l^2}{c^5}\dddot{y}_{slr}\right]
        \end{aligned}
        \hspace{0.5cm} \right\}
        \text{error from $\frac{\mathrm{d}^2}{\mathrm{d}\tau^3_{2,3}}\approx \frac{\mathrm{d}^3}{\mathrm{d}t^3}$} \\        
        & + \mathcal{O}\left[\frac{L_l^4\ddot{L}_l}{c^5}\dddot{y}_{slr}\right]
        + \mathcal{O}\left[\frac{L_l^4\dot{L}_l\ddot{L}_l}{c^6}\dddot{y}_{slr}\right]
        + \mathcal{O}\left[\frac{L_l^4}{c^4}\ddddot{y}_{slr}\right]
        \hspace{0.3cm} \text{error from Taylor series truncation after $\frac{\mathrm{d}^4}{\mathrm{d}\tau_{2,3}^4}\approx \frac{\mathrm{d}^4}{\mathrm{d}t^4}$}
    \end{aligned}
\end{equation}
where, at leading order, the thrid derivatives of $y_{slr}$ have canceled out. 
The leading-order error term originates a relative error $\mathcal{O}(10^{-4})$ (0.01\%). This 
is consistent with the residuals shown in Fig.~\ref{fig:cpu_visual_agreement_TDI2}. In this limit, the 
1.5 and 2.0 TDI Michelson variables are related as
\begin{equation}\label{eq:relation_X20_X15_nonequal_timedelays}
    X_{2.0}(t) = 2\left[\frac{L_2(t)+L_3(t)}{c}\right]\dot{X}_{1.5}\left(t-\frac{L_3(t)}{c}-\frac{L_2(t)}{c}\right),
\end{equation}
where $X_{1.5}$ is given by Eq.~\eqref{eq:X15_nonequal_timedelays}.

If we approximate the time delays to be equal and constant Eq.~\eqref{eq:X20_nonequal_timedelays} 
simplifies to
\begin{equation}\label{eq:X20_equal_timedelays}
    \begin{aligned}
        X_{2.0}(t) &= \frac{16L^2}{c^2} \left[\ddot{y}_{123}\left(t-\frac{7L}{2c}\right) - \ddot{y}_{231}\left(t-\frac{7L}{2c}\right)\right]\\
        & + \mathcal{O}\left[\frac{e L^2}{c^2}\ddot{y}_{slr}\right]
        \hspace{0.5cm}\text{error from $L_l(t)=L$} \\
        & + \mathcal{O}\left[\frac{R\Omega L^2}{c^3}\ddot{y}_{slr}\right] + \mathcal{O}\left[\frac{R\Omega L^3}{c^4}\dddot{y}_{slr}\right]
        \hspace{0.5cm}\text{error from $L_l\simeq L_{-l}$ and $y_{slr}\simeq y_{r-ls}$} \\
        & 
        \left.
        \begin{aligned}
            +& \, \mathcal{O}\left[\frac{L_l^2\ddot{L}_l}{c^3}\dot{y}_{slr}\right]
            + \mathcal{O}\left[\frac{L_l^2\dot{L}\ddot{L}_l}{c^4}\dot{y}_{slr}\right]\\
            +& \, \mathcal{O}\left[\frac{L_l^2\dot{L}_l}{c^3}\ddot{y}_{slr}\right]
            + \mathcal{O}\left[\frac{L_l^2\dot{L}_l^2}{c^4}\ddot{y}_{slr}\right]
        \end{aligned}
        \hspace{0.5cm} \right\}
        \text{error from $\frac{\mathrm{d}^2}{\mathrm{d}\tau^2_{2,3}}\approx \frac{\mathrm{d}^2}{\mathrm{d}t^2}$} \\
        & 
        \left.
        \begin{aligned}
            +& \, \mathcal{O}\left[\frac{L_l^3\ddot{L}_l}{c^4}\ddot{y}_{slr}\right]
            + \mathcal{O}\left[\frac{L_l^3\dot{L}_l\ddot{L}_l}{c^5}\ddot{y}_{slr}\right]\\
            +& \, \mathcal{O}\left[\frac{L_l^3\dot{L}_l}{c^4}\dddot{y}_{slr}\right]
            + \mathcal{O}\left[\frac{L_l^3\dot{L}_l^2}{c^5}\dddot{y}_{slr}\right]
        \end{aligned}
        \hspace{0.5cm} \right\}
        \text{error from $\frac{\mathrm{d}^2}{\mathrm{d}\tau^3_{2,3}}\approx \frac{\mathrm{d}^3}{\mathrm{d}t^3}$} \\        
        & + \mathcal{O}\left[\frac{L_l^4\ddot{L}_l}{c^5}\dddot{y}_{slr}\right]
        + \mathcal{O}\left[\frac{L_l^4\dot{L}_l\ddot{L}_l}{c^6}\dddot{y}_{slr}\right]
        + \mathcal{O}\left[\frac{L_l^4}{c^4}\ddddot{y}_{slr}\right]
        \hspace{0.3cm} \text{error from Taylor series truncation after $\frac{\mathrm{d}^4}{\mathrm{d}\tau_{2,3}^4}\approx \frac{\mathrm{d}^4}{\mathrm{d}t^4}$}
    \end{aligned}
\end{equation}
where now the leading-order error term originates a relative error of $\mathcal{O}(e)\sim 10^{-3}$ 
(0.1\%), which agrees with Fig.~\ref{fig:cpu_visual_agreement_TDI2}. 
In this approximation, the 
relation given by Eq.~\eqref{eq:relation_X20_X15_nonequal_timedelays} becomes 
\begin{equation}\label{eq:relation_X20_X15_equal_timedelays}
    X_{2.0}(t) = \frac{4L}{c} \dot{X}_{1.5}\left(t-\frac{2L}{c}\right),
\end{equation}
where $X_{1.5}$ is computed from Eq.~\eqref{eq:X15_equal_timedelays}.

In the low-frequency regime and adopting 2.0 TDI, the Michelson variables are proportional 
to third-time derivatives of the GW polarization ($\dot{\psi}_4$) since
\begin{equation}\label{eq:X20_thrid_strain_derivative}
    X_{2.0}(t) \sim \dddot{h}_{+,\times}\left(t-\frac{4L}{c}-\frac{\hat{\mathbf{k}}\cdot\mathbf{R}_0(t)}{c}\right),
\end{equation}
where we have used the approximation given by Eq.~\eqref{eq:Rs_Rr_R0}.
This behavior is shared with $Y_{2.0}$ and $Z_{2.0}$, and also with $A_{2.0},E_{2.0},T_{2.0}$.
 
\section{Comparison with the Fourier-domain LISA response in the literature}\label{sec:app:comparison_with_fd_response}

In this section, we compare the expression obtained for $y_{slr}$ in Eq.~\eqref{eq:yslr_with_tau_derivatives} with the one provided in Refs.~\cite{Marsat:2018oam, PhysRevD.103.083011}. 
In order to provide a faithful comparison, we simplify Eq.~\eqref{eq:yslr_with_tau_derivatives} according to the assumptions made in \cite{Marsat:2018oam, PhysRevD.103.083011}: we approximate LISA as a rigid equilateral constellation and force all inter-spacecraft delays to be constant and equal to $L$. 
We also adopt the same conventions (see App.~A of \cite{Marsat:2018oam} and Eq.~(5) of \cite{PhysRevD.103.083011}):\footnote{Note that the Fourier convention of Refs.~\cite{Marsat:2018oam, PhysRevD.103.083011} differs from the one defined in App.~\ref{sec:app:fourier_conventions}.}
\begin{equation}\label{eq:fd_different_convention}
    \tilde{x}(f)=\mathcal{F}[x(t)]=\int \mathrm{d}t\, e^{+2i\pi f t}x(t),
\end{equation}
which implies
\begin{subequations}
    \begin{align}
        \mathcal{F}[x(t-t_0)]&=e^{+2i\pi f t_0}\tilde{x}(f), \hspace{0.5cm} \text{and}\label{eq:fd_different_convention_shift}\\
        \mathcal{F}\left[\frac{\mathrm{d}^m}{\mathrm{d}t^m}x(t)\right] & = (-2i\pi f)^m \tilde{x}(f).\label{eq:fd_different_convention_derivative}
    \end{align}
\end{subequations}

Following this convention, the Fourier transform of Eq.~\eqref{eq:yslr_with_tau_derivatives} reads
\begin{equation}\label{eq:yslr_fd}
    \begin{aligned}
        \mathcal{F}[y_{slr}(t)] &= -\frac{L}{2c}\sum\limits_{m=0}^{\infty}\frac{1}{(2m+1)!}\left[\frac{L}{2c}\left(1-\hat{\mathbf{k}}\cdot\hat{\mathbf{n}}_l\right)\right]^{2m}\mathcal{F}\left[\frac{\mathrm{d}^{2m+1}}{\mathrm{d}t^{2m+1}}H_l\left(t-\frac{L}{2c}-\frac{\hat{\mathbf{k}}\cdot(\mathbf{R}_s + \mathbf{R}_r)}{2c}\right)\right]\\
        &= -\frac{L}{2c}\sum\limits_{m=0}^{\infty}\frac{1}{(2m+1)!}\left[\frac{L}{2c}\left(1-\hat{\mathbf{k}}\cdot\hat{\mathbf{n}}_l\right)\right]^{2m}(-2i\pi f)^{2m+1}\mathcal{F}\left[H_l\left(t-\frac{L}{2c}-\frac{\hat{\mathbf{k}}\cdot(\mathbf{R}_s + \mathbf{R}_r)}{2c}\right)\right]\\
        &= \frac{i\pi f L}{c}\text{exp}\left[\frac{i\pi f}{c}\left(L + \hat{\mathbf{k}}\cdot (\mathbf{R}_s + \mathbf{R}_r)\right)\right]\tilde{H}_l\sum\limits_{m=0}^{\infty}\frac{(-1)^m}{(2m+1)!}\left[\frac{\pi f L}{c}\left(1-\hat{\mathbf{k}}\cdot\hat{\mathbf{n}}_l\right)\right]^{2m}\\
        & = \frac{i\pi f L}{c}\text{sinc}\left[\frac{\pi f L}{c}\left(1-\hat{\mathbf{k}}\cdot\hat{\mathbf{n}}_l\right)\right]\text{exp}\left[\frac{i\pi f}{c}\left(L + \hat{\mathbf{k}}\cdot (\mathbf{R}_s + \mathbf{R}_r)\right)\right]\tilde{H}_l,
    \end{aligned}
\end{equation}
where we have applied Eqs.~\eqref{eq:fd_different_convention_shift} and~\eqref{eq:fd_different_convention_derivative}, 
and we have used $\text{sinc}(x) = \sin(x)/x = \sum_{m=0}^{\infty} \frac{(-1)^mx^{2m}}{(2m+1)!}$. For clarity, we have 
neglected the error terms coming from $\mathrm{d}/\mathrm{d}\tau \approx \mathrm{d}/\mathrm{d}t$. 
Eq.~\eqref{eq:yslr_fd} matches the transfer functions given in Refs.~\cite{Marsat:2018oam, PhysRevD.103.083011}.\footnote{Note that, as pointed out in \cite{deng2025fastdetectionreconstructionmerging}, in the equations presented in Refs.~\cite{Marsat:2018oam, PhysRevD.103.083011}, there is an extra factor of $1/2$ which is a typo.}

\section{Space-borne vs. ground-based detectors in the low-frequency limit}\label{sec:app:lisa_vs_groundbased}

In this section, we first briefly derive the standard scalar product routinely used in GW data analysis in Sec.~\ref{subsec:app:scalar_product}. 
Alternative descriptions for the GW signal employing time derivatives of the strain are described in Sec.~\ref{subsec:app:data_analysis_second_derivative} in the context of data analysis. 
Finally, in Sec.~\ref{subsec:app:space_and_ground_whitened_data} we compare the whitened data measured by LISA and a modified TDI-like ground-based detector.

\subsection{Scalar product}\label{subsec:app:scalar_product}

The probability that a set of physical parameters $\bm{\theta}$ describes a GW signal through a model $h(t;\bm{\theta})$ in a data stream $d(t)$ can be computed with Bayes' theorem as
\begin{equation}\label{eq:bayes_theorem}
    P(\bm{\theta}|d,h) = \frac{P(d|\bm{\theta},h) P(\bm{\theta}|h)}{P(d|h)},
\end{equation}
where $d(t)=n(t)+h(t;\bm{\theta})$ if the signal is present and $d(t)=n(t)$ otherwise, with $n(t)$ the noise in the detector measurement. In Eq.~\eqref{eq:bayes_theorem}, $P(\bm{\theta}|d,h)$ is known as the posterior probability distribution on the parameters of the source; 
$P(d|\bm{\theta},h)$, also known as 
the likelihood $\mathcal{L}(d|h(\bm{\theta}))$,
is the conditional probability of measuring data $d$ assuming that a GW signal described by the model $h$ and the model parameters 
$\bm{\theta}$ is present;
the prior $P(\bm{\theta}|h)$ is the \emph{a priori} probability of the parameters $\bm{\theta}$ given the model $h$; and the evidence $P(d|h)$ is a normalization constant. 
Provided our prior knowledge of the parameters and neglecting the constant evidence, to compute the posteriors $P(\bm{\theta}|d,h)$, we must first determine the likelihood $\mathcal{L}$.
Note that the conditional probability of measuring $d$ assuming a GW signal described $h(\bm{\theta})$ is present in the data stream, i.e. the likelihood, is the same as assuming there is no GW signal in $d^\prime = d-h$:
\begin{equation}\label{eq:likelihood_no_signal}
    P(d|\bm{\theta},h) \equiv \mathcal{L}(d|h(\bm{\theta})) = \mathcal{L}(d-h(\bm{\theta})|0) = \mathcal{L}(d^\prime|0).
\end{equation}
In this case, $d^\prime$ is simply a realization of the random process $n(t)$ and can therefore be sampled from the same distribution. In realistic LISA data, the noise is expected to exhibit non-stationary features and non-Gaussian profiles (see e.g.~\cite{redbook, PhysRevLett.116.231101}). For simplicity, in this work, we assume the noise follows a stationary and Gaussian process (see for instance~\cite{Burke:2025bun} for a detailed noise treatment in the presence of data gaps).
Under this assumption the likelihood given by Eq.~\eqref{eq:likelihood_no_signal} can be determined from (see~\cite{PhysRevD.46.5236} for the derivation)
\begin{equation}
    \mathcal{L}(d^\prime|0) \propto {\rm{Exp}}\left[-\frac{1}{2}\langle d^\prime|d^\prime \rangle\right],
\end{equation}
or equivalently,
\begin{equation}\label{eq:derived_likelihood}
    \ln\mathcal{L}(d|h(\bm{\theta})) \propto -\frac{1}{2}\langle d-h|d-h \rangle = -\frac{1}{2}\left( \langle d|d \rangle +\langle h|h \rangle - 2\langle d|h \rangle\right).
\end{equation}
Here the inner product $\langle x | y \rangle$ is defined in the usual way as
\begin{equation}\label{eq:derived_scalar_product}
    \langle x | y \rangle = 4\mathrm{Re}\int_{0}^{\infty} \mathrm{d} f \frac{\tilde{x}^{*}(f) \tilde{y}(f)}{S_{\mathrm{n}}(f)} = 4\mathrm{Re}\int_{0}^{\infty} \mathrm{d} f \left[\frac{\tilde{x}(f)}{\sqrt{S_{\mathrm{n}}(f)}}\right]^{*}\left[\frac{\tilde{y}(f)}{\sqrt{S_{\mathrm{n}}(f)}}\right],
\end{equation}
with $S_{\rm{n}}$ the single-sided PSD. The expressions between squared brackets $[\,\cdot\,]$ represent whitened data.

\subsection{Data analysis with second-time derivative of the strain}\label{subsec:app:data_analysis_second_derivative}

As far as the likelihood probability in Eq.~\eqref{eq:derived_likelihood} is left invariant, data analysis could be performed with alternative descriptions of the GW signal.
Any non-zero linear operator $\rm{L}$ applied to the data must then leave the likelihood in Eq.~\eqref{eq:derived_likelihood} unchanged.
This means that the scalar product of Eq.~\eqref{eq:derived_scalar_product}
must also be an invariant quantity, which ultimately implies that any transformation of the data must leave the whitened data invariant. 
As already commented on Sec.~\ref{sec:introduction}, the authors of~\cite{PhysRevX.13.041048} consider replacing the model for the GW strain, $h$ in Eq.~\eqref{eq:bayes_theorem}, by the second-time derivative ${\rm{d}}^2 h/{\rm{d}}t^2$, also referred to as $\psi_4$.
In this case, the new data stream becomes ${\rm{L}}[d]=\ddot{h}(t) + \ddot{n}(t)$, with the dot indicating time derivation.
To not alter the likelihood probability, we should modify the PSD in Eq.~\eqref{eq:derived_scalar_product} such that 
\begin{equation}\label{eq:same_whitened_data}
    \frac{\tilde{d}(f)}{\sqrt{S_{\rm{n}}(f)}}=\pm\, \frac{\tilde{\ddot{d}}(f)}{\sqrt{S_{\rm{n}}^{\rm{L}}(f)}},
\end{equation}
where $S_{\rm{n}}^{\rm{L}}(f)$ is the PSD of the transformed data.
In general, a $\pm$ sign will still leave the scalar product of Eq.~\eqref{eq:derived_scalar_product} invariant, and consequently, also the likelihood of Eq.~\eqref{eq:derived_likelihood}.
In~\cite{PhysRevX.13.041048}, $S_{\rm{n}}^{\rm{L}}$ is computed by estimating the new distribution of the noise after applying the second-order finite differences operator (see App.~B of~\cite{PhysRevX.13.041048} for details).
Here we will derive $S_{\rm{n}}^{\rm{L}}$ using the properties of the Fourier transform in the continuum, which is equivalent to the results reported in~\cite{PhysRevX.13.041048} in the low-frequency limit, that is the regime we are interested.

In the continuum, we can use the Fourier derivative property given by Eq.~\eqref{eq:FT_derivative_property} to write 
\begin{equation}
    \tilde{\ddot{d}}(f)=(i2\pi f)^2 \tilde{d}(f),
\end{equation}
which directly implies that 
\begin{equation}\label{eq:derived_psd}
    S_{\rm{n}}^{\rm{L}}(f)=(i2\pi f)^4 S_{\rm{n}}(f).
\end{equation}
Although Ref.~\cite{PhysRevX.13.041048} tackles the problem directly in the discrete, the expression for $S_{\rm{n}}^{\rm{L}}(f)$ in Eq.~\eqref{eq:derived_psd} matches the leading-order term in the low-frequency expansion of Eq.~(11) in~\cite{PhysRevX.13.041048}. 
Note that the discrete analog of Eq.~\eqref{eq:derived_psd} could have been directly derived from imposing Eq.~\eqref{eq:same_whitened_data} and using the Fourier-transformed relation between the continuous and discrete second derivative, given by Eq.~(8) of~\cite{PhysRevX.13.041048}, instead of deriving the new PSD for $\ddot{n}(t)$ as in App.~B of~\cite{PhysRevX.13.041048}.
The trigonometric functions in Eq.~(11) of~\cite{PhysRevX.13.041048} appear from the discretization of the second-order derivative in the Fourier domain and resemble those that also appear in $S_{\rm{n}}^{A,E}$ and $S_{\rm{n}}^{T}$ in Eqs.~\eqref{eq:SnAE} and~\eqref{eq:SnT}, reflecting the inherent finite-difference nature of the TDI formalism.
Overall, the LISA response is not only applied to GWs but also to all noise sources, so effectively computing derivatives of the noise.

\subsection{Space-borne and TDI-like ground-based whitened data}\label{subsec:app:space_and_ground_whitened_data}

Below, we qualitatively determine the differences between the 
whitened waveforms for ground- and space-based detectors in the low-frequency limit. 
As representative cases, we will focus on LIGO and LISA. 
For a faithful comparison, both detectors should have a similar GW response, in order to perform data analysis with the same observables.
As discussed in App.~\ref{sec:app:expansions}, the LISA response quantities can be expressed as derivatives of the GW polarizations $h_{+}$ and $h_{\times}$. 
In contrast, the LIGO response does not involve derivatives of $h_{+}$ and $h_{\times}$, but rather linear combinations of them, as commented in Sec.~\ref{sec:introduction}.
For a faithful comparison, we will follow the derivation above to modify LIGO as if it were working with TDI variables, and therefore approximating 1.5 TDI and 2.0 TDI by second- and third-time derivatives of the GW polarizations, respectively (see Secs.~\ref{subsubsec:app:lwa_1.5} and~\ref{subsubsec:app:lwa_2.0} for the relation between TDI generations and time derivatives).
Hereafter, we study the LISA and LIGO cases separately.

\subsubsection{LISA}\label{subsec:app:lisa_case}

We focus on the $A$ and $E$ channels, but a similar reasoning can be applied to the $T$ channel.
For clarity, we approximate the low-frequency power-law behavior of the LISA PSD in the 
$A$ and $E$ channels using piecewise expressions. 
The PSDs will be modeled by their power-law approximated versions, given 
in Sec.~\ref{subsec:noise_model}.
For the 1.5 TDI configuration, we model the PSD as
\begin{equation}\label{eq:lisa_low_freq_psd_tdi15}
    S_{\rm{n}}^{A_{1.5},E_{1.5}}(f) \sim 
    \begin{cases}
        f^{-2} & \mathrm{if}\, f\leq f^{\rm{LISA}}_*\,, \\
        f^{0} & \mathrm{if}\, f > f^{\rm{LISA}}_*\,,
    \end{cases}
\end{equation}
while the 2.0 TDI configuration is approximated by
\begin{equation}\label{eq:lisa_low_freq_psd_tdi20}
    S_{\rm{n}}^{A_{2.0},E_{2.0}}(f) \sim 
    \begin{cases}
        f^{0} & \mathrm{if}\, f\leq f^{\rm{LISA}}_*\,, \\
        f^{2} & \mathrm{if}\, f > f^{\rm{LISA}}_*\,,
    \end{cases}
\end{equation}
where $f^{\rm{LISA}}_*$ denotes the characteristic frequency at which the dominant power-law changes. 
This transition frequency is introduced symbolically and can be estimated from Fig.~\ref{fig:aet_psds}. 
With Eqs.~\eqref{eq:X15_with_strain_derivatives}, \eqref{eq:X20_with_strain_derivatives},
\eqref{eq:lisa_low_freq_psd_tdi15} and \eqref{eq:lisa_low_freq_psd_tdi20}, and
the Fourier derivative property given by Eq.~\eqref{eq:FT_derivative_property} we 
can qualitatively compute the whitened waveform.
For the 1.5-generation TDI, the whitened $A$ channel is
\begin{equation}\label{eq:whitened_A15_derivatives}
        \frac{\tilde{A}_{1.5}}{\sqrt{S_{\rm{n}}^{A_{1.5}}}}
        \sim \frac{\mathcal{F}[\ddot{h}^{\rm{SSB}}_{+,\times}]}{\sqrt{S_{\rm{n}}^{A_{1.5}}}}
        \sim \begin{cases}
            \frac{f^2 \tilde{h}^{\rm{SSB}}_{+,\times}}{\sqrt{f^{-2}}}
            \sim f^3\tilde{h}^{\rm{SSB}}_{+,\times}
            \sim \mathcal{F}[\dddot{h}^{\rm{SSB}}_{+,\times}] & \mathrm{if}\, f\leq f^{\rm{LISA}}_*\,, \\
            & \\
            \frac{f^2 \tilde{h}^{\rm{SSB}}_{+,\times}}{\sqrt{f^{0}}}
            \sim f^2\tilde{h}^{\rm{SSB}}_{+,\times}\sim \mathcal{F}[\ddot{h}^{\rm{SSB}}_{+,\times}] & \mathrm{if}\, f > f^{\rm{LISA}}_*\,. 
        \end{cases}
\end{equation}
Likewise, for the 2.0-generation TDI, the whitened $A$ channel is
\begin{equation}\label{eq:whitened_A20_derivatives}
        \frac{\tilde{A}_{2.0}}{\sqrt{S_{\rm{n}}^{A_{2.0}}}}
        \sim \frac{\mathcal{F}[\dddot{h}^{\rm{SSB}}_{+,\times}]}{\sqrt{S_{\rm{n}}^{A_{2.0}}}}
        \sim \begin{cases}
            \frac{f^3 \tilde{h}^{\rm{SSB}}_{+,\times}}{\sqrt{f^{0}}}
            \sim f^3\tilde{h}^{\rm{SSB}}_{+,\times}
            \sim \mathcal{F}[\dddot{h}^{\rm{SSB}}_{+,\times}] & \mathrm{if}\, f\leq f^{\rm{LISA}}_*\,, \\
            & \\
            \frac{f^3 \tilde{h}^{\rm{SSB}}_{+,\times}}{\sqrt{f^{2}}}
            \sim f^2\tilde{h}^{\rm{SSB}}_{+,\times}
            \sim \mathcal{F}[\ddot{h}^{\rm{SSB}}_{+,\times}] & \mathrm{if}\, f > f^{\rm{LISA}}_*\,. 
        \end{cases}
\end{equation}
From Eqs.~\eqref{eq:whitened_A15_derivatives} and~\eqref{eq:whitened_A20_derivatives} we observe that the whitened signal behaves as a second- or third-time derivative of the GW polarizations (depending on the frequency range), independently of the TDI algorithm employed. Therefore, we anticipate that at low frequencies the likelihood is approximately a preserved quantity when using 1.5 TDI or 2.0 TDI. We will delve into the equivalence between TDI generations in App.~\ref{sec:app:first_vs_second_tdi}.

\subsubsection{LIGO}

Now, in order illustrate the differences between the LIGO and LISA responses, we want to modify
the LIGO PSD in analogy with the TDI response.  
We need
to change $S_{\rm{n}}^{\rm{LIGO}}$ according to the order of the derivative of the GW polarizations, $N$, in Eqs.~\eqref{eq:X15_with_strain_derivatives} 
and~\eqref{eq:X20_with_strain_derivatives}. 
Motivated by Eq.~\eqref{eq:derived_psd}, we can qualitatively simplify this change as
$S_{\rm{n}}^{\rm{LIGO}}\rightarrow f^{2N} S_{\rm{n}}^{\rm{LIGO}}$.
The PSDs will be modeled by their power-law approximated versions, given 
in Sec.~\ref{subsec:noise_model}.

For TDI 1.5, according to Eq.~\eqref{eq:X15_with_strain_derivatives},
$N=2$ and the corresponding TDI-like LIGO PSD at low frequencies is modeled as
\begin{equation}\label{eq:ligo_low_freq_psd_tdi15}
    S_{\rm{n},1.5\rm{TDI}}^{\rm{LIGO}}(f) 
    \sim f^{4}S_{\rm{n}}^{\rm{LIGO}}(f)
    \sim \begin{cases}
        f^{-16} & \mathrm{if}\, f\leq f^{\rm{LIGO}}_*\,, \\
        f^{-1} & \mathrm{if}\, f > f^{\rm{LIGO}}_*\,.
    \end{cases}
\end{equation}
Likewise, $N=3$ for TDI 2.0 according to Eq.~\eqref{eq:X20_with_strain_derivatives}, 
and
\begin{equation}\label{eq:ligo_low_freq_psd_tdi20}
    S_{\rm{n},2.0\rm{TDI}}^{\rm{LIGO}}(f) 
    \sim f^{6}S_{\rm{n}}^{\rm{LIGO}}(f)
    \sim \begin{cases}
        f^{-14} & \mathrm{if}\, f\leq f^{\rm{LIGO}}_*\,, \\
        f & \mathrm{if}\, f > f^{\rm{LIGO}}_*\,,
    \end{cases}
\end{equation}
where in Eqs.~\eqref{eq:ligo_low_freq_psd_tdi15} and~\eqref{eq:ligo_low_freq_psd_tdi20}, we have defined $f^{\rm{LIGO}}_*$ as the characteristic frequency at which the dominant power-law changes.
Then, the whitened waveforms for the TDI-like LIGO detector are
\begin{equation}\label{eq:ligo_whitened_A15_derivatives}
        \frac{\tilde{A}_{1.5}}{\sqrt{S_{\rm{n},1.5\rm{TDI}}^{\rm{LIGO}}}}
        \sim \frac{\mathcal{F}[\ddot{h}^{\rm{SSB}}_{+,\times}]}{\sqrt{S_{\rm{n},1.5\rm{TDI}}^{\rm{LIGO}}}}\\
        \sim \begin{cases}
            \frac{f^2 \tilde{h}^{\rm{SSB}}_{+,\times}}{\sqrt{f^{-16}}}
            \sim f^{10}\tilde{h}^{\rm{SSB}}_{+,\times}
            & \mathrm{if}\, f\leq f^{\rm{LIGO}}_*\,, \\
            & \\
            \frac{f^2 \tilde{h}^{\rm{SSB}}_{+,\times}}{\sqrt{f^{-1}}}
            \sim f^{5/2}\tilde{h}^{\rm{SSB}}_{+,\times}
            & \mathrm{if}\, f > f^{\rm{LIGO}}_*\,,
        \end{cases}
\end{equation}
for 1.5 TDI, and
\begin{equation}\label{eq:ligo_whitened_A20_derivatives}
        \frac{\tilde{A}_{2.0}}{\sqrt{S_{\rm{n},2.0\rm{TDI}}^{\rm{LIGO}}}}
        \sim \frac{\mathcal{F}[\dddot{h}^{\rm{SSB}}_{+,\times}]}{\sqrt{S_{\rm{n},2.0\rm{TDI}}^{\rm{LIGO}}}}\\
        \sim \begin{cases}
            \frac{f^3 \tilde{h}^{\rm{SSB}}_{+,\times}}{\sqrt{f^{-14}}}
            \sim f^{10}\tilde{h}^{\rm{SSB}}_{+,\times}
            & \mathrm{if}\, f\leq f^{\rm{LIGO}}_*\,, \\
            & \\
            \frac{f^3 \tilde{h}^{\rm{SSB}}_{+,\times}}{\sqrt{f}}
            \sim f^{5/2}\tilde{h}^{\rm{SSB}}_{+,\times}
            & \mathrm{if}\, f > f^{\rm{LIGO}}_*\,,
        \end{cases}
\end{equation}
for 2.0 TDI. 

Comparing Eqs.~\eqref{eq:whitened_A15_derivatives} and~\eqref{eq:whitened_A20_derivatives}, derived in Sec.~\ref{subsec:app:lisa_case} for the LISA case, and Eqs.~\eqref{eq:ligo_whitened_A15_derivatives} and~\eqref{eq:ligo_whitened_A20_derivatives}
we observe different relative weightings of the waveform content depending on the specific frequency range, caused by the intrinsic different nature of the detector's noise sources.
For LISA, the lower end of its frequency spectrum is more sensitive than that for LIGO, allowing the measurement of second- or third-order derivatives of the GW polarizations weighted by red $\sim f^{-2}$, white $\sim f^0$, or purple $\sim f^{2}$ noise components,\footnote{See e.g.~\cite{Bakalis2023} for a mapping between noise power-laws and colors.} yielding effective weightings of $f^3\tilde{h}_{+,\times}^{\rm{SSB}}$ and $f^2\tilde{h}_{+,\times}^{\rm{SSB}}$. 
In contrast, the sharp low-frequency power law originated by seismic noise in LIGO leads to the measurement of an effective second- or third-derivative weighted by stronger black noise components of $\sim f^{-16}$ or $\sim f^{-14}$, followed by milder pink $\sim f^{-1}$ and blue $\sim f$ noise components, depending on the order of the derivative.
This distinct behavior between LISA and LIGO is displayed in Fig.~\ref{fig:whitened_waveforms_lisa_vs_ground_based}, where we show the (normalized) whitened spherical harmonic modes for both detectors and also including Einstein Telescope (ET)~\cite{bluebook} for further comparisons. 

The aforementioned difference may enhance the contribution of higher-order spherical harmonic modes during the early inspiral part of the waveform in LISA, relative to current ground-based detectors. 
The origin of this enhancement lies mainly in the different PSD behavior at low frequencies but also in the distinct frequency evolution of the $(\ell,m)$ modes: 
at leading order, their phase evolves as $\phi^{\rm{insp}}_{\ell m}(t)\approx (m/2)\phi^{\rm{insp}}_{22}(t)$.
Consequently, time derivatives of the individual modes $h_{\ell m}$ are scaled by a factor of $m/2$, which amplifies the relative contribution of higher-$m$ modes to the total SNR.
Moreover, the fact of computing derivatives will naturally further boost the contribution of these higher-order modes near the merger-ringdown phase, as shown in Fig.~\ref{fig:whitened_waveforms_lisa_vs_ground_based}, where the fast chirp of the signal results in large strain derivatives.

In summary, the LISA response function, which effectively acts as a finite difference operator, combined with the relatively mild power-law behavior of instrumental noise at low frequencies, is expected to enhance the visibility of higher-order spherical harmonic modes in GW signals. 
This contrasts with current ground-based detectors, where low-frequency regime is dominated by terrestrial noises. 
The intrinsic response amplification to higher-$m$ modes could offer significant advantages for early-warning alerts with LISA, while simultaneously necessitating waveform models that accurately incorporate these $(\ell,m)$ modes not only in the inspiral but also in the merger-ringdown phase, as already pointed out by e.g.~\cite {Hoy:2024ovd}.

\begin{figure*}[t]
    \centering
    \includegraphics[width=\textwidth]{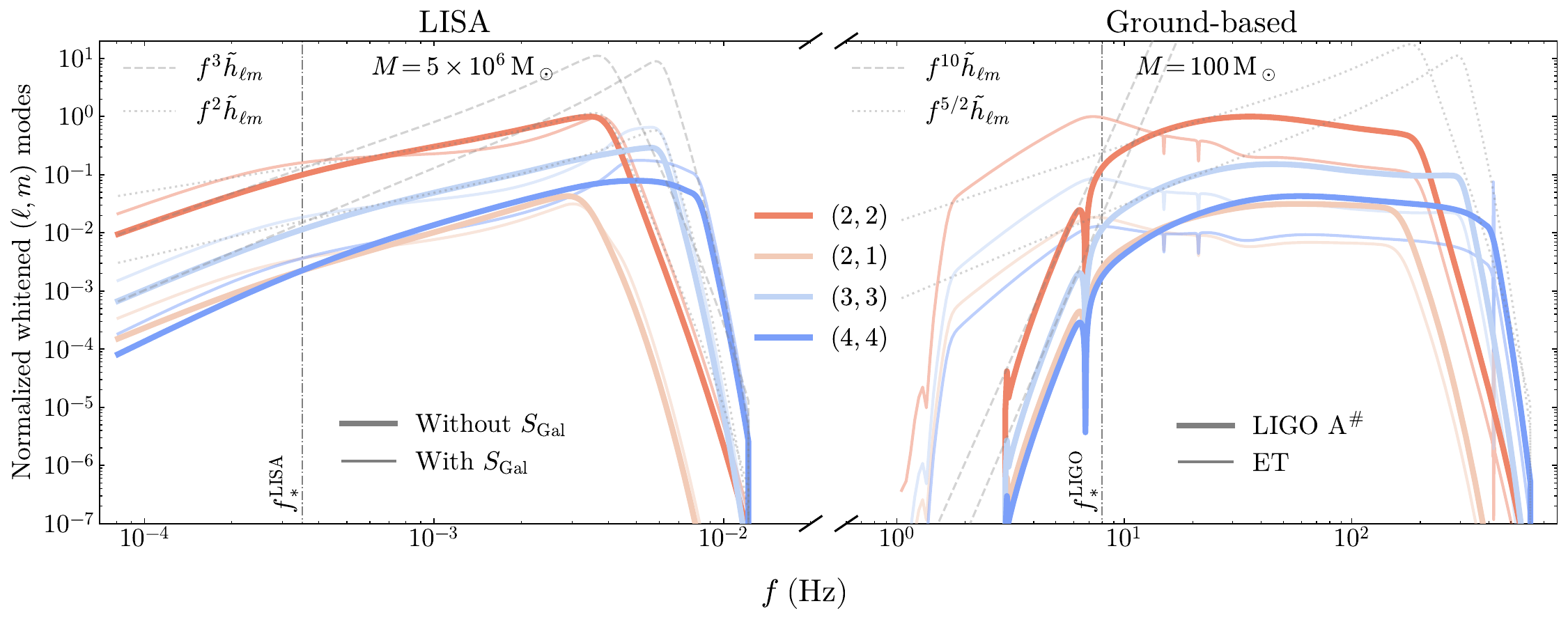}
    \caption{Normalized whitened spherical harmonic modes for two representative black hole binaries for LISA (left) and ground-based detectors (right), as a function of the GW frequency (in Hz). The whitened modes corresponding to the same detector have been normalized by the maximum of the corresponding whitened $(2,2)$-mode. The color of each curve indicates the spherical harmonic mode. 
    For LISA, we show the whitened modes for an MBHB with $M=5\times 10^6\,\rm{M}_\odot$, $q=2.5$, $\chi_1=0.5$, and $\chi_2=0.2$,  for both neglecting (thick, solid) and including (thin, solid) the galactic foreground noise $S_{\rm{Gal}}$, which was determined using the same parameters as the ones given in Sec.~\ref{subsec:noise_model}. For ground-based detectors, we display the same mode content for the same system as for LISA but with $M=100\,\rm{M}_\odot$. We use LIGO A$^\#$ (thick, solid) and Einstein Telescope (ET) (thin, solid) PSDs. For both LISA and ground-based, the dashed and dotted gray lines represent the approximated whitened (2,2) and (3,3) spherical harmonics in the low-frequency limit as given by Eqs.~\eqref{eq:whitened_A15_derivatives},~\eqref{eq:whitened_A20_derivatives},~\eqref{eq:ligo_whitened_A15_derivatives}, and~\eqref{eq:ligo_whitened_A20_derivatives}.
    Finally, the vertical dash-dotted black lines indicate the estimated power-law transition frequencies $f_{*}^{\rm{LISA}}$ and $f_{*}^{\rm{LIGO}}$. The $(\ell,m)$ modes haven been generated using the frequency-domain \phXHM~\cite{PhysRevD.102.064002} waveform model.}
    \label{fig:whitened_waveforms_lisa_vs_ground_based}
\end{figure*} 
\section{Equivalence of 1.5 and 2.0 TDI}\label{sec:app:first_vs_second_tdi}

The PSD for the channel A, given by Eq.~\eqref{eq:SnAE}, at low frequencies can be approximated by
\begin{equation}\label{eq:Sn_1tdi}
    \begin{aligned}
        S_{\rm{n}}^{A_{1.5}}(f) &= A\left(\frac{2\pi L}{c}\right)^4e^{-(f/f_1)^\alpha} \left[\vphantom{\left( \left(\frac{2\pi L}{c}\right)^2 + \frac{3}{f_2^2} \left(1-\tanh \left(\frac{f_{\rm{knee}}}{f_2}\right)\right) \tanh \left(\frac{f_{\rm{knee}}}{f_2}\right)\right)}\right.\\
        & + 3\left(1+\tanh \left(\frac{f_{\rm{knee}}}{f_2}\right)\right)f^{5/3}\\
        & -\frac{3}{f_2} \text{sech}^2\left(\frac{f_{\rm{knee}}}{f_2}\right)f^{8/3} \\
        & - \left(1+\tanh \left(\frac{f_{\rm{knee}}}{f_2}\right)\right) \left( \left(\frac{2\pi L}{c}\right)^2 + \frac{3}{f_2^2} \left(1-\tanh \left(\frac{f_{\rm{knee}}}{f_2}\right)\right) \tanh \left(\frac{f_{\rm{knee}}}{f_2}\right)\right) f^{11/3} \\
        & + \left. \mathcal{O}\left(f^{14/3}\right) \vphantom{\left( \left(\frac{2\pi L}{c}\right)^2 + \frac{3}{f_2^2} \left(1-\tanh \left(\frac{f_{\rm{knee}}}{f_2}\right)\right) \tanh \left(\frac{f_{\rm{knee}}}{f_2}\right)\right)}\right] \\
        & + \frac{L^2}{c^4}(1.54\times 10^{-5}S_{\rm{acc}})f^{-2}\\
        & + \frac{L^2}{c^6}\left[5.98\times 10^{-7} c^2 S_{\rm{oms}}+\left(96c^2 - 5.05\times 10^{-4}L^2\right) S_{\rm{acc}}\right]\\
        & + \frac{L^2}{c^8}\left[\left(0.008 L^4-3158 c^2 L^2+3750 c^4\right) S_{\rm{acc}}- 1.18\times 10^{-5} c^2 L^2 S_{\rm{oms}}\right]f^2\\
        & + \frac{L^2}{c^{10}}\left[\left(2.34\times 10^{10} c^6-123370 c^4 L^2+50289 c^2 L^4-0.0778 L^6\right)S_{\rm{acc}} + \left(37405 c^4+1.06\times10^{-4} L^4\right)c^2 S_{\rm{oms}}\right]f^4\\
        & +\mathcal{O}\left(f^6\right),
    \end{aligned}
\end{equation}
for the 1.5 generation TDI, and by
\begin{equation}\label{eq:Sn_2tdi}
    \begin{aligned}
        S_{\rm{n}}^{A_{2.0}}(f) &= A\left(\frac{2\pi L}{c}\right)^6e^{-(f/f_1)^\alpha} \left[\vphantom{\tanh \left(\frac{f_{\rm{knee}}}{f_2}\right)}\right.\\
        & + 48 \left(1+\tanh \left(\frac{f_{\rm{knee}}}{f_2}\right)\right)f^{11/3} \\
        & + \left. \mathcal{O}\left(f^{14/3}\right) \vphantom{\tanh \left(\frac{f_{\rm{knee}}}{f_2}\right)}\right]\\
        & + \frac{L^4}{c^6}(9.7\times 10^{-2} S_{\rm{acc}})\\
        & + \frac{L^4}{c^8}\left[3.78\times10^{-4} c^2 S_{\rm{oms}} +\left(60639 c^2-0.830 L^2\right) S_{\rm{acc}}\right] f^2\\
        & + \frac{L^4}{c^{10}}\left[\left(32.64 L^4-5.19\times 10^6 c^2 L^2+2.37\times 10^6 c^4\right) S_{\rm{acc}}-0.0274 c^2 L^2 S_{\rm{oms}}\right] f^4 \\
        & + \mathcal{O}\left(f^6\right),
    \end{aligned}   
\end{equation}
for the 2.0 TDI. 

\begin{figure*}[t]
    \centering
    \includegraphics[width=0.65\textwidth]{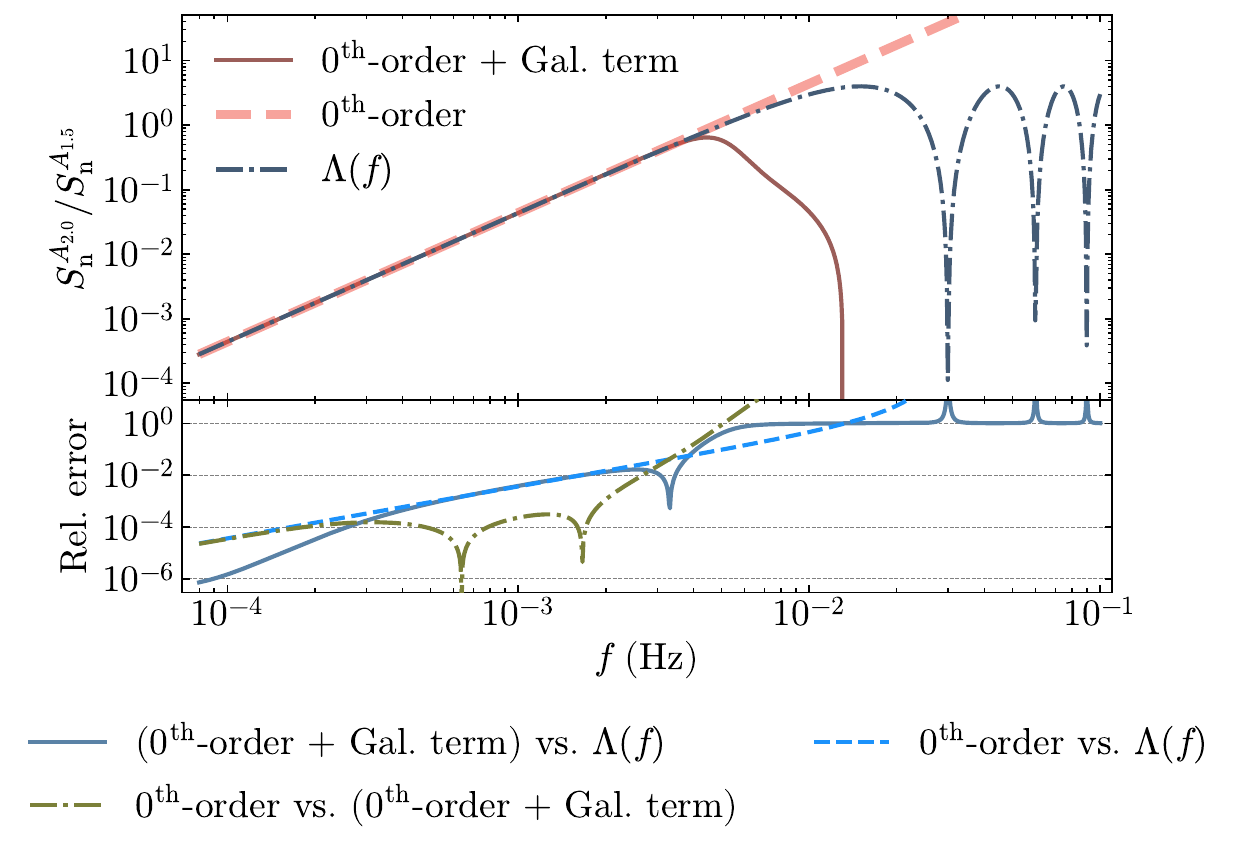}
    \caption{Approximations for the ratio between the 2.0-generation and 1.5-generation $A$-channel PSDs (upper panel), and their relative error (lower panel), as a function of the GW frequency $f$ (in Hz). The upper panel displays the $0^{\rm{th}}$-order term $(4L/c)^2(2\pi f)^2$ (red, dashed) and the galactic-noise term (brown, solid), both derived from Eq.~\eqref{eq:relationSn2.0andSn1.5}, along with the analytical ratio $\Lambda (f)$ (gray, dash-dotted). 
    The lower panel presents the relative error with respect to $\Lambda (f)$, comparing the cases with (navy, solid) and without (blue, dashed) inclusion of the galactic term, as well as the relative error incurred by omitting the galactic term in Eq.\eqref{eq:relationSn2.0andSn1.5} (dark green, dash-dotted).}
    \label{fig:psd_expansions_and_rel_err}
\end{figure*}

The previous equations imply the relation
\begin{equation}\label{eq:relationSn2.0andSn1.5}
    S_{\rm{n}}^{A_{2.0}}(f) = \left[\left(\frac{4L}{c}\right)^2\left(2\pi f\right)^2 + \mathcal{O}\left(f^4\right)+\frac{1}{\mathcal{O}\left(f^{-4}\right) e^{(f/f_1)^\alpha}+\mathcal{O}\left(f^{-2}\right)}\right] S_{\rm{n}}^{A_{1.5}}(f).
\end{equation}
This expression contains a leading-order (zeroth-order) term 
proportional to $f^2$ and 
a term that contains $e^{(f/f_1)^{\alpha}}$ and comes from the galactic foreground noise, introduced in Eq.~\eqref{eq:sgal}.
In Fig.~\ref{fig:psd_expansions_and_rel_err}, we show the 
effect of neglecting the galactic-noise term in Eq.~\eqref{eq:relationSn2.0andSn1.5}.
In the upper panel, we display different approximations for the ratio $S_{\rm{n}}^{A_{2.0}}/S_{\rm{n}}^{A_{1.5}}$ as a function of the frequency.
These include the zeroth-order term, $(4L/c)^2(2\pi f)^2$, both with (solid line) and without (dashed line) the galactic correction, as well as the exact analytical expression $\Lambda(f)$, introduced in Sec.~\ref{subsec:noise_model}, which is valid across all the frequency range
(dash-dotted line).
The lower panel shows the corresponding relative errors. 
Specifically, it compares:
\textit{i)} the approximation of $\Lambda(f)$ by the zeroth-order term $(4L/c)^2(2\pi f)^2$ with (solid line) and without (dashed line) the galactic correction, and
\textit{ii)} the error introduced by neglecting the galactic-noise term in the frequency expansion of Eq.~\eqref{eq:relationSn2.0andSn1.5} (dash-dotted line).

According to Fig.~\ref{fig:psd_expansions_and_rel_err}, neglecting the 
galactic-noise correction in the proportionality factor relating 
$S_{\rm{n}}^{A_{2.0}}$ and $S_{\rm{n}}^{A_{1.5}}$, as given in 
Eq.~\eqref{eq:relationSn2.0andSn1.5}, introduces a 
relative error of less than 1\% for frequencies below $10^{-3}\,\rm{Hz}$.
Therefore, in this limit, we can ignore this correction without significantly
affecting the approximation. Thus, at leading order,
\begin{equation}\label{eq:relationSn2.0andSn1.5_simplified}
    S_{\rm{n}}^{A_{2.0}}(f) \approx \left(\frac{4L}{c}\right)^2\left(2\pi f\right)^2 S_{\rm{n}}^{A_{1.5}}(f).
\end{equation}

To study the impact this approximation in the PSD will have on the whitened waveform, we have to take into 
account the relation between the 1.5 and 2.0 generation TDI outputs given by 
Eq.~\eqref{eq:relation_X20_X15_equal_timedelays}. With Eqs.~\eqref{eq:relationSn2.0andSn1.5_simplified} and 
\eqref{eq:relation_X20_X15_equal_timedelays}, and applying the Fourier properties of Eqs.~\eqref{eq:FT_shift_property} and~\eqref{eq:FT_derivative_property},  
the whitened $A_{2.0}$ channel reads

\begin{equation}\label{eq:whitened_A20_A15}
    \frac{\tilde{A}_{2.0}}{\sqrt{S_{\rm{n}}^{A_{2.0}}}} = \frac{\frac{4L}{c} \, (i2\pi f)e^{-4i\pi f L/c} \tilde{A}_{1.5}}{\sqrt{\left(\frac{4L}{c}\right)^2(2\pi f)^2S_{\rm{n}}^{A_{1.5}}}} = 
    ie^{-4i\pi f L/c}\frac{\tilde{A}_{1.5}}{\sqrt{S_{\rm{n}}^{A_{1.5}}}}.
\end{equation}
The oscillatory term $ie^{-4i\pi f L/c}$ preserves the scalar product defined in 
Eq.~\eqref{eq:scalar_product} provided $x$ and $y$
share the same TDI generation. 
The value of the scalar product using 1.5 and 2.0 TDI will be 
approximately the same since the complex conjugation in Eq.~\eqref{eq:scalar_product} makes the oscillatory terms in 
Eq.~\eqref{eq:whitened_A20_A15} 
equal to 1.
In the low-frequency regime the operator ``1.5-generation TDI" and the operator ``2.0-generation TDI" 
are equivalent. 
Therefore, as described in Sec.~\ref{sec:introduction} and App.~\ref{sec:app:lisa_vs_groundbased}, since the scalar product of Eqs.~\eqref{eq:scalar_product} and~\eqref{eq:derived_scalar_product} is 
preserved, the likelihood will also be preserved. 
From the parameter estimation side, as far as the 
TDI generation of the injected signal matches the one for the recovery, using 1.5 TDI or 2.0 TDI will 
give, in this limit, approximately the same posterior distributions. 
This is consistent with previous observations 
reported on \cite{Garcia-Quiros:2025usi, PhysRevD.111.044039}.

\begin{figure*}[htb]
    \centering
    \includegraphics[width=\textwidth]{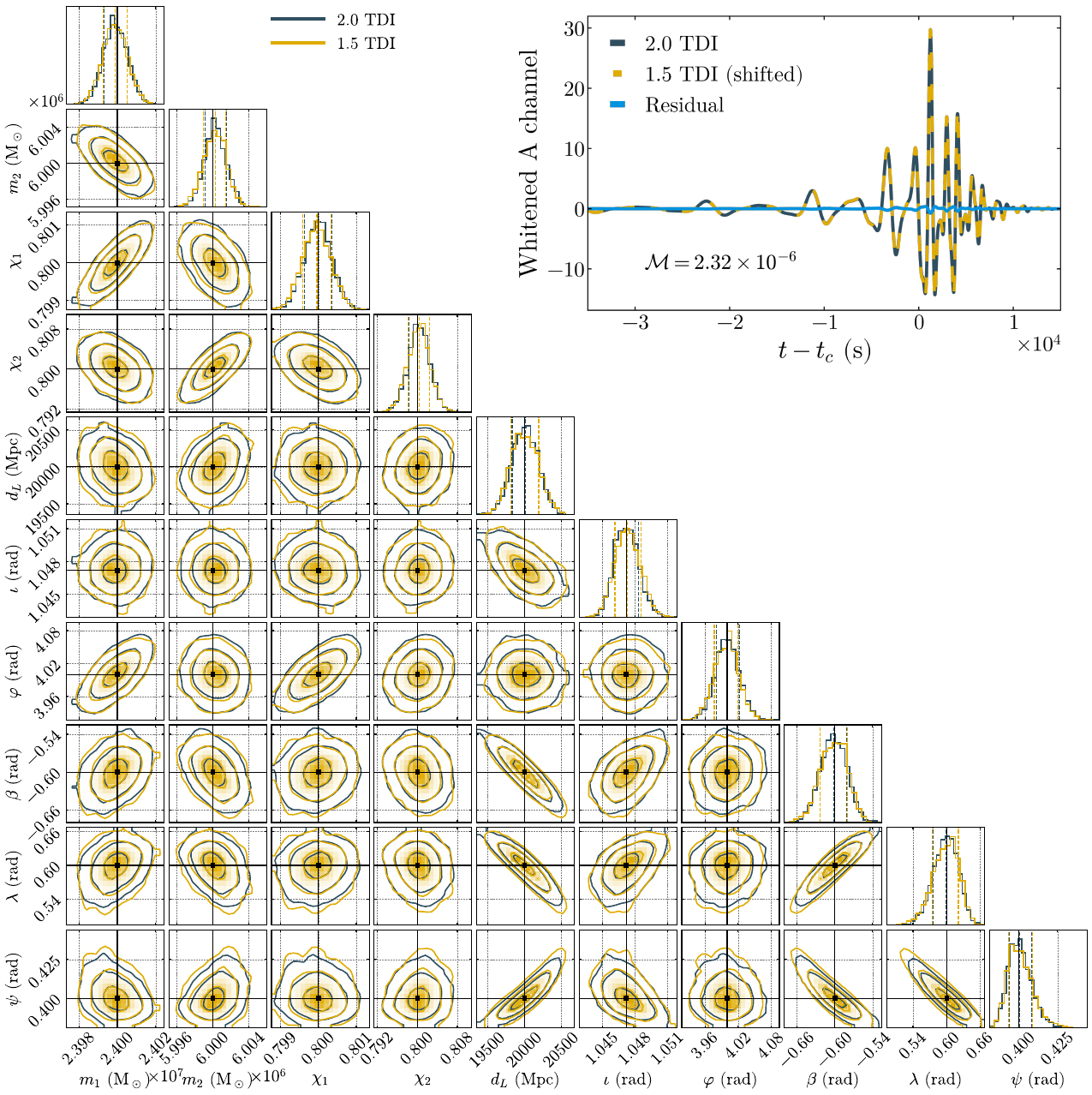}
    \caption{\textit{Corner panel}: Marginalized one- and two-dimensional parameter posterior distributions recovered with the full response adopting 2.0- (gray) and 1.5-generation TDI (gold), for an MBHB injection. The black cross indicates the true injected parameters: $m_1=2.4\times 10^7\,\rm{M}_\odot$, $m_2=6\times 10^6\,\rm{M}_\odot$, $\chi_1=\chi_2=0.8$, $d_L=20\,000\,\rm{Mpc}$, $\iota=1.04720\,\rm{rad}$, $\varphi=4.0\,\rm{rad}$, $\beta=-0.6\,\rm{rad}$, $\lambda=0.6\,\rm{rad}$, and $\psi=0.4\,\rm{rad}$. \textit{Upper-right panel}: Time-domain whitened $A$ output as a function of time to coalescence $t-t_c$, in seconds. The 2.0- (gray, solid) and 1.5-generation (gold, dashed) TDI signals correspond to the inverse FT of the left- and right-hand side of Eq.~\eqref{eq:whitened_A20_A15}, respectively. The residual is shown in blue and the mismatch $\mathcal{M}$ between both whitened waveforms is annotated on the axis.}
    \label{fig:pe_equivalence_12TDI}
\end{figure*}

To test this statement, in Fig.~\ref{fig:pe_equivalence_12TDI} we show the posterior distributions for a PE run with $M=3\times10^{7}\,\rm{M}_{\odot}$ and SNR $\rho=1600$, in which we have injected and recovered with the same TDI generation. 
Additionally, in the upper right panel, we visually demonstrate Eq.~\eqref{eq:whitened_A20_A15}. 
We plot the inverse Fourier transform of the left and right-hand sides of the equality, which correspond to the time-domain whitened signals, and we compute
the mismatch, as defined below Eq.~\eqref{eq:normalized_overlap}, between them.

The posterior distributions shown in the corner panel of Fig.~\ref{fig:pe_equivalence_12TDI} agree across both TDI generations, with similar statistical and systematic errors. 
Furthermore, as illustrated in the upper right panel, the mismatch between the 2.0 TDI and the 1.5 TDI signals, as given in Eq.~\eqref{eq:whitened_A20_A15}, is of the order of $\mathcal{O}(10^{-6})$, which quantifies the goodness of the approximation derived in this section.

\section{Performance on eccentric inspirals}\label{sec:app:non_monochromatic inspiral}

The deviations from the slowly varying pattern observed in the quasi-circular non-precessing case,
amplify the time derivatives of the GW 
strain during the inspiral.
This increases the contribution of the higher-order terms in the expansions 
derived in App.~\ref{sec:app:expansions}.

Examples of these inspirals are, for instance, caused by spin-precession or 
eccentricity. 
In this section, we will focus on eccentric binaries, which produce the emission 
of GW bursts at each periastron passage (see e.g. Fig.~3 of \cite{SEOBNRv5EHM}). 
Given that the $X$, $Y$, and $Z$ Michelson variables are, depending on the TDI generation,
sensitive to the second and third-time derivatives of the strain, 
these bursts will be significantly amplified for LISA relative to the rest of the waveform; thus
strongly breaking the monochromaticity during the inspiral.

To test the hybrid and LFA responses, we generate $h_{+}(t)$ and $h_{\times}(t)$
with \phTE~\cite{PHENOMTEHM} and follow the same projection procedure as for the previously 
studied quasi-circular examples.  
The performance of both modeling approaches is shown in Fig.~\ref{fig:ecc_cpu_visual_agreement_TDI} for 
two eccentric binaries with eccentricities $e_{\rm{gw}}=0.175$ (left panel) and $e_{\rm{gw}}=0.595$ 
(right panel).

\begin{figure*}[t]
    \centering
    \begin{subfigure}[t]{0.5\textwidth}
        \centering
        \includegraphics[width=\columnwidth]{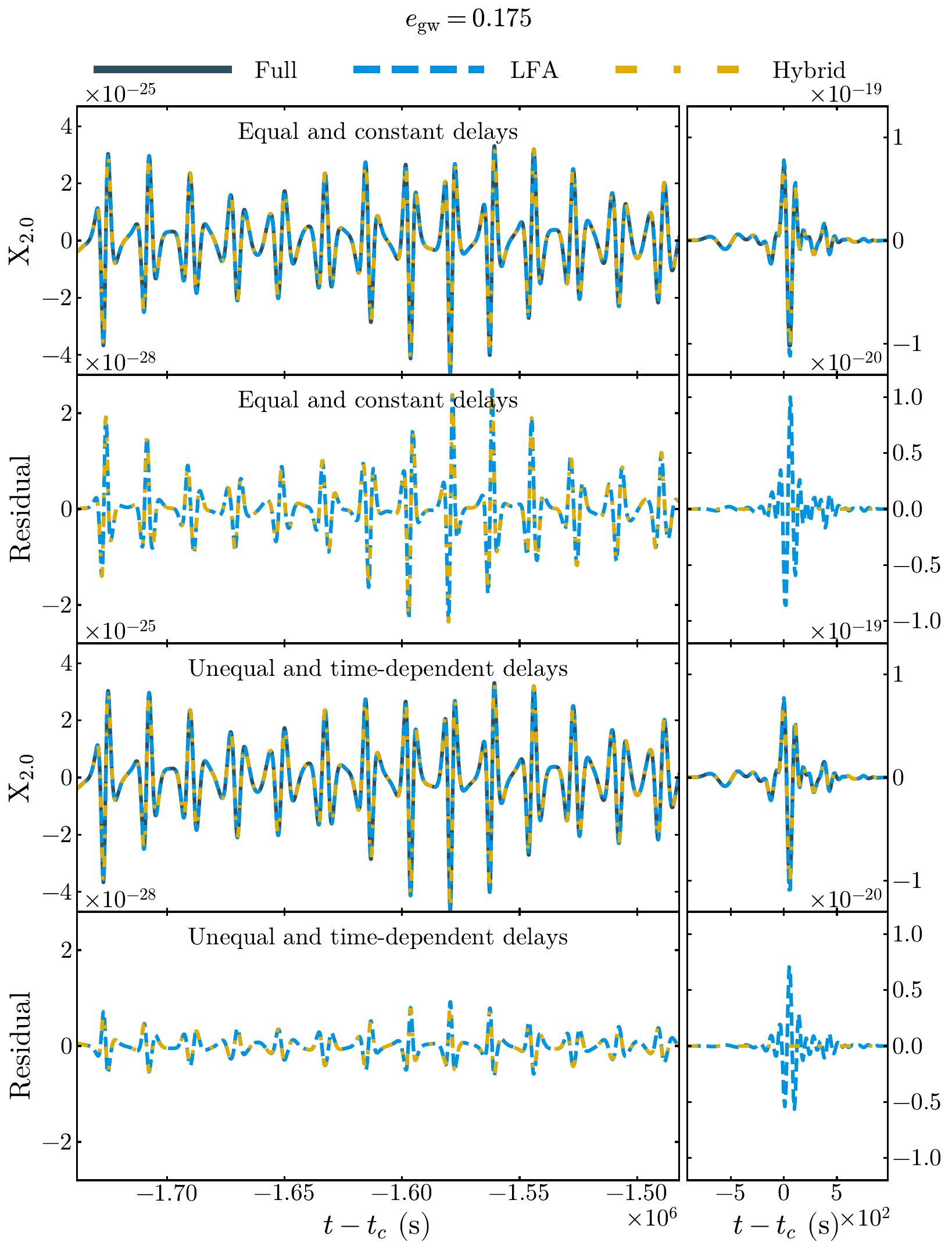}
        \label{fig:lowecc_cpu_visual_agreement_TDI2}
    \end{subfigure}~
    \begin{subfigure}[t]{0.5\textwidth}
        \centering
        \includegraphics[width=0.98\columnwidth]{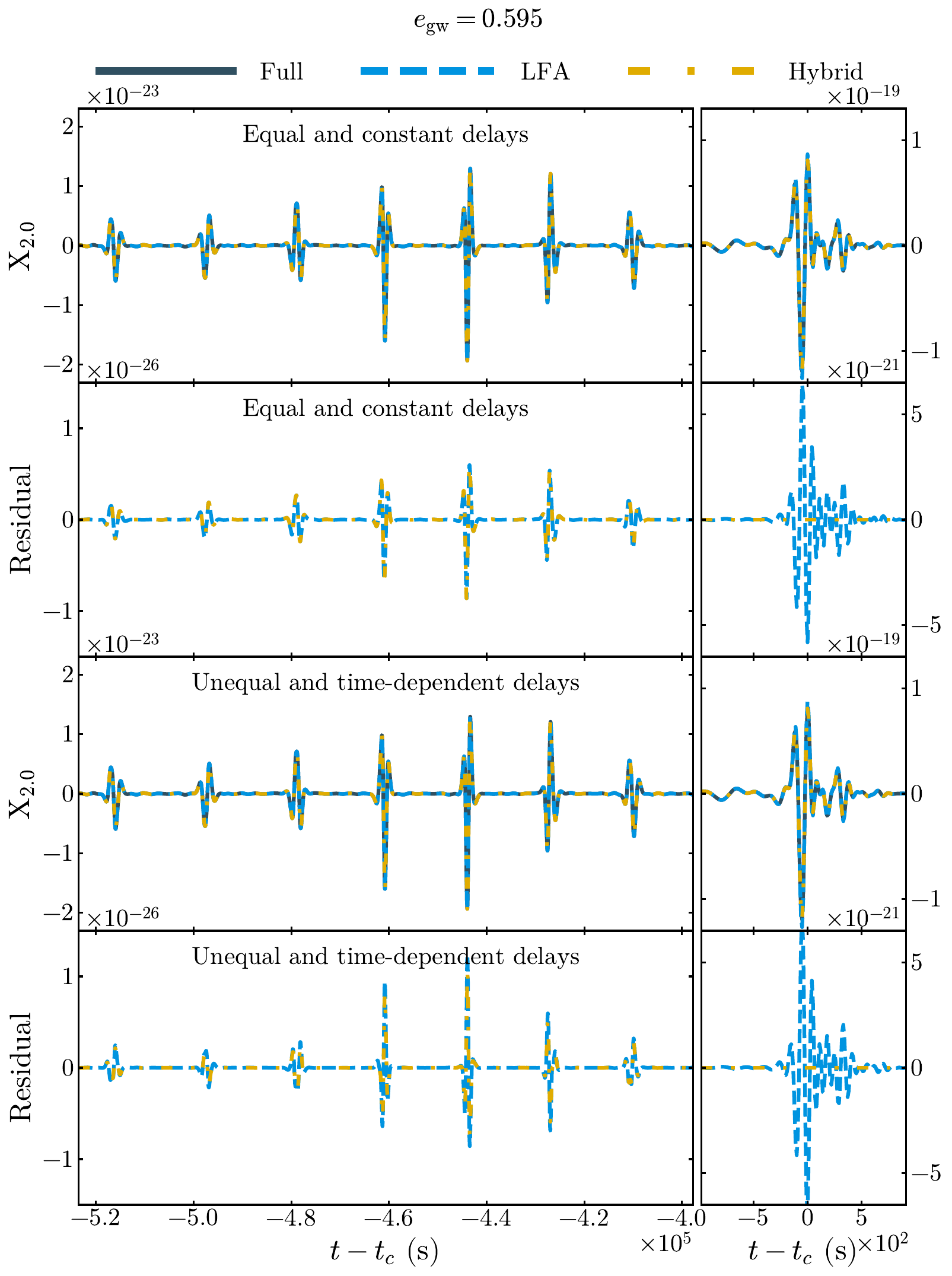}
        \label{fig:highecc_cpu_visual_agreement_TDI2}
    \end{subfigure}
    \caption{Second-generation Michelson combination $X_{2.0}$ and approximation residuals as a function of the time to coalescence $t-t_c$, expressed in seconds, for two eccentric inspiral-merger-ringdown waveforms generated with \phTE.
    The results are computed with the CPU implementation of the full (black, solid), LFA (blue, dashed), and hybrid (gold, dash-dotted) responses, for an MBHB with GW eccentricities $e_{\rm{gw}}=0.175$ (left panel) and $e_{\rm{gw}}=0.595$ (right panel), GW mean anomaly $l_{\rm{gw}}=0\,\rm{rad}$. The rest of the parameters are identical to the example shown in Fig.~\ref{fig:cpu_visual_agreement_TDI}: $M=3\times 10^6\,\rm{M}_\odot$, $q=4$, $\chi_1=\chi_2=0.8$, $\iota=\pi/3\,\rm{rad}$, $d_L=50\,\rm{Gpc}$, $\varphi=4.0\,\rm{rad}$, $\beta=0.6\,\rm{rad}$, $\lambda=\pi\,\rm{rad}$, and $\psi=0.25\,\rm{rad}$. 
    For the hybrid response we have used $f_{\rm{Hyb}}=3\times10^{-4}\,\rm{Hz}$, $T=800\delta t$, and $\sigma=T/40$, with $\delta t=10\,\rm{s}$.
    In the first two rows, the LFA and hybrid $X$ output is computed using an equal and constant delay configuration (Eq.~\eqref{eq:X20_equal_timedelays}), whereas the last two rows adopt an unequal and time-dependent delay configuration (Eq.~\eqref{eq:X20_nonequal_timedelays}). For all cases, the full response is consistently computed using unequal and time-dependent delay configuration (Eq.~\eqref{eq:X2.0}). The corresponding residuals, defined as $X_{\rm{Full}}-X_{\rm{LFA/Hybrid}}$, are displayed below each panel showing the $X$ observable. 
    }
    \label{fig:ecc_cpu_visual_agreement_TDI}
\end{figure*} 

In Fig.~\ref{fig:ecc_cpu_visual_agreement_TDI}, for the low eccentricity example (left panel) and under 
the assumption of equal and constant delays, 
the approximations can reproduce the full LISA response during the early inspiral 
phase with an accuracy of approximately $0.1\%$. 
As opposed to the quasi-circular case (compare with Fig.~\ref{fig:cpu_visual_agreement_TDI}), 
the adoption of a more realistic, time-varying delay configuration does not yield 
an order-of-magnitude reduction in the error. 
Instead, the improvement is limited to roughly a factor of two.
This reduced gain in accuracy arises from the increased influence of error terms,
which are proportional to time derivatives of the strain.
Consequently, error contribution associated with assuming a static arm length (see Eq.~\eqref{eq:X20_nonequal_timedelays}) start to 
become subdominant as the non-monochromaticity of the waveform in the inspiral phase accentuates.

In the right panel of Fig.~\ref{fig:ecc_cpu_visual_agreement_TDI}, 
we show the results for a higher eccentricity binary with $e_{\rm{gw}}=0.595$.
We observe the projected signal has a pulse-like morphology. 
This is because, high-eccentricity configurations, $e_{\rm{gw}}\gtrsim 0.4$,
are characterized by high angular velocities close to the periastron;
thus, concentrating the GW emission into much shorter duration bursts.
As a result, according to Eq.~\eqref{eq:X20_thrid_strain_derivative},
the action of the third-time derivative operator on the GW polarizations
produces a significantly larger amplification of the GW bursts compared 
to that of the waveform segments between successive periastron passages, 
which evolve on a much slower time scale.
Although this pulse-like signal is far from being monochromatic, we observe the residual errors
from the two algorithms are around three orders of magnitude below the projected signal.
In contrast to the low eccentricity case, here we do not see any accuracy improvement between
different time-delay configurations.
This is because the error from assuming unequal and constant arm lengths has been dominated by the rest of the errors, which are proportional to strain derivatives.

We expect to reduce the errors by including next-to-leading order terms in the expansions derived in 
App.~\ref{sec:app:expansions}.

\end{widetext}

\bibliography{references}
\end{document}